\documentclass[12pt]{article}
\usepackage{amsmath}
\usepackage{graphicx}

\def\s#1{\setbox0=\hbox{$#1$}%
\rlap{\ifdim\wd0>.7em\kern.22\wd0\else\kern.1\wd0\fi /}#1}

\newcommand{\beq}{\begin{equation}}
\newcommand{\eeq}{\end{equation}}
\newcommand{\bea}{\begin{eqnarray}}
\newcommand{\eea}{\end{eqnarray}}

\newcommand{\tto}{\!\to\!}

\newcommand{\gsim}{\lower.7ex\hbox{$
\;\stackrel{\textstyle>}{\sim}\;$}}
\newcommand{\lsim}{\lower.7ex\hbox{$
\;\stackrel{\textstyle<}{\sim}\;$}}

 \newcommand{\bibit}[1]{\bibitem{#1}}

\newcommand{\aver}[1]{\langle #1\rangle}

\newcommand{\La}{\overline{\Lambda}}

\newcommand{\mhad}{\mu_{\rm hadr}}

\newcommand{\as}{\alpha_s}
\newcommand{\GeV}{\,\mbox{GeV}}
\newcommand{\MeV}{\,\mbox{MeV}}
\newcommand{\matel}[3]{\langle #1|#2|#3\rangle}
\newcommand{\state}[1]{|#1\rangle}

\newcommand{\eod}{\end{document}}

\newcommand{\msp}[1]{\mbox{\hspace*{#1mm}~}}

\begin{document}

\begin{titlepage}
\begin{flushright}
SI-HEP-2011-15 \\UND-HEP-12-BIG\hspace*{1.5pt}10 \\[0.2cm]
\end{flushright}

\vspace{1.7cm}
\begin{center}
{\LARGE\bf 
 \boldmath $B\tto D^*$ zero-recoil formfactor and the  \vspace*{12pt} heavy
quark expansion in QCD: a systematic study}
\end{center}

\vspace{0.5cm}
\begin{center}
{\large {\sc Paolo Gambino}} \\[7pt]
{\sf  Universit\`a di Torino, Dipartimento di Fisica \\ [-1.pt] 
{\sf and} \\[-1pt] INFN Torino, I-10125 Torino, Italy} \\[25pt]
{\large{\sc Thomas Mannel, ~Nikolai Uraltsev}}\raisebox{4pt}{$^{a,*}$} \\[8pt]
{\sf Theoretische Physik 1, Fachbereich Physik,
Universit\"at Siegen,  D-57068 Siegen, Germany}\\[10pt]
\scalebox{.935}{\sf  ~$^a$\,{\rm also}\, Department of Physics, 
University of Notre Dame  du Lac, Notre Dame, IN 46556, U.S.A.}
\end{center}

\vspace{0.8cm}
\begin{abstract}
\vspace{0.2cm}\noindent 
We present a QCD analysis of  heavy quark mesons focussing on
the $B \tto D^*$ formfactor at zero recoil, ${\cal F}_{D^*}(1)$. 
An advanced treatment of the perturbative corrections in the Wilsonian
approach is presented. We estimate the higher-order power corrections 
to the OPE sum rule and describe a refined analysis of the
nonresonant continuum contribution.
In the framework of a model-independent approach,
we show that the inelastic contribution in the phenomenological part of
the OPE is related to the $m_Q$-dependence of the hyperfine splitting
and conclude  that the former is large, 
lowering the
prediction for ${\cal F}_{D^*}(1)$ down to about $0.86$. 
This likewise implies an enhanced  yield of radial and
$D$-wave charm excitations in semileptonic $B$ decays and alleviates the problem with
the inclusive yield of the wide excited states.  
We also apply the approach to the expectation values of dimension 7 and 8
local operators and to a few other issues in the heavy quark
expansion. 

\end{abstract}

\vfill

\noindent
\hrulefill\hspace*{290pt} \\[1pt]
$^*$\,\scalebox{.82}{On leave of absence from Petersburg Nuclear 
Physics Institute, Gatchina, St.\,Petersburg 188300, Russia}

\thispagestyle{empty}

\setcounter{page}{0}

\end{titlepage}

\tableofcontents

\section{Introduction}

A precision determination of the parameters of the Standard Model (SM) is
mandatory for  a stringent test of the theory and eventually to identify the effects of
new physics. Decades of large-scale experimental efforts have
been devoted to testing the SM. At high energies LEP has examined the gauge
structure of the SM with great  
precision without any serious hint of effects beyond the SM.
The flavor structure of the SM has been verified in many ways, 
and in particular at the $B$ factories of SLAC and KEK. In both respects, 
the flavor and the gauge structure, the SM has so far passed all the tests.

A  sound theoretical foundation is needed to perform such tests along with
precise data.  Methods have been developed in heavy flavor physics over the
last two decades that allow systematic and controllable calculations. As an
example, the determination of the CKM matrix element $V_{cb}$ employing the
heavy mass expansion has reached a relative theoretical uncertainty of about
two percent; its extraction from inclusive semileptonic decays is presently
believed to be most precise.

The determination of $V_{cb}$ from the exclusive decay $B \to D^* \ell
\bar{\nu}$ requires the knowledge of the form factor at the  zero-recoil point
where the velocities of the initial and final states are equal, $v\cdot v' \!=\! 1$.
In the heavy quark limit the form factor 
is normalized to unity owing to
the heavy quark symmetries; the precision determination of $V_{cb}$ requires
to control the deviation from the symmetry limit.

The zero-recoil version of the heavy quark sum rules was proposed back
in the 1990s to estimate the scale of the nonperturbative corrections, which
turned out significant, even considering that the effects are 
driven by the moderate charm mass. 
As soon as the first experimental $B \to D^* \ell
\bar{\nu}$ data were available, these early analyses suggested a 
value of  $V_{cb}$ in agreement with the inclusive value, but 
were unable to make precise predictions because of the poor knowledge of the
important heavy quark hadronic parameters.

A model-independent accurate calculation of the formfactor may be expected from QCD
lattice simulations 
provided they measure directly its  deviation 
from unity.   The value for the form 
factor ${\cal F}(1)$ obtained in existing lattice calculations
leads to smaller $V_{cb}$ compared to the values from the inclusive
determination. The latter meanwhile has become quite mature both
experimentally and theoretically. 

 The significant progress in our  understanding of heavy mesons 
brought about by all the precision data in the $B$ sector
calls for a reappraisal of the 
heavy quark expansion for  ${\cal F}(1)$. 
In particular, it allows us to obtain a prediction for  ${\cal F}(1)$ with an informative and defendable error estimate.

In the present paper we discuss the technical 
details of the analysis that led to the results reported in Ref.~\cite{f0short}.
Our estimate for the central value of the formfactor at
$v\cdot v' \!=\!1$ is appreciably  
lower than  lattice estimates,  which in turn appear
marginally  compatible with the  unitarity bound.
Consequently, the value of
$V_{cb}$ extracted using our formfactor is larger and  happens to be 
close to the inclusive one, although 
it still suffers from larger theory uncertainty. 

In the course of the analysis of the zero-recoil sum rule 
we have found a fruitful link among three apparently unrelated  topics 
in  heavy meson phenomenology: the value of ${\cal F}(1)$,  the
hyperfine splitting in $B$ and $D$, and the `$\frac{1}{2}\!>\!\frac{3}{2}$'
puzzle.  In all three cases the experimental data are naturally described  
if the transition amplitudes into the
`radial' and/or $D$-wave states are more significant than in the naive quark
models.

The paper is divided in two parts. The first part  
describes the principal ingredients required for the evaluation of
${\cal F}(1)$. Our general approach, outlined in Sect.~\ref{approach},
is essentially 
an implementation of the Heavy Quark sum rules laid down in
Ref.~\cite{optical}. Technically, we formulate it  in  a somewhat different way
which has several advantages.
To eventually sharpen the estimate, a number of points had to be improved. 

First,  the perturbative corrections 
should be calculated with the hard cutoff consistent with the Wilsonian
`kinetic' renormalization scheme. This is a nontrivial point which 
caused confusion in the past literature. We present a method to obtain 
the leading $\alpha_s$-corrections together with all
BLM-improvement terms as a 
function of the cutoff scale $\mu$, avoiding an expansion in $\mu/m_Q$. This is the
subject of Sect.~\ref{pert} where the numeric results 
are also presented, including known second-order non-BLM terms.

Section~\ref{powercor} discusses power corrections in the short-distance
expansion of the scattering amplitude off the heavy quark; these correspond to
the power corrections in the sum rules. We find a noticeable impact of the
$1/m_c^3$ corrections lowering ${\cal F}(1)$, and estimate higher-order effects in
the framework of the recent analysis \cite{hiord} of $1/m_Q^4$ and
$1/m_Q^5$ corrections in the inclusive decays. 

Section~\ref{winel} presents the novel evaluation of the inelastic contributions
to the unitarity relation, which uses as an input the 
hyperfine splitting of $B$ and $D$ mesons. 
We found that the usual nonlocal $D\!=\!3$ correlators for heavy quark
mesons are rather large, and this yields a large inelastic contribution which 
significantly exceeds the naive estimates of the past. This in turn lowers the expected
central value of ${\cal F}(1)$ down to about $0.86$. We have also refined the
estimate of the $D^{(*)}\pi$ 
continuum contribution to the inelastic contribution,
which supports its overall numerical significance. 
Independently of the hyperfine constraints, the continuum contributions
lowers  the unitarity upper bound for ${\cal F}(1)$. 

The conclusions  of the first part focussing on ${\cal F}(1)$ are  
    summarized in Sect.~\ref{conc}, with the numeric values given in Eqs.~(\ref{c20}), (\ref{c21}) and (\ref{c24}).

Section~\ref{alook} opens the second, more theoretical part, where  we
scrutinize, in a model-independent way, the higher excited heavy quark states
contributing to the inelastic transitions. In the heavy quark limit these
belong to the radial or $D$-wave states. This is the 
framework which allowed us to relate the mass dependence of
the hyperfine splitting to ${\cal F}(1)$ in an informative way.
Within the same framework we could link the observed
enhancement of the nonlocal effects to the significant inclusive yield of
higher-excited states beyond $D$, $D^*$ and their $P$-wave excitations,
thereby substantiating a possible  resolution of the so-called `$\frac{1}{2} \! > \!
\frac{3}{2}$' puzzle. As a byproduct of the model-independent description we
analyze the nonfactorizable effects in the higher-dimension expectation
values, and we give  numeric estimates for some representative combinations.
The non-resonant $D^{(*)}\pi$ continuum has been studied in the same
heavy-quark limit setting, relying on the soft-pion approximation.

In Section~\ref{disc} we discuss certain theoretical aspects related to both the
present and the lattice analyses of ${\cal F}(1)$. We also briefly mention what can
be gained applying our approach to the vector $B\tto D$ transitions.
The main conclusions are summarized in  Sect.~\ref{conc}.

The Appendices contain several of the details omitted from the main text. They mostly concern the perturbative calculations; there we discuss some of the conceptual aspects 
and also  give the concrete expressions required in the analysis of ${\cal F}(1)$.

\section{The framework} 
\label{approach}

\begin{figure}[t]
\begin{center}
 \includegraphics[width=12.cm]{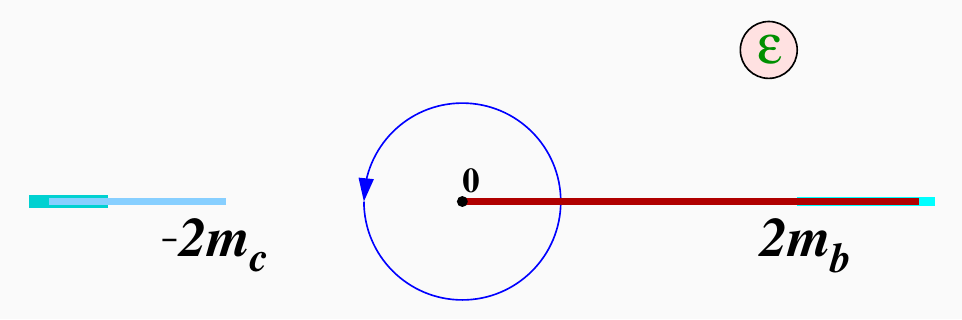}
\end{center}
\caption{The analytic structure of $T^{\rm zr}(\varepsilon)$ 
and the integration contour yielding the sum rule. Distant cuts are
shown along with the physical cut. The radius of the circle is 
$\varepsilon_{M}$.}
\label{fig1}
\end{figure}

We consider the zero-recoil ($\vec q \!=\!0$) forward scattering amplitude
$T^{\rm zr}(\varepsilon)$ of the flavor-changing axial current $\bar{c}\vec\gamma
\gamma_5 b$ off a $B$ meson at rest:
\beq
T^{\rm zr}(\varepsilon)=\int\! {\rm d}^3x
\int\! {\rm d}x_0\; e^{-i x_0 (M_B - M_{D^*} - \varepsilon) }
\frac{1}{2M_B} \matel{B}{\mbox{$\frac{1}{3}$}\:{ iT}\,
  \bar{c}\gamma_k\!\gamma_5b(x)\:\bar{b}\gamma_k\!\gamma_5 c(0)}{B}\, , 
\label{80}
\eeq
where $\varepsilon$ is the excitation energy above $M_{D^*}$
in the $B \tto X_c $ 
transition (the point $\varepsilon=0$ corresponds to 
the elastic $B\to D^*$ transition). 
The amplitude $T^{\rm zr}(\varepsilon)$ is
an analytic function of $\varepsilon$ and has a physical decay cut at
$\varepsilon\!\ge\!0$, and other `distant' singularities at 
$|\varepsilon|\!\gsim\!2m_c$. The 
analytic structure of $T^{\rm zr}(\varepsilon)$ 
is illustrated by Fig.~\ref{fig1}. 

We consider the contour integral 
\beq
I_0(\varepsilon_{M})= -\frac{1}{2\pi i}\;
\oint_{\raisebox{-3pt}{\hspace*{-5pt}\scalebox{.7}{$|\varepsilon|\!=\!
\varepsilon_M$}}\hspace*{-5pt} }T^{\rm zr}(\varepsilon) \,{\rm d}\varepsilon
\label{82}
\eeq
with the contour running counterclockwise from the upper side of the
cut, see Fig.~\ref{fig1}. 
Using the analytic properties of $T^{\rm zr}(\varepsilon)$
the integration contour can be shrunk onto the decay cut; the
discontinuity there is related to the weak transition amplitude
squared of the axial current into the final charm state with mass
$M_X\!=\!M_{D^*}\!+\!\varepsilon$. If we explicitly single out the
elastic transition contribution $B\tto D^*$ at $\varepsilon\!=\!0$ then
\beq
I_0(\varepsilon_{M})= {\cal F}^2(1)+I_{\rm inel}(\varepsilon_{M}), 
\qquad I_{\rm inel}(\varepsilon_{M}) \!\equiv\!
\frac{1}{2\pi i}\;  \int_{\varepsilon>0}^{\varepsilon_{M}}
{\rm disc}\,  T^{\rm zr}(\varepsilon) \,{\rm d}\varepsilon
\label{84}
\eeq
holds, where $I_{\rm inel}(\varepsilon_{M})$ is related to the sum of the
differential decay probabilities, in the zero recoil kinematics, 
into the excited states with mass up to $M_{D^*}\!+\!\varepsilon_{M}$. 

The OPE allows us to calculate the amplitude in (\ref{80}) -- and hence
$I_0(\varepsilon_{M})$ -- in the short-distance expansion provided
$|\varepsilon|$ is sufficiently large compared to   the ordinary 
hadronic mass scale. It
should be noted that strong interaction corrections are driven not only by
$|\varepsilon|$, but also by the proximity to distant singularities.
Therefore, $\varepsilon_{M} $ cannot be taken too large either, and the hierarchy
$\varepsilon_{M}\!\ll\!2m_c$ has to be observed.

The sum rule Eq.~(\ref{84}) can be cast in the form
\beq
 {\cal F}(1)= \sqrt{I_0(\varepsilon_{M})\!-\!I_{\rm inel}(\varepsilon_{M})}
\label{86}
\eeq
which is the master identity for the considerations to follow. Since 
$I_{\rm  inel}(\varepsilon_{M})$    is strictly positive, we get an upper bound 
on the formfactor 
\beq
{\cal  F}(1) \le \sqrt{I_0(\varepsilon_{M})}
\label{86a}
\eeq
which relies only on the OPE calculation of $I_0$. Note that this bound
depends on the parameter $\varepsilon_{M}$, while  Eq.~(\ref{86}) is
independent of $\varepsilon_{M}$ since the dependence in $I_0$ and 
$I_{\rm  inel}$ cancel.  Furthermore, including an estimate 
of $I_{\rm inel}(\varepsilon_{M})$ we obtain an evaluation of ${\cal F}(1)$.

The correlator in  Eq.~(\ref{80}) can be computed using the OPE, resulting
in an expansion of $T^{\rm zr}(\varepsilon)$ in inverse powers of the masses
$m_c$ and $m_b$. This yields  the corresponding expansion of 
$I_0(\varepsilon_{M})$. This OPE takes the following general form: 
\bea 
\label{88}
I_0(\varepsilon_{M}) &=& \xi_A^{\rm pert}(\varepsilon_{M},\mu) + 
\sum_{k} C_k(\varepsilon_{M},\mu)
\, \frac{\mbox{$\frac{1}{2M_B}$}\matel{B}{O_k}{B}_\mu}{m_Q^{d_k-3}} \\ 
\nonumber 
&=&
\xi_A^{\rm pert}(\varepsilon_{M},\mu)-\Delta_{1/m_Q^2}(\varepsilon_{M},\mu)
-\Delta_{1/m_Q^3}(\varepsilon_{M},\mu)-
\Delta_{1/m_Q^4}(\varepsilon_{M},\mu)-...\\ \nonumber
&\equiv& \xi_A^{\rm pert}(\varepsilon_{M},\mu)-\Delta^A(\varepsilon_{M},\mu), 
\eea
where $O_k$ are local $b$-quark operators $\bar{b}...b$ of increasing dimension
$d_k\!\ge\!5$,  $\;C_k(\varepsilon_{M},\mu)$ are Wilson coefficients 
for power-suppressed terms, 
and $\xi_A^{\rm pert}$ is the short-distance  renormalization (corresponding to the 
Wilson coefficient of the unit operator), which is unity at tree level. 
We have also introduced a Wilsonian cutoff $\mu$ used to separate long and short 
distances. The complete result  does not depend 
on $\mu$ since the $\mu$-dependence cancels between the Wilson coefficients 
and the matrix elements of the operators. 
At tree level $\Delta_A$ does not depend on $ \varepsilon_{M}$. 
The choice of $\mu$ is subject to the same general constraints as that 
of $\varepsilon_M$, and therefore we will often  set $\mu\!=\!\varepsilon_{M}$, 
in which case we will also use $\xi_A^{\rm pert}(\mu)\equiv \xi_A^{\rm pert}(\mu,\mu)$. 

\section{Perturbative corrections}
\label{pert}

The leading perturbative renormalization factor $\xi_A^{\rm  pert}(\varepsilon_M,\mu)$
can be expanded in powers of
$\alpha_s$. In the Wilsonian OPE all infrared physics is removed
from perturbative corrections; the perturbative series for
$\xi_A^{\rm  pert}$ is then free from infrared renormalons. The
exact form of the perturbative coefficients depends on the concrete
definition of the higher-dimension operators used in the OPE; we
consistently assume the scheme of Refs.~\cite{five,dipole} often
referred to as ``kinetic'' (or Small Velocity); see also
Ref.~\cite{chrom}. Here we describe a compact method to calculate 
$\xi_A^{\rm  pert}$ in this scheme; the details as well as the
justification of the method can be found in  Appendix~\ref{oneloopwils}.

For the theoretical analysis we need to keep $\mu$ and $\varepsilon_M$
distinct. Their role is different: $\varepsilon_M$ specifies the observable
$I_0(\varepsilon_M)$ under study, the energy integral of the spectral density.
$I_0(\varepsilon_M)$ depends on $\varepsilon_M$, and the same applies to the
Wilson coefficient $\xi_A^{\rm pert}$ for its leading term in the OPE. On the
contrary, $\mu$ is a technical tool employed in the OPE to separate `soft' and
`hard' physics. Since $I_0(\varepsilon_M)$ should not depend on $\mu$, the
same applies to the right-hand side of Eq.~(\ref{88}).  While 
$\xi_A^{\rm pert}$ does depend on $\mu$ once the  perturbative 
corrections are included, this
dependence is canceled by the $\mu$-dependence of the power-suppressed
expectation values.

The $\varepsilon_M$-dependence of $I_0$ 
is governed by the inelastic spectral density $w_{\rm inel}$: 
\beq
\frac{{\rm d}I_0(\varepsilon_M)}{{\rm d}\varepsilon_M} = 
\frac{{\rm d} I_{\rm
    inel}(\varepsilon_M)}{{\rm d}\varepsilon_M}\equiv w_{\rm inel}(\varepsilon_M).
\label{51a.2}
\eeq
The same holds for the $\varepsilon_M$-dependence of  
$\xi_A^{\rm  pert}(\varepsilon_M, \mu)$ 
as long as only the perturbative part of
the inelastic spectral density is considered:\footnote{This is true in 
general for  $\varepsilon_M \!\gg \mu$, or if we neglect $O(\as)$ contributions to the 
Wilson coefficients of the power suppressed operators. }
\beq
\frac{{\rm d}\xi_A^{\rm pert}(\varepsilon_M, \mu)}{{\rm d}\varepsilon_M} = 
w^{\rm pert}(\varepsilon_M )  ,
\label{51a.4}
\eeq
where $w^{\rm pert}(\varepsilon)\equiv w^{\rm pert}_{\rm inel}(\varepsilon)$.

Although both $\varepsilon_M$ and $\mu$  should be sufficiently large, 
in a fixed-order perturbative expression one can formally set them 
equal to zero in $\xi_A^{\rm  pert}(\varepsilon_M, \mu)$. It is evident from the
definition of $I_0(\varepsilon_M)$ that this would yield 
a `purely perturbative'  renormalization factor $\eta_A$
for the zero-recoil axial current to this order, as it is 
commonly defined in HQET:
\beq
\xi_A^{\rm pert}(0,0)=\eta_A^2.
\label{51a.8}
\eeq
We need, however to keep both mass parameters nonvanishing.
 Eq.~(\ref{51a.4}) allows one to pass to non-vanishing $\varepsilon_M$. In general, the 
  $\mu$-independence of the overall result is used to fix $\xi_A^{\rm pert}$ at non-vanishing $\mu$.

 A relation similar to Eq.~(\ref{51a.8}) holds for a physical
choice of the arguments in $\xi_A^{\rm pert}$ if the perturbative
calculation of the renormalization factor $\eta_A$ is performed with an infrared
cutoff $\mu$, and provided $\mu$ is close to  $\varepsilon_M$. 
There is also a perturbative contribution to $I_{\rm inel}$, 
whose soft part is removed by the same cutoff. The
precise relation is not trivial, however, even at one loop. 
An extensive discussion is given in Appendix~\ref{oneloopwils}.

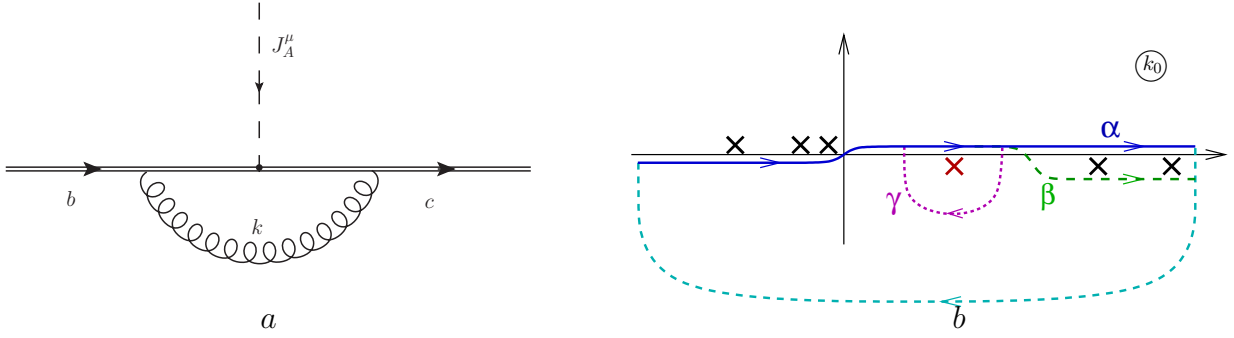
\begin{figure}[t]
\vspace*{10pt}
 \begin{center}
\raisebox{17pt}{\includegraphics[width=7cm]{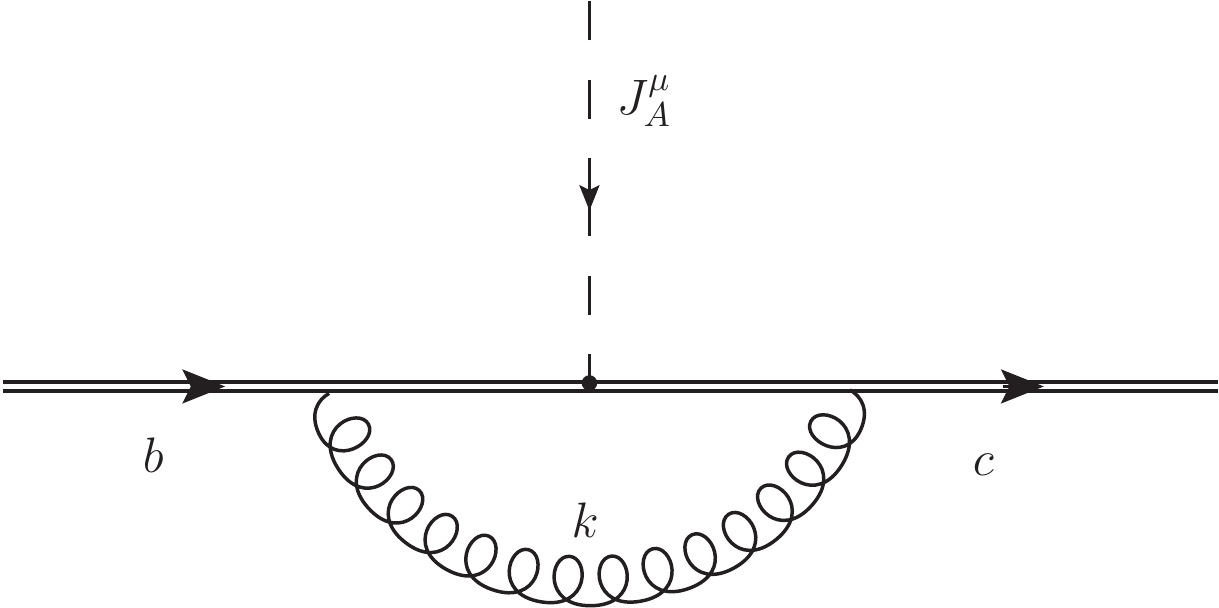}
}\hspace*{30pt}
\includegraphics[width=8cm]{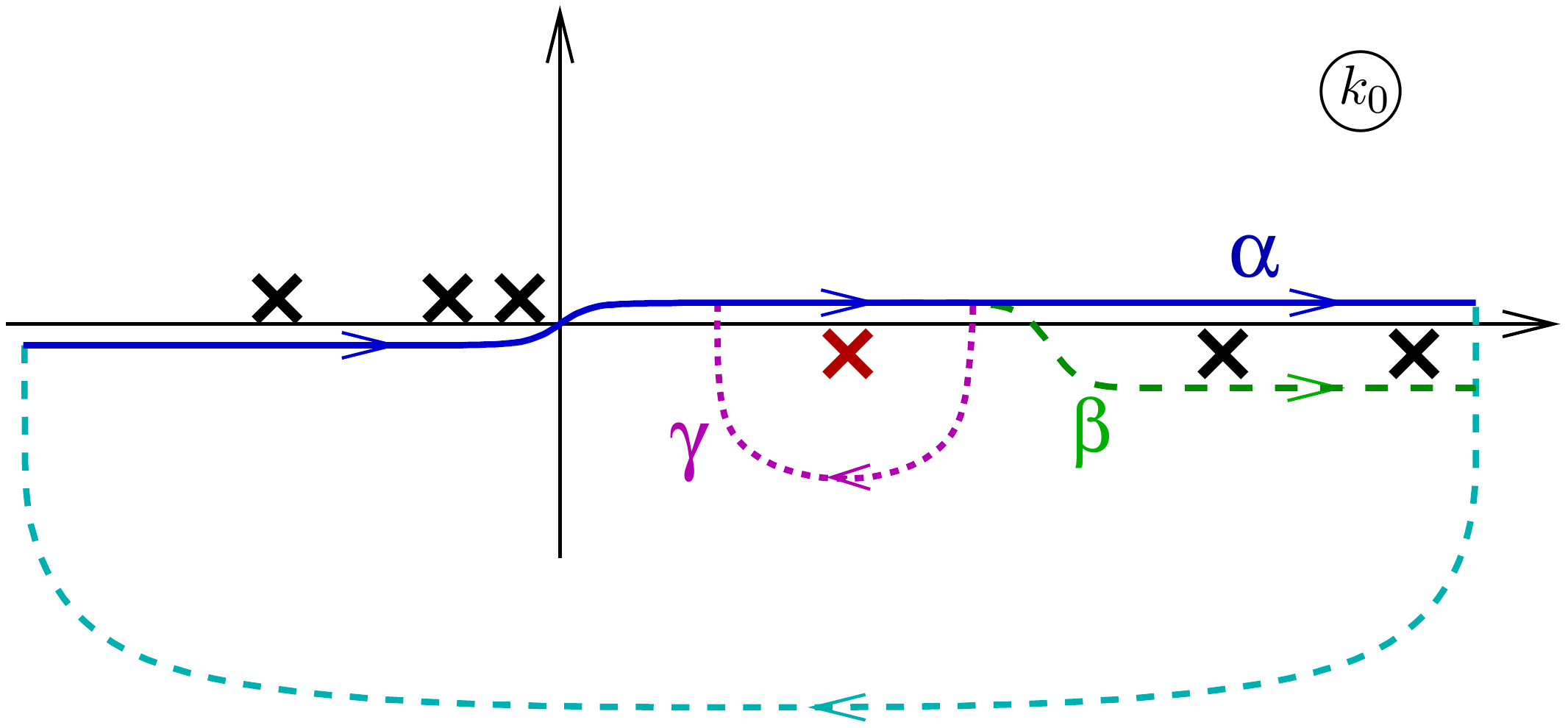}\begin{picture}(0,0)(0,0)
\put(-129, -2.5){$a$}\put(-37, -2.5){$b$}
\end{picture}
 \end{center}
\caption{ \small $a)$ Vertex diagram contributing to $\eta_A$.
$b)$ poles in the complex $k_0$-plane encountered in the one-loop diagrams. The
standard Feynman propagators lead to contour $\alpha$ (blue); upon
closing it into
the lower half-plane the integral equals  the sum of the three
residues. The calculation of $\delta \eta_A^{\rm soft}(\mu)$ requires
to pick up only the
`near' gluon pole (red), and is obtained by
integration over contour $\beta$ (the bypass in green).}
\label{diagr}
\end{figure}

A direct  calculation of the Wilson coefficients implies an infrared cutoff on
the internal gluon momentum in the Feynman diagrams. The contribution of soft
gluons is subtracted from $\eta_A$ by a 
term $\delta \eta_A^{\rm soft}$.
In one-loop diagrams,  see Fig.~\ref{diagr}, the kinetic scheme cuts off the Feynman integral at $|\vec
k|\!<\!\mu$ yet the integration over $k_0$ is always performed from $-\infty$ to
$\infty$. While the integral is dominated by small $k_0\!\lsim \!|\vec k|$ , a
power-suppressed contribution comes also from large $k_0\!\sim\! m_Q$. This
piece is not soft and would not be properly accounted for if attributed to the
matrix elements in the OPE.

This brings us to an important point: the  subtraction of the full
$|\vec k|\!<\!\mu$  one-loop contribution 
to $\eta_A$ 
would not yield the correct Wilson
coefficient, leading to spurious terms, formally 
of order ${\cal O}(\mu^3/m_Q^3)$.
The correct subtraction includes only the residue of
the `near' gluon pole at $k_0\!=\!|\vec k|\!-\!i0$  in the calculation 
of the integral over
$k_0$. The other two residues encountered in the conventional Feynman diagram
calculation upon closing the $k_0$ contour into the lower half-plane, see
 Fig.~\ref{diagr}b, correspond in fact to hard physics with  gluon
virtuality  $\sim \!2m_Q$ and must be left to the hard Wilson 
coefficients.

\thispagestyle{plain}

The different physics associated with the `distant' poles in the diagram
located near $k_0\!=\!2m_Q\!+\!\vec k^{\,2}/2m_Q$ can be intuitively understood:
picking up the corresponding residue would leave the gluon propagator hard, 
$k^2 \!\approx \! 4m_Q^2$ for small $\vec k$. These poles are actually related to
the divergence of the power expansion in $\vec k/m_Q$ when the soft scale
$\mu$  is increased towards $m_Q$.  They contribute terms  
$\sim \!{\rm d}^3 \vec{k}/m_Q^3 \propto \mu^3/m_Q^3$ and are seen starting at
order $1/m_Q^3$ in the OPE.
A detailed discussion of this technical point is given in
Appendix~\ref{oneloopwils}. 
The validity of this
prescription in the kinetic scheme is explicitly checked by the $\mu$-independence
of the OPE relations  beyond order $1/m_Q^2$. The matching of the 
$\mu$-dependence of $\xi_A$ to that of the OPE expectation values is
demonstrated in Appendix~\ref{pertdetails}. We have also verified  this for the
$1/m_Q$ expansion of the heavy hadron mass. 

\begin{figure}[t]
\vspace*{-7pt}
 \begin{center}
\includegraphics[width=8cm]{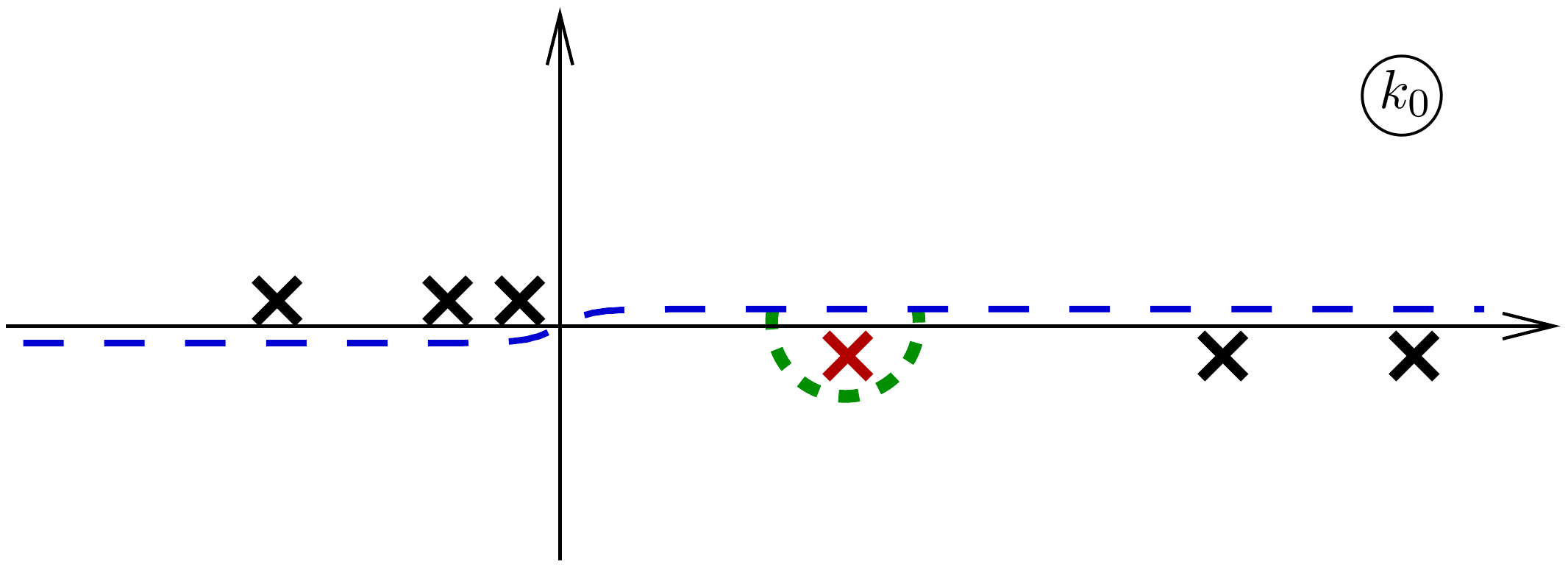}
 \end{center}\vspace*{-5pt}
\caption{ \small
The poles in the integrand and the integration contour in the complex
$k_0$ plane for calculating the Wilson coefficient. At $\vec{k}^{\,2}\!+\!\lambda^2 \!>\!\mu^2$ the
integration is performed in the standard way, dashed line. At
$\vec{k}^{\,2}\!+\!\lambda^2 \!<\!\mu^2$ the `near' gluon pole (red)
is moved above the real axis, and the integration contour changes to
pass below it, along the green short-dashed path. $\lambda$ stands for the gluon mass.}
\label{contour}
\end{figure}

As a result, the Feynman integration must be modified
to exclude the residues of the distant poles in the $k_0$ plane from 
the calculation of $\delta\eta_A^{\rm soft}$.
The complete one-loop expression for $\xi_A^{\rm pert}(\varepsilon_M,\mu)$ 
accounting for the $\varepsilon_M$-dependence, Eq.~(\ref{51a.4}), takes the
following form: 
\beq
\xi_A^{\rm pert}(\varepsilon_M,\mu)= \eta_A^2 -2\, \delta \eta_A^{\rm soft}(\mu)-
\int_{\varepsilon_M}^{\mu'} {\rm d}\varepsilon \,
w^{\rm pert}(\varepsilon), 
\label{40.2}
\eeq
where $\mu'$ corresponds to the excitation energy of the inelastic transition with emission
of a  gluon with energy $\omega\!=\!|\vec k|\!=\!\mu$,
\beq 
\mu'\!=\! \mu\!+\!\sqrt{m_c^2\!+\!\mu^2}\! -\! m_c.
\label{30.2}
\eeq
The last term in Eq.~(\ref{40.2}) describes, at 
$\mu\!=\!\varepsilon_M$,  
the recoil correction in the relation between the normalization point and the
upper limit of energy integration in $I_0$.  Let us note that the
recipe for calculating the combined contribution of the first two
terms in Eq.~(\ref{40.2}) can also be formulated as  changing, for
$|\vec k|\!<\!\mu$, the bypass prescription for the gluon pole at
$k_0\!=\!|\vec k|$ from $k_0\!+\!i0$ to $k_0\!-\!i0$, see Fig.~\ref{contour}. In other words,
the proper Wilsonian prescription consists in  discarding the gluon pole
contribution  for gluons softer than the cutoff.

\thispagestyle{plain}

The one-loop renormalization factor $\eta_A$ without a cutoff is well known:
\beq
\eta_A=1+\frac{3}{4}C_F\frac{\alpha_s}{\pi}\left(
\frac{m_b\!+\!m_c}{m_b\!-\!m_c}\ln{\frac{m_b}{m_c}}-\frac{8}{3}\right)
 + 
{\cal O}(\alpha_s^2)\,;
\label{42.2}
\eeq
the explicit calculations yield for $\delta \eta_A^{\rm soft}(\mu)$
\beq
\delta \eta_A^{\rm soft}(\mu) \!=\!  -\frac{C_F g_s^2}4 \! \int_{|\vec{k}\,|\!<\!\mu} 
\frac{{\rm d}^3 \vec k}{(2\pi)^3 |\vec k|}
\left(\frac{1}{m_c^2}\!+\!\frac{2}{3m_cm_b}\!+\!\frac{1}{m_b^2} \right)
\!=\! -\frac{C_F\alpha_s}{\pi} \frac{\mu^2}{4}
\left(\frac{1}{m_c^2}\!+\!\frac{2}{3m_cm_b}\!+\!\frac{1}{m_b^2} \right) 
\label{44.2}
\eeq
and the $O(\alpha_s)$ inelastic spectral density is
\beq
w^{\rm pert}(\varepsilon)  =
C_F\frac{\alpha_s}{\pi}\,\frac{M^2\!-\!m_c^2}{12\,M^3m_b^2} 
\left(2M^2+3m_b^2+2m_bm_c+m_c^2\right)\,, \qquad M\!=\!m_c\!+\!\varepsilon;
\label{46a.2}
\eeq
$M$ is the invariant hadronic mass in the final state.
The integral of the one-loop spectral function that appears in (\ref{40.2}) is
\bea
&&\int^{\varepsilon_M}_{\mu'} { d}\varepsilon \,
w^{\rm pert}(\varepsilon)=C_F\frac{\alpha_s}{\pi}\,\Big\{
\frac{ (\varepsilon_M\!-\!\mu') (\varepsilon_M \!+\!2 m_c\!+\!\mu')}
{24\, m_b^2\,  (\varepsilon_M\!+\!m_c)^2}  
{\Big [} 2 \varepsilon_M(\varepsilon_M \!+\!2m_c) \\
&&
+\frac{ m_c^2 \left(m_c^2 \!-\!3 m_b^2\!-\!2 m_b
   m_c+4 {m_c} {\mu'}+2 {\mu'}^2\right)}{ ({m_c}+\mu')^2}\Big]
  -\frac{(3m_b\!-\!m_c) (m_b\!+\!m_c)}{12\, m_b^2}
\ln{\frac{m_c\!+\!\mu'}{m_c\!+\!\varepsilon_M}} \Big\} .\nonumber
\eea
Combining these in Eq.~(\ref{40.2}) we obtain the  one-loop 
$\xi_A^{\rm pert}(\varepsilon_M,\mu)$ as an explicit function of 
$\mu/m_Q$. The numeric dependence on $\mu$ can be seen in the first-order 
plot in Fig.~\ref{xipert}.

Our analysis of the one-loop corrections can be readily extended to include
higher-order BLM corrections which describe the effect of the running of
$\alpha_s$ in one-loop diagrams; a complete BLM-summation is also possible. A
detailed discussion of the technique, in the context of the Wilsonian OPE, can
be found in the Appendix of Ref.~\cite{imprec}.\footnote{There was a typo in
  Eq.~(A.6) of paper \cite{blmvcb} for the BLM-resummed expression which
  unfortunately propagated into the later paper \cite{imprec}, Eq.~(A.20); in
  that equation $\Lambda_V^2$ must be replaced by $\Lambda_{\rm QCD}^2$ (the
  conventional $\overline{{\rm MS}}$ one) in the denominator of the power
  term. The correct expression is given here in Appendix~\ref{blmstuff},
  Eq.~(\ref{blmresum}).}  We have recapitulated the salient points in
Appendix~\ref{blmstuff}.
In practice, a  fictitious gluon mass  $\lambda$ is introduced in the one-loop diagrams,  
and eventually a weighted integral over $\lambda^2$ is taken. 
Consequently, the above discussion concerning  the Wilsonian cutoff and the OPE 
applies also to the BLM corrections of any order.

The case of a massive gluon requires  the following kinematic modifications. 
Since the gluon energy is now 
$$
\omega=\sqrt{\vec k^{\,2}\!+\!\lambda^2}, 
$$
the range of soft gluon momenta $\vec k$ shrinks, 
$|\vec k| \!<\! \sqrt{\mu^2\!-\!\lambda^2}$ and the cutoff
in the diagrams is now triggered by 
\beq
\theta(\mu^2\!-\!\lambda^2 \!-\!\vec k^{\,2}).
\label{45.2}
\eeq
Notably, no subtraction is necessary at $\lambda>\mu$. 
The recoil energy is also modified and $\mu'$ in Eqs.~(\ref{30.2}) is given by 
\beq
\mu'\!=\!\sqrt{m_c^2\!+\!\mu^2\!-\!\lambda^2}\!-\!m_c\!+\!\mu;
\label{30a.2}
\eeq
the perturbative inelastic spectral density 
is given in Appendix~\ref{pertdetails}, Eq.~(\ref{46}). 

In the Feynman diagrams the relevant gluon pole is now located at 
$k_0\!=\!\sqrt{\vec k^{\,2}\!+\!\lambda^2}$ rather than at $k_0\!=\!|\vec
k\,|$; this makes the explicit expression  for 
$\delta \eta_A^{\rm soft}$ more cumbersome:
\bea
\nonumber
\delta \eta_A^{\rm soft}(\mu) \msp{-4}& = &\msp{-4}
C_F g_s^2 \int_{k_0\!<\!\mu}
\frac{{\rm d}^3 \vec k}{(2\pi)^3 2k_0} 
\left(
\frac{2m_b^2\!-\!2m_b k_0\!-\!2k_0^2\!+\!\lambda^2}{(2m_b k_0\!-\!\lambda^2)^2} 
\right.\\
&& \msp{12} \left.
+\frac{2m_c^2\!-\!2m_c k_0\!-\!2k_0^2\!+\!\lambda^2}{(2m_c
  k_0\!-\!\lambda^2)^2} 
- 2\frac{2m_b m_c \!-\!(m_b\!+\!m_c) k_0 \!+\! \frac{2}{3}\vec k^{\,2}\!+\!\lambda^2}
{(2m_b k_0\!-\!\lambda^2)(2m_c k_0\!-\!\lambda^2)} 
\right)\!;\quad
\label{44a.2}
\eea
the integral can be solved analytically, resulting in a lengthy
expression. The expression for $\eta_A$ at  nonvanishing gluon mass is given
in Eq.~(\ref{etaalam}) of Appendix~\ref{pertdetails}. 

Combining these elements in Eq.~(\ref{40.2}) we obtain the one-loop correction
to $\xi_A^{\rm pert}$ at arbitrary gluon mass $\lambda$, 
$\xi_A^{\rm pert}(\varepsilon_M,\mu; \lambda^2)$. Assuming 
$\mu\!=\!\varepsilon_M$,  for $\lambda^2\!>\!\mu^2$ the last two
terms in Eq.~(\ref{40.2}) are absent at $\lambda^2\!>\!\mu^2$ and 
one simply has $\xi_A^{\rm pert}(\mu; \lambda^2)\!=\!\eta_A^2(\lambda^2)$. 
Using the formulas of Appendix~\ref{blmstuff} one readily obtains the BLM corrections of
arbitrary order or the resummed result. We performed the final integration 
over the gluon mass numerically.

The explicit expression for $\eta_A(\lambda^2)$ at small $\lambda^2$ shows
non-analytic terms in $\lambda^2$, starting with $\frac{\lambda^2}{m_Q^2}
\ln{\lambda^2}$. They signal the infrared sensitivity of $\eta_A$ and the
emergence of infrared renormalons, 
which in turn makes it impossible to assign a definite value to the purely
perturbative $\eta_A$, and leads to a significant numerical instability from
 the higher-order corrections.  Non-analytic terms are also present in
$\delta \eta_A^{\rm soft}$, and they precisely offset those in
$\eta_A(\lambda^2)$: the combined $\xi_A^{\rm pert}(\varepsilon_M,\mu)$ in
Eq.~(\ref{40.2}) is an analytic function of $\lambda^2$ in the vicinity of
zero at any positive $\mu$. The radius of convergence of the Taylor series in
$\lambda^2$ is precisely $\mu^2$. The cancellation of all the non-analytic
pieces is most simply seen directly in the integral representation for
$\eta_A\!-\!\delta\eta_A^{\rm soft}(\mu)$, upon closing the contour over the
residues at positive $k_0$. The recoil integral describes the effect of
shifting the normalization point and is purely short-distance. Indeed, the
one-loop spectral density ${ w}^{\rm pert}(\varepsilon; \lambda^2)$ is
explicitly analytic at small $\lambda^2$ for 
$\varepsilon \!> \!\mu$ where $|\vec{k}\,|$ is of order $\mu$.  Consequently,
there are no formal obstacles in deriving either any higher-order BLM
coefficient or the fully resummed BLM value for the Wilson coefficient
$\xi_A^{\rm pert}(\varepsilon_M,\mu)$; apart from the ultraviolet domain, its
perturbative series has a finite radius of convergence, at least in the BLM
approximation.

The advantage of the method described in this section is that it allows to
calculate the full $\mu$-dependence of the Wilson coefficient.  
It does not apply to non-BLM corrections, starting 
with ${\cal O}(\alpha_s^2)$.
In that case, the $\mu$-dependence of $\xi_A^{\rm pert}$  has to be
determined,  order by order in $1/m_Q$, applying the normalization 
point independence of the OPE
relations,  and using the initial condition Eq.~(\ref{51a.8}).  
This was done for the non-BLM ${ O}(\alpha_s^2)$ corrections in Ref.~\cite{xi2}, through 
$O(1/m_Q^2)$; in particular, the two-loop spectral density 
$ w^{\rm pert}(\varepsilon)$ was calculated there to this accuracy.  The
corresponding non-BLM coefficient was found to be small numerically, which
suggests that omitted terms ${\cal O}(\as^2\,\mu^3/m_Q^3)$ and higher should not
have a noticeable impact. 

\subsection{Numerical analysis}

The perturbative corrections   $\xi_A^{\rm  pert}(\varepsilon_M)$ appear to be 
small for values of $\varepsilon_M$ between $0.6\GeV$ and $1\GeV$.
Taking, for instance, $\varepsilon_M\!=\!\mu\!=\!0.75\GeV$,
$m_c\!=\!1.2\GeV$, $m_b\!=\!4.6\GeV$ and
assuming $\alpha_s(m_b)\!=\!0.22$ we get 
\beq
\sqrt{\xi_A^{\rm pert}}=1-0.022+(0.005-0.004)+0.002-0.0015 + \:... 
\label{96}
\eeq
Here the first term is the tree-level  value, the second is the ${\cal O}(\alpha_s)$
term evaluated with $\alpha_s\!=\!0.3$,  which corresponds to the strong coupling evaluated
at an intermediate scale  between $m_c$ and $m_b$. The two values in brackets
show the shift, relative the  one-loop evaluation with $\alpha_s\!=\!0.3$,
due to  ${\cal O}(\alpha_s^2)$ corrections (positive for the BLM part
and negative from the non-BLM contribution); the last two terms are the
${\cal O}(\beta_0^2\alpha_s^3)$ and ${\cal O}(\beta_0^3\alpha_s^4)$, which may serve 
as an estimate of even higher-order perturbative corrections.

\thispagestyle{plain}
\begin{figure}[t]
\vspace*{-7pt}
 \begin{center}
\includegraphics[width=8.5cm]{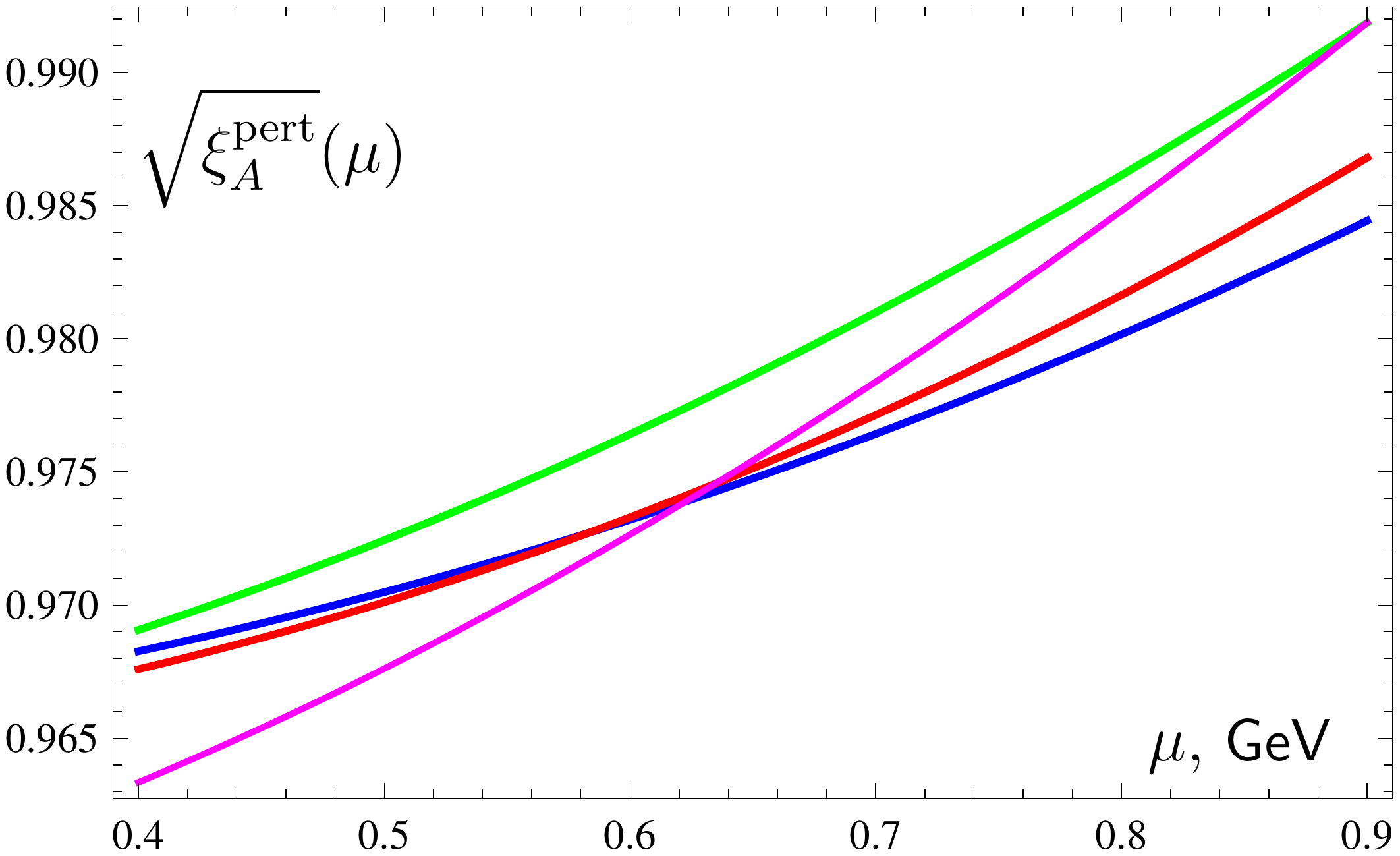}
 \end{center}\vspace*{-5pt}
\caption{ \small $\sqrt{\xi_A^{\rm pert}}(\mu)$ as a function of $\mu$ for $m_b=4.6\GeV$, $m_c=1.2\GeV$, $\as(m_b)=0.22$. The curves represent  the one-loop result
evaluated with  $\as=0.3$ (blue),  one-loop plus first order BLM
(green), complete ${\cal{O}}(\as^2)$ (red), two-loop plus third-order BLM (maroon).   
}
\vspace*{-2.5pt}
\label{xipert}
\end{figure}

Fig.~\ref{xipert} shows the dependence of
$\sqrt{\xi_A^{\rm pert}(\mu)}$ on $\mu$
at different orders in $\alpha_s$
assuming $\alpha_s^{\overline{\rm MS}}(m_b)\!=\!0.22$.
For $\mu$ between $0.7$ and $0.8\GeV$ the value of
$\sqrt{\xi_A^{\rm pert}(\mu)}$ is close to $0.98$, and we associate to it a
rather conservative 1\% uncertainty:
$$
\sqrt{\xi_A^{\rm pert}}(0.75\GeV)=0.98\pm 0.01\, .
$$
 We emphasize that the observed
stability of the perturbative expansion applies only to the
perturbative renormalization factor in the
Wilsonian OPE, and that the quoted value refers to the specific
renormalization scheme which is used in the present analysis.

Let us note here an observation that is relevant when the terms ${\cal O}(1/m_Q^3)$
and higher  are accounted for.  As mentioned in
the previous subsection, the  calculation  of the perturbative 
factor $\xi_A^{\rm pert}(\mu)$
which neglects the subtleties related to the `distant' poles
yields the correct result  through order
$\mu^2/m_Q^2$, and one needs an improved method  only to account for the  terms $
\mu^3/m_Q^3$ and higher.  Numerically the two methods yield very close results to order $
\alpha_s$ and  $\beta_0\alpha_s^2$, but the difference  starts to accumulate
systematically from order $\beta_0^2\alpha_s^3$ on, even though
parametrically the difference between the two methods is still only of order $1/m_Q^3$. 
For instance, the fully resummed results for the two prescriptions to calculate
$\xi_A^{\rm pert}(\mu)$ differ by about $2\%$. 
This is in contrast with the case of the total $b\tto c \,\ell\nu$ width, where there is no
visible difference between the result truncated at order $\alpha_s^2$ 
and the fully resummed one. The reason is that the current 
renormalization at zero recoil has an intrinsically lower scale driven by 
the charm mass, in contrast to  $m_b$ for $\Gamma_{\rm sl}(B)$.

\section{Power corrections to \boldmath $I_0$}
\label{powercor}

In this section we  investigate the power corrections in the 
right-hand (OPE) side of the sum rule Eq.~(\ref{88}).
The leading power corrections to $I_0$ were calculated in
Refs.~\cite{vcb,optical} to order $1/m_Q^2$ and in
Ref.~\cite{rev} to order $1/m_Q^3$  and read
\bea
\Delta_{1/m^2} &\msp{-5}=\msp{-5}& \frac{\mu_G^2}{3m_c^2} +
\frac{\mu_\pi^2 \!-\!\mu_G^2 }{4}
\left(\frac{1}{m_c^2}+\frac{2}{3m_cm_b}+\frac{1}{m_b^2}
\right), 
\label{102} 
\\
\Delta_{1/m^3} &\msp{-5}=\msp{-5}&
\frac{\rho_D^3 - \frac{1}{3}
\rho_{LS}^3}{4m_c^3}\;+\;
\frac{1}{12m_b}\left(\frac{1}{m_c^2}+\frac{1}{m_c m_b} +\frac{3}{m_b^2}\right)
\,(\rho_D^3 +
\rho_{LS}^3)\,.
\label{103}
\eea
The nonperturbative parameters  $\mu_\pi^2$, $\mu_G^2$, $ \rho_D^3$ and
$\rho_{LS}^3$ all depend on the hard Wilsonian cutoff.  
In the  renormalization scheme we have adopted the inequalities 
$\mu_\pi^2(\mu)\!\ge\!\mu_G^2(\mu)$, $\:\rho_D^3(\mu)\!\ge\!-\rho_{LS}^3(\mu)$
hold at arbitrary normalization point $\mu$. The nonperturbative contributions
in Eqs.~(\ref{102}), (\ref{103}) are therefore positive.

To remain on the conservative side for numeric estimates we can adopt the 
low values
$\mu_\pi^2 (0.75\GeV)\!=\!0.4\GeV^2$,  $\rho_D^3(0.75\GeV)\!=\!0.15\GeV^3$,
and  the quark masses $m_c\!=\!1.2\GeV$ and $m_b\!=\!4.6\GeV$ (the
scale dependence of the latter plays a role here only at the level
formally beyond the accuracy of the calculation).
The dependence on $\mu_G^2$ and on  $\rho_{LS}^3$ is minimal and their
precise values do not matter; we use  for them  $0.3\GeV^2$ and
$-0.12\GeV^3$, respectively.  We then get
\beq
\Delta_{1/m^2}= 0.091, \qquad \Delta_{1/m^3}= 0.028\,.
\label{106}
\eeq

It is interesting to compare these estimates with those 
derived from the constraints on the expectation values of dimension 5
and 6 operators from the semileptonic $B$-decay moments.
If we employ the values of the OPE parameters
extracted from the latest official HFAG  fit to inclusive semileptonic and radiative decay
distributions \cite{HFAG,fit},  for $\mu=0.75\GeV$  we find 
\beq
 \Delta_{1/m^2}+\Delta_{1/m^3} = 0.102 \pm 0.017 \, ,\nonumber
\eeq
which is consistent with (\ref{106}). 
As discussed in \cite{fit}, this HFAG fit to semileptonic moments
depends  on several  assumptions
and does not yet incorporate certain higher-order effects that may be
important, including the complete $\alpha_s^2$-corrections 
\cite{NNLOmoments}. In particular, more realistic ans\"atze for the 
theoretical correlations have been considered in  \cite{newfit},  
leading to larger values of the $\Delta^A$, with bigger errors. Typically, one then has
 \beq
 \Delta_{1/m^2}+\Delta_{1/m^3} = 0.11\pm 0.03 \, .
\label{110a}
\eeq
On the other hand, combining the semileptonic moments alone  with 
a high precision determination  of the charm mass \cite{mcdet} yields \cite{newfit}
\beq
\Delta_{1/m^2}= 0.090\pm0.013, \qquad \Delta_{1/m^3}= 0.029\pm 0.008\, , 
\label{110b}
\eeq
in remarkable agreement with Eq.~(\ref{106}).

The  important question is how well the power expansion for the sum
rule converges. Recently, the OPE for the semileptonic
$B$-meson structure functions has been extended to order $1/m_Q^4$ and
$1/m_Q^5$ \cite{icsieg,hiord}.
Applying the analysis to the structure functions mediated by the 
axial and  by the vector currents separately, we 
find\footnote{We thank S.\,Turczyk for providing us with the 
input needed for this calculation.} 
\bea
\nonumber
16m_c^4 \,\Delta_{1/m^4}&\msp{-5}=\msp{-5}&- (3\!+\!\mbox{$\frac{4}{3}$}y
\!+\!y^4)m_1+
(\mbox{$\frac{2}{3}$}y \!-\!\mbox{$\frac{2}{3}$}y^2\!-\!
\mbox{$\frac{2}{3}$}y^3 \!-\!2y^4)(m_2\!+\!m_5)-
(1+\mbox{$\frac{4}{9}$}y+\mbox{$\frac{1}{3}$}y^4)m_4\\
\nonumber
&\msp{-5}&
+(1\!+\!\mbox{$\frac{4}{3}$}y\!-\!y^4)m_6-\mbox{$\frac{4}{3}$}y\, m_7 + 
(\mbox{$\frac{1}{4}$}\!+\!\mbox{$\frac{1}{3}$}y\!-\!\mbox{$\frac{1}{4}$}y^4)m_8,
\rule[-10pt]{0pt}{10pt}
\\
\nonumber
16m_c^5\, \Delta_{1/m_Q^5}&\msp{-5}=\msp{-5}& 
-(2 \!+\! \mbox{$\frac{4}{3}$}y \!-\!\mbox{$\frac{2}{3}$} y^2)r_1 + 
(6 \!+\! 4y \!+\!\mbox{$\frac{1}{2}$} y^2\!+\! \mbox{$\frac{2}{3}$}y^3 
\!+\!\mbox{$\frac{4}{3}$} y^4\!+\! 2y^5 )(r_2\!-\!r_3) \\
\nonumber
&\msp{-5}& 
+(6 \!+\! 4y \!-\!\mbox{$\frac{16}{3}$} y^2\!+\! \mbox{$\frac{2}{3}$}y^3 
\!+\! 2y^5 )r_4 -
(2 \!+\! 4y \!+\!\mbox{$\frac{1}{3}$} y^3\!+\! \mbox{$\frac{1}{3}$}y^4 
\!+\! y^5 )r_5\\
\nonumber
&\msp{-5}&
-(2 \!+\! \mbox{$\frac{4}{3}$}y \!+\!\mbox{$\frac{23}{6}$} y^2\!+\! 
\mbox{$\frac{1}{3}$}y^3 \!+\!y^4\!+\! y^5 )r_6+
(2 \!+\! \mbox{$\frac{4}{3}$}y \!+\!\mbox{$\frac{7}{6}$} y^2\!+\! 
\mbox{$\frac{1}{3}$}y^3 \!+\!y^4\!+\! y^5 )r_7\\
\nonumber
&\msp{-5}&
+(\mbox{$\frac{2}{3}$} \!-\!\mbox{$\frac{4}{3}$} y\!+\! 
\mbox{$\frac{2}{3}$}y^2)r_8-
(2 \!+\! y \!+\!\mbox{$\frac{14}{3}$} y^2\!-\! 
\mbox{$\frac{2}{3}$}y^3 \!-\!\mbox{$\frac{2}{3}$}y^4\!-\! 3y^5 )(r_9\!-\!r_{12})
\\
\nonumber
&\msp{-5}&
-(2 \!+\! y \!+\!\mbox{$\frac{7}{2}$} y^2\!-\! 
\mbox{$\frac{2}{3}$}y^3 \!-\!\mbox{$\frac{4}{3}$}y^4\!-\! 3y^5
)(r_{10}\!-\!r_{11})\!-\!
(2 \!+\! 6y \!+\!\mbox{$\frac{26}{3}$} y^2\!-\! 
\mbox{$\frac{2}{3}$}y^3 \!-\!\mbox{$\frac{2}{3}$}y^4\!-\! 4y^5) r_{13}\! \\
\nonumber
&\msp{-5}&
-(2 \!-\! 4y \!+\!\mbox{$\frac{16}{3}$} y^2\!-\! 
\mbox{$\frac{2}{3}$}y^3 \!-\! 2y^5) r_{14} +
(\mbox{$\frac{4}{3}$} \!-\!\mbox{$\frac{1}{3}$} y \!+\!4y^2\!-\! 
\mbox{$\frac{2}{3}$}y^3 \!-\! \mbox{$\frac{2}{3}$}y^4\!-\! 3y^5) r_{15} \\
\nonumber
&\msp{-5}&
+(\mbox{$\frac{2}{3}$} \!+\!\mbox{$\frac{11}{3}$} y \!+\!4y^2\!-\! 
\mbox{$\frac{1}{3}$}y^3 \!-\! \mbox{$\frac{1}{3}$}y^4\!-\! 2y^5) r_{16} +
(\mbox{$\frac{2}{3}$} \!-\!\mbox{$\frac{4}{3}$} y \!-\!\mbox{$\frac{23}{6}$}
y^2\!-\! \mbox{$\frac{1}{3}$}y^3 \!-\! y^4\!-\! y^5) r_{17} \\
&\msp{-5}&
-(\mbox{$\frac{4}{3}$} \!+\!\mbox{$\frac{7}{3}$} y \!+\!\mbox{$\frac{17}{6}$}y^2\!-\! 
\mbox{$\frac{2}{3}$}y^3 \!-\! \mbox{$\frac{4}{3}$}y^4\!-\! 3y^5) r_{18} ,
\label{107}
\eea
where $y\!=\!m_c/m_b$ and 
the $D\!=\!7$ and $D\!=\!8$ expectation values $m_{1-9}$ and $r_{1-18}$
are defined in Ref.~\cite{hiord}. These can be evaluated in the 
ground-state factorization 
approximation. Using the expressions given in Ref.~\cite{hiord} we obtain 
the estimates 
\beq
\Delta_{1/m^4}\simeq -0.023\,, \qquad 
\Delta_{1/m^5}\! \simeq\!-0.013 \,.
\label{108}
\eeq
It is worth noting
that retaining only the terms suppressed by the powers of $1/m_c$ (i.e.,
evaluating the higher-order corrections in the limit $m_b\tto \infty$) yields
a perfect numeric approximation to the full expression. 
In the ground state saturation approximation 
the dominant contributions to $\Delta_{1/m^4}$ and $\Delta_{1/m^5}$
are those of $m_{4,8}$  and
$r_{2,10}$, respectively, without significant cancellations.
We then observe that the
power series for $I_0$ appears well-behaved at the required level of precision.

For what concerns the  loop corrections to $\Delta^A$,
the  $\alpha_s$-correction to the Wilson coefficient for
the kinetic operator in Eq.~(\ref{102}) was calculated in
Ref.~\cite{xi2} and turned out numerically insignificant. Generally
larger $\alpha_s$-corrections are expected  in the
chromomagnetic and Darwin coefficient functions. However, the
dependence on $\mu_G^2$ in  Eq.~(\ref{102}) turns out negligible;
therefore perturbative corrections are not expected to introduce
significant numerical changes in the estimate of $\Delta_{1/m^2}$.
At  order $1/m_Q^3$, even if radiative corrections change the coefficient
for the Darwin term by $30\%$ the effect on the sum rule would still be
small.

Taking into account all the available information, our estimate for
 the total power correction at $\varepsilon_M\!=\!0.75\GeV$ is 
\beq
\Delta^A = 0.105
\label{110}
\eeq
with a $0.015$ uncertainty due to higher orders. On the theoretical grounds,
larger values of  $\mu_\pi^2$ and/or $\rho_D^3$ are actually favored; they tend
to increase  $\Delta^A$.
Combining the above with the perturbative corrections we arrive at an
estimate for $I_0$ and, according
to Eq.~(\ref{86a}),  at a bound on the form factor, which in terms
of the central values
at $\varepsilon_M \!=\!0.75\GeV$ is 
\beq
{\cal F}(1) < 0.925   \,.
\label{139}
\eeq

As stated above, the upper bound in Eq.~(\ref{86a}) depends on $\varepsilon_M$,
becoming stronger for smaller $\varepsilon_M$. 
It is  advantageous to choose  the minimal value of
$\varepsilon_M$ for which the OPE-based short-distance  expansion 
of the integral (\ref{82}) for $I_0(\varepsilon_M)$ sets in. This directly
depends on how low one can push the renormalization scale $\mu$ 
while still observing the expectation values actual $\mu$-dependence
in the kinetic
scheme approximated by the perturbative one. Since in this scheme
$\mu_\pi^2(\mu)\!\ge \!\mu_G^2(\mu)$ holds for arbitrary $\mu$, in essence this
boils down to the question at which scale 
$\mu_{\rm min}$ the spin sum rule and the one
for $\mu_G^2$ get approximately saturated, e.g.\  
$\mu_G^2(\mu_{\rm min})\!\simeq\! 0.3\GeV^2$. 
The only vital assumption in the analysis is that the
onset of the short-distance regime is not unexpectedly delayed in actual QCD 
and hence does not require $\varepsilon_M \!>\! 1\GeV$. This principal question 
can and should be verified on the lattice. This will complement already
available evidence from 
preliminary lattice 
data \cite{tau32} as well as from the  successful experimental 
confirmation \cite{Belle} in nonleptonic $B$ decays of the 
predicted  $\frac{3}{2}^-$-dominance.

\section{Estimates of \boldmath $I_{\rm inel}$}
\label{winel}

We now turn to the actual estimate for the inelastic
contribution. 
On general grounds  \cite{vcb} $I_{\rm inel}$ is expected to be
comparable to the power correction 
$\Delta^A$  considered above; therefore the inelastic contributions should
be important numerically.

Our starting point is the first moment of the 
scattering amplitude spectral density given by the contour integral
\beq
I_1(\mu)= -\frac{1}{2\pi i}\;
\oint_{\raisebox{-8pt}{\hspace*{-12pt}\scalebox{.7}{$|\varepsilon|\!=\!\varepsilon_M$}}\hspace*{-5pt} }
T^{\rm zr}(\varepsilon) \, \varepsilon\,{\rm d}\varepsilon \,;
\label{120}
\eeq
 we can write
\beq
 I_{\rm inel}(\varepsilon_M) =
 \frac{I_1(\varepsilon_M)}{\tilde\varepsilon}\, ,
\label{122}
\eeq
where $\tilde\varepsilon$ is an average excitation energy which depends on
$\varepsilon_M$. For moderate $\varepsilon_M$  the integral 
is expected to be dominated by the lowest `radial' 
excitations\footnote{These excited states play an important role in
the analysis of power corrections in the HQE, and we clarify
our terminology. In the heavy quark limit these are
either the true radial excitations of the ground state, or the
counterparts of the $D$-waves. The former have spin-parity of the 
light cloud $\frac{1}{2}^+$ while for the latter it is
$\frac{3}{2}^+$. 
At finite quark masses by "radial excitation"  we refer to the
descendants of any hyperfine multiplet member of these heavy-quark states. 
More details are addressed in Sect.~\ref{alook}.
} 
of the ground state, with $\tilde\varepsilon \approx
\varepsilon_{\rm rad}\approx 700\MeV$. The first moment
$I_1(\varepsilon_M)$  can also be calculated in the OPE
\cite{optical};  the result reads
\beq
I_1 \!=\! \frac{-(\rho_{\pi G}^3\!+\!\rho_A^3)}{3m_c^2}
+ \frac{-2\rho_{\pi\pi}^3\!\!-\!\rho_{\pi G}^3}{3m_c m_b} +
\frac{\rho_{\pi\pi}^3\!\!+\!\rho_{\pi G}^3\!+\!\rho_S^3
\!+\!\rho_A^3}{4} \!\left(\!
\frac{1}{m_c^2} \!+\!  \frac{2}{3m_c m_b}\!+\! \frac{1}{m_b^2} \!\right)
 + {\cal O}\!\left(\!\frac{1}{m_Q^3}\!\right)\! .
\label{124}
\eeq
  The nonlocal zero momentum transfer correlators $\rho_{\pi
\pi}^3$, $\rho_{\pi G}^3$, $\rho_S^3$ and $\rho_A^3$ have been introduced in
\cite{optical} and are given by
$$
\rho_{\pi\pi}^3=
 \int {\rm d}^4 x\,
\; \frac{1}{4M_{B}}\langle B| i T\{\bar b \vec\pi\,^2b(x),
\;\bar b\vec\pi\,^2b(0)\}|B\rangle ',
$$
$$
\rho_{\pi G}^3=
 \int {\rm d}^4 x\,
\; \frac{1}{2M_{B}}\langle B| i T\{\bar b \vec\pi\,^2b(x),
\;\bar b\vec\sigma \vec B b(0)\}|B\rangle ',
$$
\begin{equation}
\frac{1}{3}\rho_{S}^3\delta_{ij}\delta_{kl}+\frac{1}{6}
\rho_{A}^3(\delta_{ik}\delta_{jl}-
\delta_{il}\delta_{jk})=
 \int {\rm d}^4 x\,
\; \frac{1}{4M_{B}}\langle B| i T\{\bar b \sigma_i B_kb(x),\;
\bar b \sigma_j B_l b(0)\}|B\rangle ' , 
\label{66a}
\end{equation}
where $\vec B$  denotes the chromomagnetic field strength operator. 
The prime indicates that the ground-state contribution is subtracted -- otherwise
the integral diverges at large $x_0$ where the correlators approach a 
constant determined by the ground-state factorization contribution.

At higher orders in $1/m_Q$ the expansion (\ref{124}) 
 will include, along with the local expectation values of
higher dimensional operators, more intricate nonlocal $T$-products. The latter are poorly
known.  Moreover, the quantum-mechanical interpretation of these relations
tells us that there will be significant cancellations among different terms in
higher orders; say, the factorizable terms must drop out.

Since our goal is only to obtain a reasonable
estimate, we discard higher-order corrections  and keep only the leading
${\cal O}(1/m_Q^2)$ terms. This implies that the expectation values can be
considered in the static theory.  The nonlocal correlators 
appear in $I_1$ because the energy variable
$\varepsilon$ is defined with respect to the physical threshold
$q_0\!=\!M_B\!-\!M_{D^*}$ rather than relative to the parton-level threshold
at $q_0\!=\!m_b\!-\!m_c$ native to the OPE. The terms given by the
local operators cancel out in $I_1$ to the leading order, as becomes transparent in the
quantum-mechanical interpretation of Ref.~\cite{optical}, Sect.~6. 
The latter also explains 
the explicit  field-theoretic expression for $w_{\rm inel}(\varepsilon)$, which is given in Sec.~6.1.
It is important to stress that the third term in (\ref{124}) is positive
since the combination $\rho_{\pi\pi}^3\!\!+\!\rho_{\pi G}^3\!+\!\rho_S^3
\!+\!\rho_A^3$ is actually equivalent to the correlator
of two identical operators $\bar{b}(\vec \sigma \vec \pi)^2 b$.

An important  piece of information is provided by the heavy quark mass dependence
of the hyperfine splitting $\Delta M^2$, which  allows us to estimate the overall
magnitude of the nonlocal correlators in Eqs.~(\ref{124}). 
The $1/m_Q$ scaling of the  vector-pseudoscalar mass splitting is 
nonperturbatively affected by the values of two $D\!=\!3$ 
spin-triplet parameters, $\rho_{LS}^3$ and $\rho_{\pi G}^3\!+\!\rho_A^3$:
\beq
M_{B^*}-M_B= \frac{2}{3}\frac{\mu_G^2}{m_b}+
\frac{\rho_{\pi G}^3\!+\!\rho_A^3\!-\!\rho_{LS}^3}{3\,m_b^2}+ 
{\cal O}\left(\frac{1}{m_b^3}\right) , 
\label{70.2}
\eeq
and likewise for charm. Since the spin-orbit expectation value is reasonably
constrained by the heavy quark sum rules, (\ref{70.2}) yields 
information on the combination 
$$
-(\rho_{\pi G}^3\!+\!\rho_A^3).
$$
A preliminary  analysis was outlined in Ref.~\cite{chrom} and  indicated a 
large value exceeding the naive expectations. We reconsider it carefully
in Sect.~\ref{alook} and confirm the observation.
This combination enters directly the expression  (\ref{124}) for 
$I_{\rm inel}$ and is particularly important, as will become clear in the following 
subsection, where we
combine the theoretical expressions for $I_{\rm inel}$ with the numerical
analysis of the hyperfine splitting to arrive at our estimate for $I_{\rm inel}$.
 Additional technical details are given in Sect.~\ref{hyper}.

\subsection{Numerical estimate of \boldmath $I_{\rm inel}$}
\label{winelnumer}

It turns out advantageous to analyze $I_{\rm inel}$ 
from the perspective of the BPS approximation
for $B$ mesons \cite{BPS}. 
In the BPS limit $\mu_\pi^2\!=\!\mu_G^2$ and, since 
$\mu_\pi^2\!-\!\mu_G^2\!=\!\aver{(\vec\sigma\vec\pi)^2}$
in the $B$ meson, one  concludes that $\:\bar{b}(\vec\sigma\vec\pi) b
\state{B}\!=\!0$. This would also imply 
$-\rho_{LS}^3\!=\!\rho_D^3$. 
The deviation from the BPS limit is quantified by the smallness
of the difference $\mu_\pi^2\!-\!\mu_G^2$ compared to $\mu_\pi^2$ 
itself \cite{BPS}.   
Many remarkable relations hold in the BPS limit; for instance, among the 
spectral densities of the correlators 
of $\bar b \vec\pi^2 b$ and $\bar b \vec \sigma \vec B b$
that we will introduce in Sec. 6,
\beq
\rho_{p}^{(\frac{1}{2}^+)}(\omega)=\rho_{pg}^{(\frac{1}{2}^+)}(\omega)=
\rho_{g}^{(\frac{1}{2}^+)}(\omega), \qquad
\rho_{f}^{(\frac{3}{2}^+)}(\omega)=\rho_{fg}^{(\frac{3}{2}^+)}(\omega)=
\rho_{g}^{(\frac{3}{2}^+)}(\omega)
\label{170},
\eeq
and among  their integrals determining the  $\rho^3$ 
correlators:
\beq
\rho_{\pi G}^3=-2\rho_{\pi\pi}^3, \qquad 
\rho_{\pi G}^3 \!+\!\rho_A^3 = -(\rho_{\pi\pi}^3 \!+\!\rho_S^3). 
\label{82f0}
\eeq
In the BPS limit  $I_1$ in Eq.~(\ref{124}) 
is then given  
by the same combination of the nonlocal correlators that drives, 
besides $\rho_{LS}^3$, the hyperfine 
splitting to order $1/m_Q^2$, cf.\ Eq.~(\ref{70.2}): 
\beq
I_1\,\stackrel{\scalebox{.55}{BPS}}{=} \,\frac{-(\rho_{\pi G}^3+\rho_A^3)}{3m_c^2}
 + {\cal O}\left(\frac{1}{m_c^3}\right) .
\label{172}
\eeq

The second term in Eq.~(\ref{124}) for $I_1$ is of the first
order in the deviation from BPS; as such it is not sign-definite. However, it
is suppressed by the $b$-quark mass. The third term is of the second order in
the BPS violation and is positive; it comes with a large
coefficient. Therefore, the
full expression develops only a shallow minimum where the whole sum differs
from the BPS value by a factor of no less than $0.93$, see Section
\ref{I1est}. In fact, $I_1$ may 
exceed the BPS value by a  larger amount, although our analysis 
favors a negative sign for the second term. This typically results, at small 
deviations,  in an overall slight decrease. 

A simple minimal -- and  most physical -- ansatz for the spectral densities
determining the correlators elucidates the role of the constraints
the correlators obey to. It assumes that they are saturated by a single
multiplet of excited states, for each angular momentum of light
degrees of freedom.
Apart from the excitation mass gap the relevant contributions are then determined
by three residues; they are introduced in Sect.~\ref{modind} and are denoted
by $P,\,G$ (for the radially excited  $\frac{1}{2}^{+}$) and  by $g$ (for the 
$\frac{3}{2}^{+}$ state). In the BPS limit $P\!=\!G$. 

\begin{figure}[t]
\vspace*{-7pt}
 \begin{center}
 \includegraphics[width=8cm]{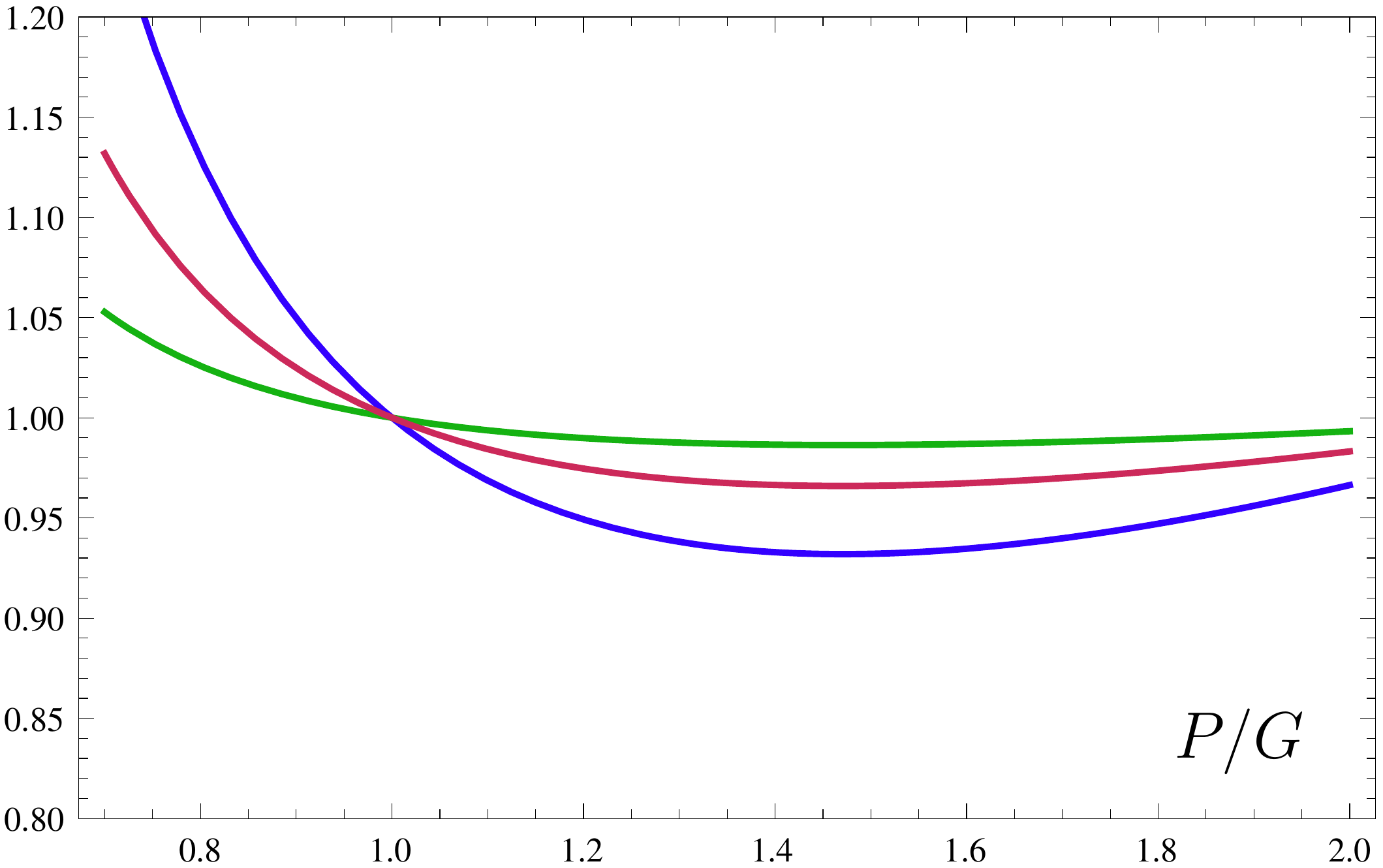}
  \end{center}\vspace*{-5pt}
\caption{  \small  
Variation of $I_1/I_1^{\rm BPS}$ with $P/G$ for
different relative contribution $\nu$ of $\frac{3}{2}^{+}$ to 
$\frac{1}{2}^{+}$ states to the hyperfine splitting, $\nu=0$ (blue), $0.5$
(red), 0.8 (green).
}
\label{I1}
\end{figure}

At a fixed hyperfine constraint on $-(\rho_{\pi G}^3\!+\!\rho_A^3)$ the full
expression for $I_1$ in Eq.~(\ref{124}) depends on two dimensionless ratios,
 $P/G$ and on the relative contribution of the $\frac{3}{2}^{+}$
state  proportional to $g^2$. The minimum value for $I_1$ occurs where the latter
vanishes, $g\!=\!0$; whenever $\frac{3}{2}^{+}$ dominates, $I_1$ 
uniformly approaches its BPS value.
The value of  $P/G$ for which there is a minimum depends only on the ratio of the 
quark masses, see Eq.~(\ref{173.04}). 
Fig.~\ref{I1} shows the variation of  $I_1/I_1^{\rm BPS}$
with $P/G$ for a few values of the relative 
contribution $\nu$ of the $\frac{3}{2}^{+}$ 
state into the combination determining the hyperfine splitting. More
details of the analysis are given in Sect.~\ref{I1est}.

\thispagestyle{plain}

Neglecting a possible few percent relative decrease in $I_1$ we should, 
therefore, adopt the BPS
relation Eq.~(\ref{172}) as a reasonably accurate lower bound estimate;
this leads to the estimate 
\beq
I_{\rm inel}(\varepsilon_M\!\approx\! 0.75\GeV) \gsim \frac{0.45\GeV^3 + (\kappa\!+\!0.2) \cdot 0.35\GeV^3}{3m_c^2\,
\tilde\varepsilon},
\label{174}
\eeq
where the dimensionless $\kappa$, introduced in Eq.~(\ref{76}), 
parameterizes the exact value of the correlator $-(\rho_{\pi G}^2\!+\!\rho_A^3)$. 
$\kappa$ is
uncertain due to unknown higher power corrections and due to the limited accuracy of the 
perturbative renormalization. 
Here $\varepsilon_M$ is assumed to be around $0.75\GeV$ to include the families of the
lowest `radial' excitations. Our analysis in Sect.~\ref{hyperold} suggests that
$\kappa$ is relatively small, between  $-0.4$ and $0$.

Equating $\tilde\varepsilon$ in  Eq.~(\ref{174}) with
$\epsilon_{\rm rad}\simeq 700\MeV$ we estimate
\beq
I_{\rm inel} \gsim 0.14\,.
\label{176}
\eeq
We recall  that,
in contrast to the OPE for $I_0(\mu)$, this estimate assumes 
only the leading $\mhad/m_Q$ pieces in the transition amplitudes. 
The subleading $1/m_c$ corrections can be numerically significant --
this is illustrated, for instance, by Eq.~(\ref{106}) -- 
and can potentially modify the actual $I_{\rm inel}$ by as much as $30\%$ of
the estimate, even though the inclusive sums like 
$I_{\rm inel}$ are usually affected less than the individual transitions.

The precise value of $\mu_\pi^2$ is not yet well known; it mainly affects
the degree of proximity of actual $B$ mesons to the BPS regime.  The BPS
expansion would become more qualitative than quantitative if
$\mu_\pi^2$ eventually exceeds $0.45\GeV^2$ by a significant
amount. At the same time, as illustrated in this section, this would not affect
significantly our estimates. 
Moreover, larger  $\mu_\pi^2$ lowers
the model-independent upper bound which only assumes 
positivity of the inelastic contribution. Complementarily, from
the full set of  the heavy quark sum rules we should expect larger
transition amplitudes to the excited states at larger $\mu_\pi^2$. This
conforms the physical intuition which suggests $\sqrt{\mu_\pi^2}$ to
quantify the mass scale $\mu_{\rm hadr}$ governing 
the strength of the suppressed transition amplitudes 
$\propto \mu_{\rm hadr}/m_Q$. 
There is no a priori reason  to have small power
corrections in ${\cal F}(1)$  at large $\mu_\pi^2$.

\subsection{\boldmath Continuum   $D^{(*)}\pi$  contribution}
\label{sectdpi}

The resonant states are expected to dominate the inelastic transitions at low
excitation energy. Continuum effects are formally $1/N_c$ suppressed and
are usually numerically small,  unless the chiral singularity for the soft pion is
strong enough;  in the case of $I_{\rm inel}$ it is only logarithmic.
Here we give the continuum states $D^{(*)}\pi$ a dedicated consideration since
they populate the lowest energy domain and are characterized by an 
average excitation gap that can be noticeably lower than $\varepsilon_{\rm rad}$.
Moreover, these states contribute to the deviation from
the BPS regime, possibly dominating the deviation in the low-energy regime.
In this subsection we compute their contribution to the zero recoil sum rule with the soft pion technique (see \cite{chiral} for a review). Complementary theoretical considerations can be found in
Sect.~\ref{pionloop}. Here we follow and extend the analysis of Ref.~\cite{vcb}.

\begin{figure}[t]
\vspace*{10pt}
 \begin{center}
\includegraphics[width=12.0cm]{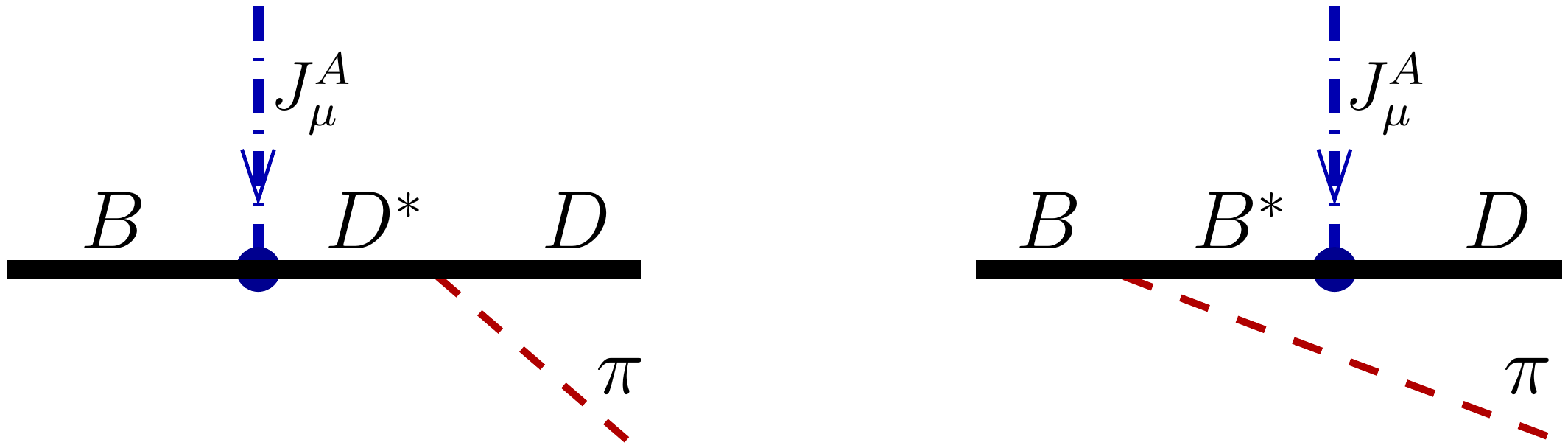}
 \end{center}
\caption{ \small
Pole diagrams for the $B\tto D\!+\!\pi$  amplitude.
 }
\label{figdpi}
\end{figure}

Both $D\pi$ and $D^*\pi$ channels  contribute to $I_{\rm inel}$. 
The $D\pi$ amplitude is given by the sum of the two pole graphs in
Fig.~\ref{figdpi}. 
The pion vertex is parameterized by the 
effective  Lagrangian 
\beq
{\cal L}_{\chi}= 2\, g_{D^*D\pi}(M_D D^*_\mu D\partial^\mu \pi + r\, M_B
B^*_\mu B\partial^\mu \pi).
\label{1130}
\eeq
where $r\!=\!{g_{B^*B\pi}}/{g_{D^*D\pi}}$; heavy quark symmetry implies $r\!=\!1$.
In the heavy quark limit the two diagrams  in Fig.~\ref{figdpi} cancel each other;
all inelastic  transitions  vanish due to heavy quark symmetry.
A nonvanishing result emerges once the mass shifts in the virtual
propagators of heavy mesons are accounted for, or due to 
$r\!\neq\! 1$. The amplitude, for a charged pion, then becomes 
\beq
\frac{1}{2\sqrt{M_B M_D}} \,\langle D^-\pi^+ |\,\vec A\,|B^+\rangle =
-g_{D^*D\pi} \;{\vec p}_\pi\left(
\frac{1}{\varepsilon}\!-\!\frac{r}{\varepsilon \!+\!\Delta}\right) ,
\label{1132}
\eeq
where $(E_\pi,\vec{p}_\pi)$ is the pion four-momentum and we have neglected subleading
 terms.
We have set the weak vertex
to unity, which is legitimate to the first order in the deviations of ${\cal F}(1)$ from
unity. In terms of the $D$ meson energy, $E_D$, the excitation energy  is
$\varepsilon\!=\! E_D\!+\!E_\pi\!-\! M_{D^*}
\!\simeq\!E_\pi\!+\!M_D-\!M_{D^*}\! +\! \vec p_{\pi}^{\,2}/2M_D$.
 $\Delta$ represents a power correction: $ \Delta \!\simeq\! 2/3 \,\mu_G^2 \left(
1/m_c\!+\! 1/m_b\right)\!+\! {\cal O}(\vec p_{\pi}^{\,2}/m_Q)$.
Expanding the amplitude in powers of $1/m_Q$ we would get, for $r\!=\!1$,
$$
\frac{1}{2\sqrt{M_B M_D}} \,\langle D^-\pi^+ |\vec A|B^+\rangle =
-g_{D^*D\pi} \:{\vec p}_\pi \,
\frac{\Delta}{\varepsilon^2},
$$
 which has the correct scaling $\propto 1/m_Q$ for non-diagonal transitions
that violate the heavy quark symmetry. 
However,  the hadronic mass scale in the denominator of the amplitude
is peculiar:  it is the pion energy $E_\pi$,
 which for a light pion  can be significantly lower than the typical QCD mass gap
$\varepsilon_{\rm rad}$. 

The amplitude of  Eq.~(\ref{1132}) at $r\!=\!1$  gives rise to the  spectral density
\beq
 w_{\rm inel}^{D\pi}
=\frac{g_{D^*D\pi}^2}{12\pi^2}\,|{\vec p}_\pi |^3
\frac{\Delta^2 }{\varepsilon^2 (\varepsilon+\Delta)^2};
\label{1136}
\eeq
the logarithmic chiral singularity that emerges upon integration in 
the heavy quark limit is regulated by the spin symmetry breaking term
$2/3 \,\mu_G^2 \left(1/m_c\!+\! 1/m_b\right)$ in $\Delta$, even for a massless pion.
As noted  in \cite{vcb}, however, the constant term is numerically larger than the
chiral logarithm, $\ln \Delta$. The contribution due to neutral pions
is half of the one related to charged pions.

The $B\tto D\pi $ amplitude in Eq.~(\ref{1132}) in fact  receives
additional relativistic corrections;
it is unique because of the soft pion enhancement we have discussed.
Since this enhancement is mild in the integrated
probability, the regular contribution to the amplitude should be included.
In the actual calculations we use the complete relativistic
propagators 
for $B^*$ and $D^*$ and invariant vertices, and do not rely on an expansion in $1/m_Q$.
In other words, we only use  the soft-pion
Lagrangian to model the pion emission amplitude,
assuming that the  couplings do
not vary significantly with energy.  In effect, this implies 
a certain form for the contact terms  which appear in the chiral Lagrangian
at  the subleading $1/m_Q$ order.

In the calculation of the integrated inelastic probability,
there is a subtlety that requires some  care and  was discussed already in  \cite{vcb}.
Since  $M_{D^*}\!>\! M_D\!+\!m_\pi$, the point $\varepsilon\!=\!0$ corresponds to
$|{\vec p}_\pi |\!\simeq \! 39 \MeV$ and the  integration extends to small negative
$\varepsilon$.
At $\varepsilon=0$ the integral has a singularity related to the
$D\pi$ decay of the unstable $D^*$, which should be distinguished from
the actual continuum contribution, and has to be removed.
In practice, the physical regularization is to
introduce the Breit-Wigner factor , replacing  $1/\varepsilon^2$ in Eq.~(\ref{1136}) by
$1/(\varepsilon^2\!+\!\Gamma^2/4)$, where $\Gamma$ is the decay width of $D^*$.
In actuality $\Gamma$ is small compared even to the energy release in $D^*\tto D\pi$.
Therefore including the width serves only to
regularize the integral. Adopting it, integration around
$\varepsilon\!=\!0$ yields unity, the probability of $B\tto D^*$ we start with,
where $D^*$ is represented by the $D\pi$-resonance. The integration over $\varepsilon$
is then carried out  with $|\varepsilon|\!>\!\varepsilon_{\rm min} \!\gg\!\Gamma$.
The resulting inelastic integral does not depend on the choice of 
$\epsilon_{\rm min}$ as long as 
$\Gamma\!\ll\!\epsilon_{\rm min}\!\ll\!M_{D^*}\!-\!M_D\!-\!m_\pi$ holds. 
An accurate analysis shows that for all practical purposes the integral 
can simply be evaluated by setting $m_\pi\!=\! M_{D^*}\!-\!M_D$.

\thispagestyle{plain}
\begin{figure}[t]
\vspace*{-.7pt}
 \begin{center}
\includegraphics[width=8.5cm]{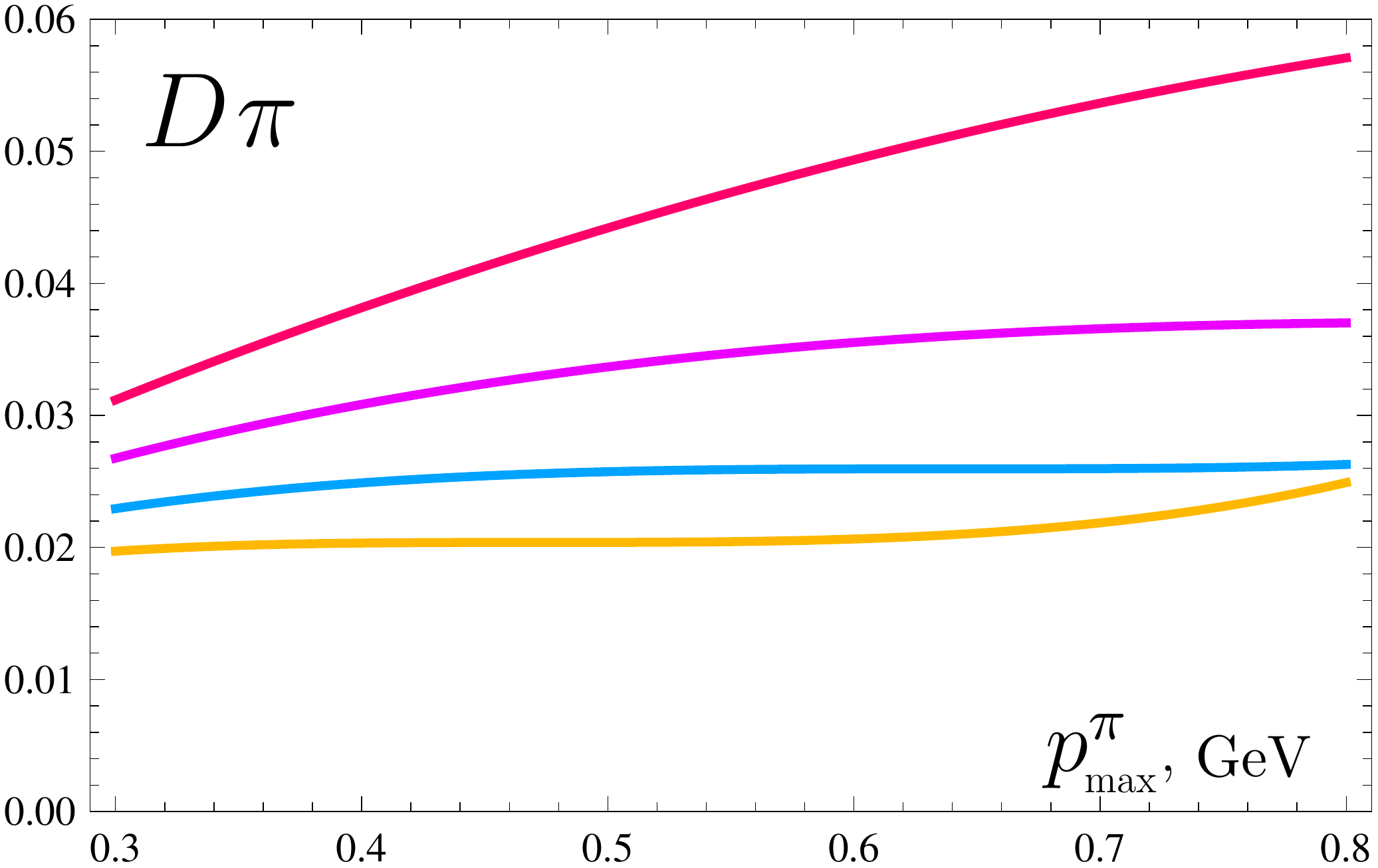}
 \end{center}\vspace*{-5pt}
\caption{ \small
Nonresonant $D\pi$ contribution including both charged and neutral pion
depending on the maximal pion momentum, at
$g_{D^*D\pi}\!=\!4.9\GeV^{-1}$ which corresponds to
$\Gamma_{D^{*+}}\!=\!96\,{\rm keV}$. The plots show, from bottom to top,
$r\!=\!1$, $0.8$,
 $0.6$ and $0.4$. }
\label{dipicont}
\end{figure}

It turns out that numerically the most 
important effect comes from  the difference in the pion-meson
couplings in the charm and beauty sectors, $r\!\neq\! 1$.  Various studies suggest 
$r\lsim 1$ \cite{khod}. 
Fig.~\ref{dipicont} shows the integrated $w_{\rm inel}^{D\pi}$ as a function of
the upper cutoff on the pion momentum, $p_\pi^{\rm max}$, for a few values of $r$.
 Formally, $p_\pi^{\rm max}$ is related to the maximum excitation energy $\varepsilon_M$:
$$
\varepsilon_M= \sqrt{(p_\pi^{\rm max})^2+m_\pi^2}+ 
\sqrt{(p_\pi^{\rm max})^2+M_D^2}-M_{D^*}\,;
$$
however, a lower cutoff may effectively be set by the domain of
applicability of the soft-pion approximation. It is reasonable to stop
at least somewhat below the expected `radial' resonance domain, 
around $p_\pi^{\rm max}\!\approx\! 0.6\GeV$: even if the
amplitude grows with $\vec p_\pi$, at some point this contribution starts to
belong to the resonant radial excitations and should be excluded to avoid
double counting.

\begin{figure}[t]
\vspace*{-7pt}
 \begin{center}
\includegraphics[width=8.5cm]{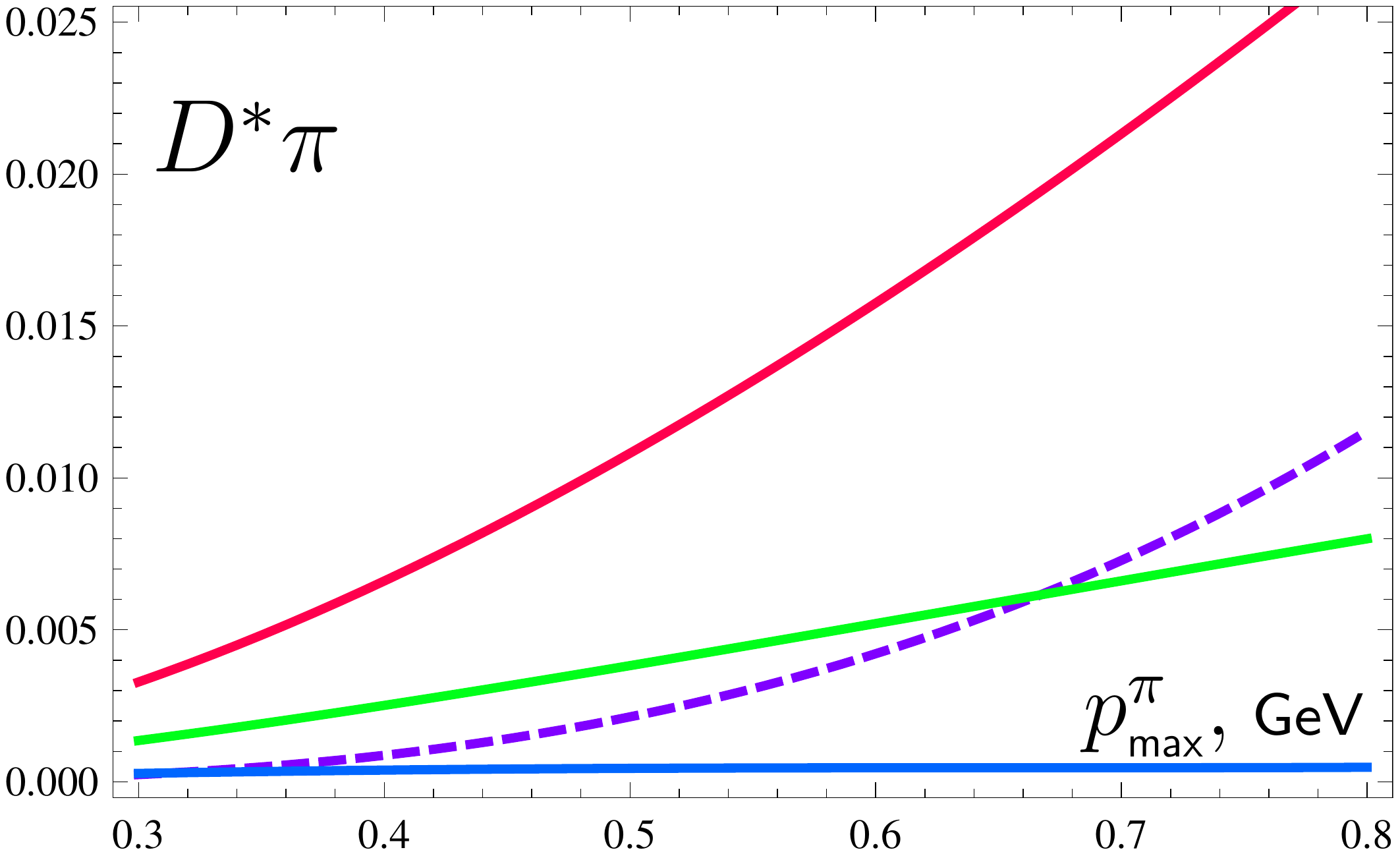}
 \end{center}\vspace*{-5pt}
\caption{ \small
Nonresonant { $D^*\pi$} contribution depending on the maximal pion
momentum, at
$g_{D^*D^*\pi}\!=\!4.9\GeV^{-1}$. The solid lines correspond, from bottom to top,
to $\tilde r\!=\!1$, $0.8$ and $0.6$, while the dashed line refers to
$\tilde r\!=\!1.3$.
 }
\label{distarpicont}
\end{figure}

The  $D^*\pi$ channel  has been considered only in 
Ref.~\cite{f0short}, even though it is required for consistency with the spin-symmetry
structure of the corrections already at $1/m_Q^2$. The pattern
of the $1/m_Q$ mass corrections in the intermediate poles is now reversed,
and the contribution at low pion momentum is suppressed. Altogether it
turns out very small unless the difference between $B^*B\pi$ and $D^*D\pi$ vertices is
appreciable. The numerical results are shown in
Fig.~\ref{distarpicont}.\footnote{Due to a typo in the plotting
notebook the plot in Ref.~\cite{f0short} showed values larger by a
  factor of $\pi$.}

\thispagestyle{plain}

The calculation for $D^*\pi$ proceeds in a way similar to the $D\pi$ case. 
The `bare' weak vertices for the mesons are now
\bea
\nonumber
\matel{D^*(\varepsilon)}{\bar{c}\gamma_\mu\gamma_5 b}{B}
\msp{-5}&=& \msp{-5}
2\sqrt{M_B M_{D^*}}\;\varepsilon_\mu^* \,, \\
\matel{D^*(\varepsilon)}{\bar{c}\gamma_\mu\gamma_5 b}{B^*(\varepsilon')}
\msp{-5}&=& \msp{-5}
i\sqrt{M_{B^*} M_{D^*}}\;\epsilon_{\mu\nu\rho\lambda}\:\varepsilon_\nu^*
\varepsilon'_\rho \left(\frac{P_\lambda}{M_{D^*}}+ 
\frac{P'_\lambda}{M_{B^*}}\right),
\label{1160}
\eea
where $P$ and $P'$ are the four-momenta of $D^*$ and $B^*$ mesons,
respectively. The pion Lagrangian paralleling Eq.~(\ref{1130}) is 
\beq
{\cal L}_{\rm D^*D^*\pi}= 2\, g_{D^*D^*\pi}\; 
\epsilon_{\mu\nu\rho\lambda}\:
D^*_\mu \,\partial_\nu D^*_\rho\,  \partial_\lambda \pi, \qquad
g_{D^*D^*\pi}= \tilde r\, g_{D^*D\pi},
\label{1162}
\eeq
from which the $D^*\tto D^*\pi$ vertex is derived.
The required diagrams mirror Figs.~(\ref{figdpi}),
with $D$ replaced by $D^*$. The point $\varepsilon\!=\!0$ is now below
the threshold and no subtleties in the integration occurs. 
Again, the neutral pion contribution to $w_{\rm inel}$ is half of that for the
charged pion.

The pion couplings $g_{B^*B\pi}$ and  $g_{D^*D^*\pi}$ have not been
measured experimentally. There is theoretical evidence that $1/m_c$
corrections in the coupling should indeed be significant
\cite{khod}; similar conclusions have been reported from the lattice 
studies \cite{latticeDpi}.
The  spin symmetry violating difference
between $g_{D^*D^*\pi}$ and  $g_{D^*D\pi}$ has not been addressed yet. On
 general grounds it may likewise be thought to be significant; we expect to have
its estimate from the  QCD sum rules.

We finally  arrive at a few percent continuum contribution to $I_{\rm inel}$:
$$
I_{\rm inel}^{D^{(*)}\pi} \approx 3 \mbox{ to } 5\%\,.
$$
This is about a fourth of the resonance estimate Eq.~(\ref{176}) and therefore
conforms to the general expectation.  A clarifying comment is in order in this
respect.

The above estimate of the $D^{(*)}\pi$ effects should  not be
added to the principal numeric estimate of Eq.~(\ref{176}). The corresponding
hadronic states contribute to the nonlocal correlators along with the resonant
states and therefore have been accounted for in the observed meson masses used
there as inputs. We will come back to this subject in Section 6.3.
 The $D^{(*)}\pi$  states  effectively lower the average excitation
energy $\tilde \varepsilon$ compared to $\varepsilon_{\rm rad}$; however,
the continuum contribution is relatively suppressed and this is not a prominent
effect.
On the other hand, should one discard the estimate of the inelastic contribution of
Eq.~(\ref{176}),  it is reasonable to include at least
 $ I_{\rm inel}^{D^{(*)}\pi}$ in the   unitarity upper bound for ${\cal F}(1)$; this 
lowers it by about $2\%$ down to
approximately  ${\cal F}(1)\!< \!0.90$.

   \vspace*{3pt}
  {\sl To conclude} the first part of the paper, we refer the reader to  Sect.~\ref{conc}, where 
  our  numeric conclusions  for ${\cal F}(1)$ can be found, see   Eqs.~(\ref{c20}), (\ref{c21}) and   (\ref{c24}).

\section{A closer look at the heavy quark excited states}
\label{alook}

The zero-recoil transition amplitudes between the $B$ meson and the excited
charm states appeared in  Sec.~\ref{winel}, where they gave 
the part of the  power-suppressed correction to 
${\cal F}(1)$ associated with the overlap of nonrelativistic wavefunctions. 
They are encountered in various  applications of the heavy quark expansion,
together with the spectrum of the
corresponding states, and deserve a dedicated analysis.

So far mostly the $P$-wave excitations of the ground-state  
mesons have been considered in the literature. They play a primary
role in the small velocity regime and enter the sum rules which
strongly constrain possible values of the main nonperturbative heavy
quark parameters \cite{rev,ioffe}. Here we extend the analysis to
higher states, in particular  to those which are commonly 
referred to as radial or $D$-wave  excitations, 
assuming the heavy quark (static) limit. 
In the actual heavy mesons  we may have
additional excited states associated with the interacting spin and
gauge field degrees of freedom. 

In our analysis we will use
heavy quarks deprived of their spin \cite{rev,ioffe}; this is 
physically motivated since in
the static limit heavy quark spin degrees of freedom decouple, and 
this  greatly simplifies all the expressions, as long as no velocity change
is considered. This formalism leads in a simple and transparent manner 
to the trace formalism of the rest frame;
a discussion of the connection  to the conventional formalism can 
be found in Re.~\cite{hiord}.
We remind that with spinless heavy quarks
the ground-state heavy mesons are spin-$\frac{1}{2}$  particles; the
corresponding states are denoted by $\state{\Omega_0}$ and 
their two-component spinor
wavefunctions, when needed explicitly, 
are denoted by $\Psi_0$. The Pauli matrices $\vec \sigma$ act on the spinor
indices of the heavy hadron wavefunctions. 
To specify our convention we set the parity of $\Omega_0$ positive. 

Translating the relations in this formalism to the case of expectation values in 
actual $B$ mesons is
straightforward. The spin-singlet quantities are in a one-to-one
correspondence. For the spin-triplet ones 
the fact that $\vec \sigma_Q \!=\! -\vec
\sigma$ for the spinless state like $B$ meson ($\vec \sigma_Q$ is the spin
matrix acting on the heavy quark indices) 
immediately gives the required translation.

\subsection{Model-independent spectral representation}
\label{modind}

Heavy quark theory requires the tower of heavy quark transition 
matrix elements with zero
spatial momentum transfer, ordered according to the number of covariant
derivatives (heavy quark momentum). They enter 
both the expansion in velocity and the non-relativistic $1/m_Q$ expansion. 
In practice we are interested in the matrix elements of the ground state.

The unit operator $\bar{Q} Q$ has trivial matrix elements at vanishing
velocity change. The operator $\bar{Q} \pi_k Q$  with a single
derivative describes the dipole transitions which connect $\Omega_0$ to
 the $\frac{1}{2}$ and $\frac{3}{2}$ $\;P$-wave states with negative
parity. At the  lowest order in the expansion we need the general inclusive two-point
correlator with two derivatives, or its absorptive part:
\bea
\nonumber
\frac{1}{\pi}{\rm Im}\, P_{jk}(\omega) &\msp{-4.5}= \msp{-4.5}& \frac{1}{2\pi}\int 
{\rm d}^3x \int\!\! {\rm d}x_0\; 
e^{-i\omega x_0} \frac{1}{2M_Q} \matel{\Omega_0} 
{Q^\dagger \pi_j Q(x)\;  Q^\dagger \pi_k Q(0)}{\Omega_0}  \\
&& \msp{20}
={\cal T}^{(\frac{1}{2}^-)}(\omega)\Psi_0^\dagger(\delta_{jk}\!+\!\sigma_{jk}) \Psi_0+
{\cal T}^{(\frac{3}{2}^-)}(\omega)\Psi_0^\dagger
(2\delta_{jk}\!-\!\sigma_{jk})\Psi_0\, ;\qquad
\label{2270.02} 
\eea 
here and in what  follows we use the notation
$$ \sigma_i \sigma_j= \delta_{ij} + \sigma_{ij}, \qquad\quad \sigma_{ij}=i \epsilon_{ijk} \sigma_k \, .$$
The two spectral densities generalize the conventional $\frac{1}{2}$ and 
$\frac{3}{2}$ SV amplitudes to continuum states:
\beq
{\cal T}^{(\frac{1}{2}^-)}(\omega)
= \omega^2\, 
\frac{{\rm d}|\tau_{1/2}(\omega)|^2} {{\rm d} \omega}, \qquad\quad
{\cal T}^{(\frac{3}{2}^-)}(\omega)= \omega^2\,
\frac{{\rm d}|\tau_{3/2}(\omega)|^2} {{\rm d} \omega};
\label{2272.02} 
\eeq 
their relation to the inelastic part of the SV structure functions 
$W_{\pm}(\omega)$ introduced in 
Ref.~\cite{newsr} is 
\beq
W_+(\omega)= \frac{2{\cal T}^{(\frac{3}{2}^-)}(\omega)\!+\! 
{\cal T}^{(\frac{1}{2}^-)}(\omega)}{\omega^2}, \qquad
W_-(\omega)= \frac{2{\cal T}^{(\frac{3}{2}^-)}(\omega)\!-\! 
2{\cal T}^{(\frac{1}{2}^-)}(\omega)}{\omega^2}, \qquad 
\omega \!>\!0.
\label{2272.04} 
\eeq

To extend the analysis to the relevant radially excited states  
we consider the spectral density of the general 
two-point correlation function of the products of two spatial 
covariant derivatives acting on the heavy quark:
\beq
R_{ijkl}(\omega)=\frac{1}{2\pi}\int\! {\rm d}^3x \int\! {\rm d}x_0\; 
e^{-i\omega x_0} \frac{1}{2M_Q} \matel{\Omega_0} 
{Q^\dagger \pi_i\pi_j Q(x)\: Q^\dagger \pi_k\pi_l Q(0)}{\Omega_0}\;.
\label{128} 
\eeq
It can be decomposed into
the invariant structures corresponding to three possible classes of the 
intermediate heavy quark
states with $j\!=\!\frac{1}{2}$,
$j\!=\!\frac{3}{2}$ and $j\!=\!\frac{5}{2}$, respectively: 
\bea
\nonumber
R_{ijkl}(\omega)&\msp{-4}=\msp{-4}&
\Psi_0^\dagger \left[\frac{1}{9}\rho_p^{(\frac{1}{2}^+)}(\omega) \delta_{ij}
\delta_{kl}- \frac{1}{18}\,\rho_{pg}^{(\frac{1}{2}^+)}(\omega)(
\delta_{ij} \sigma_{kl}\!+\! \sigma_{ij} \delta_{kl} ) +
\frac{1}{36}\rho_{g}^{(\frac{1}{2}^+)}(\omega)\sigma_{ij}
\sigma_{kl}
\right]\Psi_0\\
\nonumber
&\msp{-4}+\msp{-4}& 
\Psi_0^\dagger \left[\frac{1}{200}\rho_f^{(\frac{3}{2}^+)}(\omega)
\left(
\delta_{ik} \delta_{jl}+\delta_{il} \delta_{jk} -
\mbox{$\frac{2}{3}$}\delta_{ij} \delta_{kl} +
\frac{1}{2} (\sigma_{ik} \delta_{jl}+\sigma_{il} \delta_{jk}+
\sigma_{jk} \delta_{il}+\sigma_{jl} \delta_{ik})
\right) \right.
\\
\nonumber
&&\msp{4}
+\frac{1}{80}\,\rho_{fg}^{(\frac{3}{2}^+)}(\omega)\left(
i\epsilon_{ijk}\sigma_{l} + i\epsilon_{ijl}\sigma_{k}+
i\epsilon_{jkl}\sigma_{i}+i\epsilon_{ikl}\sigma_{j} -\frac{2}{3}
(\delta_{ij}\sigma_{kl} +\delta_{kl}\sigma_{ij})
\right) \\
&& \msp{3}\left.
-\frac{1}{16}\,\rho_{g}^{(\frac{3}{2}^+)}(\omega)
\left(\frac{2}{3}(\delta_{ik}\delta_{jl}-\delta_{il}\delta_{jk})
+\frac{1}{3}(i\epsilon_{ijk}\sigma_l  -
i\epsilon_{ijl}\sigma_k )
\right) \label{130} 
\right]\Psi_0 \\
&\msp{-4}+\msp{-4}& \frac{1}{10}\rho^{(\frac{5}{2}^+)}(\omega)\,
\Psi_0^\dagger \left[ 3(\delta_{ik} \delta_{jl}+\delta_{il} \delta_{jk} -
\mbox{$\frac{2}{3}$}\delta_{ij} \delta_{kl}) - (\sigma_{ik}
\delta_{jl}+\sigma_{il} \delta_{jk}+ \sigma_{jk}
\delta_{il}+\sigma_{jl} \delta_{ik}) \right]\Psi_0. \nonumber
\eea 
 The meaning of the three invariant functions for  $j\!=\!\frac{1}{2}$ and
$j\!=\!\frac{3}{2}$ will become clear shortly.

The nonlocal correlators which are relevant to 
our analysis of $I_{\rm inel}$ have been introduced   in Eqs.~(\ref{66a}).
In addition, it is also useful to introduce  
$$
\tilde\rho_{\pi\pi}=
\int {\rm d}^4 x\,i|x_0|
\; \frac{1}{4M_{B}}\langle B| i T\{\bar b \vec\pi\,^2b(x),
\;\bar b\vec\pi\,^2b(0)\}|B\rangle ',
$$
$$
\tilde\rho_{\pi G}=
 \int {\rm d}^4 x\,i|x_0|
\; \frac{1}{2M_{B}}\langle B|iT\{\bar b \vec\pi\,^2b(x),
\;\bar b\vec\sigma \vec B b(0)\}|B\rangle ',
$$
\begin{equation}
\frac{1}{3}\tilde\rho_{S}\delta_{ij}\delta_{kl}+\frac{1}{6}
\tilde\rho_{A}(\delta_{ik}\delta_{jl}-
\delta_{il}\delta_{jk})=
\int {\rm d}^4 x\,i|x_0|
\; \frac{1}{4M_{B}}\langle B|iT\{\bar b \sigma_i B_kb(x),\;
\bar b \sigma_j B_l b(0)\}|B\rangle ' .
\label{66}
\end{equation}
The extra factor $i|x_0|$ compared to Eqs.~(\ref{66a}) is simply an extra 
power of excitation energy $\varepsilon$ in the denominator in the spectral
representation, see  Eqs.~(\ref{150}) below. 
Using the $\tilde{\rho}$ correlators one can directly write the 
inelastic contribution to the sum rule (\ref{84}),  cf.\ Eq.~(\ref{124}):
\beq
I_{\rm inel}\!=\!\!\frac{-(\tilde\rho_{\pi G}\!+\!\tilde\rho_A)}{3m_c^2}
- \frac{2\tilde\rho_{\pi\pi}\!+\!\tilde \rho_{\pi G}}{3m_c m_b}+
\frac{\tilde \rho_{\pi\pi}\!+\!\tilde \rho_{\pi G}\!+\!
\tilde\rho_S\!+\!\tilde\rho_A}{4} 
\left(\!\frac{1}{m_c^2}\!+\!  \frac{2}{3m_c m_b}\!+\! \frac{1}{m_b^2} \right)
 + {\cal O}\!\left(\!\frac{1}{m_Q^3}\!\right)\!.\!\!
\label{125}
\eeq

All these nonlocal correlators are readily expressed in terms of 
the above  spectral densities:
\begin{alignat}{2}
\nonumber
\rho_{\pi\pi}^3 &=
\int {\rm d} \omega\;  \frac{\rho_{p}^{(\frac{1}{2}^+)}(\omega)}{\omega}
& \qquad 
\tilde\rho_{\pi\pi} &=
\int {\rm d} \omega\;  \frac{\rho_{p}^{(\frac{1}{2}^+)}(\omega)}{\omega^2}
\\
\nonumber
-\rho_{\pi G}^3  & =
\int {\rm d} \omega\;  \frac{2\rho_{pg}^{(\frac{1}{2}^+)}(\omega)}{\omega}
& \qquad -\tilde \rho_{\pi G}  & =
\int {\rm d} \omega\;  \frac{2\rho_{pg}^{(\frac{1}{2}^+)}(\omega)}{\omega^2}
\\
\nonumber
\rho_S^3\; & =
\int {\rm d} \omega  \left( \frac{\rho_{g}^{(\frac{1}{2}^+)}(\omega)}{3\omega} +
\frac{\rho_{g}^{(\frac{3}{2}^+)}(\omega)}{2\omega}\right)
& \qquad 
\tilde \rho_S\; & =
\int {\rm d} \omega  \left(\frac{\rho_{g}^{(\frac{1}{2}^+)}(\omega)}{3\omega^2} +
\frac{\rho_{g}^{(\frac{3}{2}^+)}(\omega)}{2\omega^2}\right)
\\
\rho_A^3 \; & =
\int {\rm d} \omega \left(
\frac{2\rho_{g}^{(\frac{1}{2}^+)}(\omega)}{3\omega} -
\frac{\rho_{g}^{(\frac{3}{2}^+)}(\omega)}{2\omega} \right),
& \qquad  
\tilde \rho_A \; & =
\int {\rm d} \omega \left(
\frac{2\rho_{g}^{(\frac{1}{2}^+)}(\omega)}{3\omega^2} -
\frac{\rho_{g}^{(\frac{3}{2}^+)}(\omega)}{2\omega^2} \right) \,.
\label{150} 
\end{alignat}
The integration runs over positive $\omega$; the point $\omega\!=\!0$ is
excluded. The integration is also cut at large $\omega$ for $\omega\!>\!\mu$
according to the normalization prescription in the kinetic scheme.

Note that neither $\rho_{f,fg}^{(\frac{3}{2}^+)}$ nor
$\rho^{(\frac{5}{2}^+)}$ can contribute above. Using
these relations one can  express the $1/m_Q^2$ inelastic 
transition probabilities for actual $B$ mesons. The following form appears
particularly convenient for the analysis: 
\bea
\nonumber
w_{\rm inel}(\omega) \msp{-4.5}
& = & \msp{-5.5} \left(\!\frac{1}{2m_c}\!-\!\frac{1}{2m_b}\!\right)^2 
\! \frac{\rho_{p}^{(\frac{1}{2}^+)}(\omega)}{\omega^2} \!+\! 
\left(\!\frac{1}{2m_c}\!-\!\frac{1}{2m_b}\!\right) \!
\left(\!\frac{1}{3m_c}\!+\!\frac{1}{m_b}\!\right) 
\frac{\rho_{pg}^{(\frac{1}{2}^+)}(\omega)}{\omega^2}\msp{-.2}
\\
 && 
+\, \frac{1}{4}\left(\frac{1}{3m_c}\!+\!\frac{1}{m_b}\right)^2 
\frac{\rho_{g}^{(\frac{1}{2}^+)}(\omega)}{\omega^2} + 
\frac{1}{6m_c^2} \frac{\rho_{g}^{(\frac{3}{2}^+)}(\omega)}{\omega^2} .
\label{152.02}
\eea
The analogous representation for the vector current-induced transitions 
is given in Eq.~(\ref{152v}).

\subsubsection{Intermediate state contributions}
\label{interm}

Here we consider the contribution of an individual positive-parity state
to the above correlators.
 Again, we first briefly remind what happens for the $P$-wave states.

Following Ref.~\cite{rev} we denote the $\frac{1}{2}$ and 
 $\frac{3}{2}$ $P$-wave states by $\phi$
and $\chi$ and describe them with the two-dimensional spinor $\phi$ and the non-relativistic
 Rarita-Schwinger 
spinors $\chi_j$, respectively.
There are successive families of these states which differ by their excitation
energy $\epsilon_n\!=\!M_n\!-\!M_0$; we typically
omit the index marking the excitation number. The
$\frac{3}{2}$-spinors $\chi_j$ satisfy $\sigma_j \chi_j=0$ and the
normalization fixes the sum over their polarizations:
\begin{equation}
\sum_{\rm \lambda} \chi _i (\lambda)\chi^{\dagger}_j(\lambda) =
\delta _{ij} -
\frac{1}{3} \sigma _i \sigma _j \;.
\label{d720}
\end{equation}
The dipole amplitudes are related to the conventional small-velocity
amplitudes $\tau$ by  
\begin{equation}
\matel{\phi ^{(n)}}{\pi_j}{\Omega _0} =
\epsilon_n \tau_{1/2}^{(n)} \phi ^{(n)\dagger}\sigma _j \Psi _0 \; ,\qquad
\matel{\chi^{(m)}}{\pi_j}{\Omega _0} =
\sqrt{3}\,\epsilon_m \tau_{3/2}^{(m)} \chi_j^{(m)\dagger} \Psi _0.
\label{d726}
\end{equation}

The two-derivative heavy quark operators at vanishing total spatial
momentum acting on $\Omega_0$ may create $\frac{1}{2}^+$,  $\frac{3}{2}^+$ or 
$\frac{5}{2}^+$ states. The ground state itself has $j^P\!=\!\frac{1}{2}^+$; we
will always assume however that  $\frac{1}{2}^+$ refers to an excited
multiplet. We describe the $\frac{1}{2}^+$,  $\frac{3}{2}^+$ and $\frac{5}{2}^+$ states
by a conventional spinor $\chi$ and
by the Rarita-Schwinger spinors  $\chi_j$ and  $\chi_{jl}$,
respectively, with the following constraints:\footnote{The $P$-wave 
wave-functions will no longer  appear in what follows and we use the same notation  for the
$\frac{3}{2}$-hadrons of  opposite parity.}
\beq
\sigma_j\chi_j=0\,; \qquad \chi_{jk}= \chi_{kj}, \quad  \chi_{jj}=0, \quad
\sigma_j\chi_{jk}=0\,.
\label{2095b}
\eeq
The sum over polarizations $\lambda$ giving the spin part of the propagator 
equals to
\bea
\nonumber
\sum_{\lambda=1}^2 \chi(\lambda) \chi^{\dagger}(\lambda) &\msp{-5}=\msp{-5}&
1\\
\nonumber
\sum_{\lambda=1}^4 \chi_{i}(\lambda) \chi^{\dagger}_{j}(\lambda) &\msp{-5}=\msp{-5}&
\delta_{ij}-\frac{1}{3}\sigma_i\sigma_j\\
\sum_{\lambda=1}^6 \chi_{ij}(\lambda) \chi_{kl}^\dagger(\lambda)  &\msp{-5}=\msp{-5}&
\frac{3}{10}(\delta_{ik} \delta_{jl}\!+\!\delta_{il} \delta_{jk} \!-\!
\mbox{$\frac{2}{3}$}\delta_{ij} \delta_{kl}) \!-\!
\frac{1}{10}(\sigma_{ik} \delta_{jl}\!+\!\sigma_{il} \delta_{jk}+
\sigma_{jk} \delta_{il}\!+\!\sigma_{jl} \delta_{ik}).
\label{2512}
\eea

Following the standard notation for the 
diagonal matrix element of the ground state,
\beq
\matel{\Omega_0}{\pi_k\pi_l}{\Omega_0}= 
\frac{\mu_\pi^2}{3} \,\delta_{kl}\,\Psi_0^\dagger\Psi_0 - \frac{\mu_G^2}{6} 
\,\Psi_0^\dagger \sigma_{kl}\Psi_0 \,,
\label{2093}
\eeq
we parameterize 
\beq
\matel{\mbox{$\frac{1}{2}^+$}}{\pi_k\pi_l}{\Omega_0}= 
\frac{P}{3} \,\delta_{kl}\,\chi^\dagger\Psi_0 - \frac{G}{6} 
\,\chi^\dagger \sigma_{kl}\Psi_0 \,.
\label{2095g}
\eeq
The transitions amplitude into $\frac{3}{2}^+$-states have a symmetric and
an antisymmetric structure parameterized by constants $f$ and $g$:
\beq
\matel{\mbox{$\frac{3}{2}^+$}}{\pi_k\pi_l}{\Omega_0}= 
\frac{f}{20} (\chi_k^\dagger \sigma_l+ \chi_l^\dagger \sigma_k) \Psi_0 + 
\frac{g}{4} \,i\epsilon_{klm}\,\chi_m^\dagger \Psi_0 \,,
\label{2095}
\eeq
while the $\frac{5}{2}^+$  amplitude depends on a single parameter $h$:
\beq
\matel{\mbox{$\frac{5}{2}^+$}}{\pi_k\pi_l}{\Omega_0}= 
h \, \chi_{kl}^\dagger \Psi_0 \,.
\label{2095c}
\eeq
The residues $P,G$, $f,g$ and $h$ are different for each
multiplet of the excited heavy state. 

A particular hadronic state with excitation energy
$\epsilon_n$ is associated with the following factorized contributions to the 
invariant spectral densities $\rho_{p, pg, g}^{(\frac{1}{2}^+)}(\omega)$, 
$\rho_{f, fg,  g}^{(\frac{3}{2}^+)}(\omega)$ and $\rho^{(\frac{5}{2}^+)}(\omega)$
in Eq.~(\ref{130}):
\begin{alignat}{3}
\nonumber
\delta \rho_{p}^{(\frac{1}{2}^+)}(\omega) &=
P^2\,\delta(\omega\!-\!\epsilon_n),&  \qquad \delta
\rho_{g}^{(\frac{1}{2}^+)}(\omega) &=
G^2\,\delta(\omega\!-\!\epsilon_n),& \qquad \delta
\rho_{pg}^{(\frac{1}{2}^+)}(\omega) &= PG\,\delta(\omega\!-\!\epsilon_n)\\
\nonumber
\delta \rho_{f}^{(\frac{3}{2}^+)}(\omega) &=
f^2\,\delta(\omega\!-\!\epsilon_n),&  \qquad \delta
\rho_{g}^{(\frac{3}{2}^+)}(\omega) &=
g^2\,\delta(\omega\!-\!\epsilon_n),& \qquad \delta
\rho_{fg}^{(\frac{3}{2}^+)}(\omega) &= fg\,\delta(\omega\!-\!\epsilon_n)\\
\delta \rho^{(\frac{5}{2}^+)}(\omega) &=
h^2\,\delta(\omega\!-\!\epsilon_n). & & & & 
\label{2140} 
\end{alignat}
The ground-state contribution to $R_{ijkl}$  is
located at $\omega\!=\!0$ 
and  is excluded from the nonlocal
correlators $\rho^3$ and $\tilde \rho$ of Eqs.~(\ref{150}); it
is obtained by taking $P_{(0)}\!=\!\mu_\pi^2$ and $G_{(0)}\!=\!\mu_G^2$
with $\epsilon_0\!=\!0$.

 The spin-$\frac52$ contribution $\delta \rho^{(\frac{5}{2}^+)}(\omega)$ is given by the
square of a single residue $h$ and is non-negative. The factorization
relations lead to the Cauchy inequalities 
\beq 
(\rho_{pg}^{(\frac{1}{2}^+)})^2 \le
\rho_{p}^{(\frac{1}{2}^+)} \rho_{g}^{(\frac{1}{2}^+)}\!, \qquad
(\rho_{fg}^{(\frac{3}{2}^+)})^2 \le
\rho_{f}^{(\frac{3}{2}^+)}\rho_{g}^{(\frac{3}{2}^+)} 
\label{2142}
\eeq 
which turn into equalities for a particular single state
contribution. Furthermore, one has the general inequalities
\beq 
2|\rho_{pg}^{(\frac{1}{2}^+)}(\omega)| \le \rho_{p}^{(\frac{1}{2}^+)}+
\rho_{g}^{(\frac{1}{2}^+)}, \qquad
2|\rho_{fg}^{(\frac{3}{2}^+)}(\omega)| \le
\rho_{f}^{(\frac{3}{2}^+)}+ \rho_{g}^{(\frac{3}{2}^+)}, 
\label{2143}
\eeq 
the first of which will soon be useful.

For completeness we mention the constraints imposed by the BPS limit for
the ground state, which assumes  $\vec \pi \vec \sigma 
\state{\Omega_0}\!=\!0$. Similar to the ground-state expectation values, for
transitions into 
$\frac{1}{2}^+$ states the BPS condition implies $P\!=\!G$; however the relations for 
the excited states are accurate only to first order in the
deviation from the BPS limit. No constraint emerges on $h$, while for the
transitions into $\frac{3}{2}^+$ states BPS implies $f\!=\!g$: 
\beq
\matel{\mbox{$\frac{3}{2}^+$}}{\pi_k \vec \pi \vec \sigma }{\Omega_0}= 
(\mbox{$\frac{1}{4}$}f\!-\!\mbox{$\frac{1}{4}$}g)\chi_k^\dagger\Psi_0 \,.
\label{2095h}
\eeq
For the invariant structures in the spectral density $R_{ijkl}(\omega)$ 
in Eq.~(\ref{130}) the BPS condition leads to  the relations 
\beq
\rho_{p}^{(\frac{1}{2}^+)}(\omega)= \rho_{pg}^{(\frac{1}{2}^+)}(\omega)=
\rho_{g}^{(\frac{1}{2}^+)}(\omega), \qquad
\rho_{f}^{(\frac{3}{2}^+)}(\omega)= \rho_{fg}^{(\frac{3}{2}^+)}(\omega)=
\rho_{g}^{(\frac{3}{2}^+)}(\omega).
\label{2095BPS}
\eeq

 Let us note that when the spin-$\frac{1}{2}$ degrees of freedom associated with
light antiquark in the meson decouple, as in 
the case of purely perturbative calculations,  the
correlator decomposition can be based on the angular 
momentum only, $L\!=\!0, 1, 2$, of the intermediate states composed of the QCD degrees of
freedom still interacting with heavy quark:
\beq
R_{ijkl}^{\rm Fact}(\omega) \!=\! 
\left[\frac{\rho_0(\omega)}{9} \delta_{ij}\delta_{kl}+
\frac{\rho_{1}(\omega)}{16}\,(\delta_{il}\delta_{jk}\!-\!\delta_{ik}\delta_{jl})
+ \frac{\rho_2(\omega)}{10}(\delta_{ik} \delta_{jl}\!+\!\delta_{il}
\delta_{jk} \!-\!
\frac{2}{3} \delta_{ij} \delta_{kl})\right]\! \Psi_0^\dagger \Psi_0, 
\label{130dec} 
\eeq 
with a trivial spinor structure. This would mean 
\bea
\nonumber
\rho_{p}^{(\frac{1}{2}^+)}(\omega) \msp{-4} & = & \msp{-2.5}\rho_0(\omega), \qquad 
\rho_{g}^{(\frac{1}{2}^+)}(\omega)=\mbox{$\frac{3}{4}$}\,\rho_1(\omega), \qquad 
\rho_{g}^{(\frac{3}{2}^+)}(\omega)=\rho_1(\omega),
\quad \\
\rho_{f}^{(\frac{3}{2}^+)}(\omega) \msp{-4} & = & \msp{-4} 8\rho_2(\omega), \qquad 
\rho^{(\frac{5}{2}^+)}(\omega)=\mbox{$\frac{1}{5}$}\,\rho_2(\omega), \qquad
\rho_{pg}^{(\frac{1}{2}^+)}(\omega)=\rho_{fg}^{(\frac{3}{2}^+)}(\omega)=0.
\label{3130} 
\eea 
We will use these relations when address the perturbative $\mu$-dependence 
later on.

\subsection{Hyperfine splitting and the estimate of the correlators}
\label{hyperold}

The analysis of the spin structure and of the factorization properties 
does not constrain the overall scale of the residues and the
significance of the radially excited states, which can however be estimated by considering 
the heavy quark mass dependence of the hyperfine splitting $\Delta M^2$.
This will allow us to fix the magnitude of the nonlocal  correlators in 
Eqs.~(\ref{124}), (\ref{125}).

The $B\!-\!B^*$ mass splitting basically
fixes the value of $\mu_G^2$.  The mass difference between the beauty and charm
states mainly reflects $m_b\!-\!m_c$.
A comparison of $M_{D^*}\!-\!M_D$ and $M_{B^*}\!-\!M_B$ hyperfine
splittings gives information on higher dimensional correlators, and in particular
on the values of two $D\!=\!3$ 
spin-triplet parameters, $\rho_{LS}^3$ and $\rho_{\pi G}^3+\rho_A^3$:
\beq
\Delta M_B=M_B^*-M_B= \frac{2}{3}\frac{\mu_G^2}{m_b}+
\frac{-\rho_{LS}^3+\rho_{\pi
G}^3+\rho_A^3}{3\,m_b^2}+ {\cal O}\left(\frac{1}{m_b^3}\right) 
\label{70}
\eeq
and likewise for charm. 
Following Ref.~\cite{chrom}, we explore this relation cast in  a 
somewhat different form.

As dictated by the heavy quark expansion, for sufficiently heavy quarks the
difference $\Delta M_Q^2\!=\!M_{Q^*}^2\!-\!M_Q^2$ for the vector and the pseudoscalar mesons 
approaches a constant. Yet it has been noticed
long ago that such a relation works well even for light quarks:
\beq
M^2_\rho\!-\!M_\pi^2\simeq M^2_{K^*}\!-\!M_K^2 \simeq M^2_{D^*}\!-\!M_D^2
\simeq M^2_{B^*}\!-\!M_B^2. 
\label{72}
\eeq
Clearly, the heavy quark expansion cannot explain why such a relation extends
down to the light quarks. The approximate equality resembles the universality
of the slope for Regge trajectories and may root in the peculiarities of the
strong dynamics.  Moreover, actual QCD predicts that such a relation must be
violated for sufficiently heavy quarks due to the perturbative renormalization
of the chromomagnetic operator of the heavy quark which has a nontrivial
anomalous dimension. As was also  noted long ago, the observed $12\%$
decrease in the mass square difference in the $B$ system compared to $D$
mesons fits reasonably well the naive estimates of this perturbative
evolution.

It is therefore tempting to assume the independence of the mass-square 
splitting as a phenomenological property of {\sl soft} strong 
dynamics yet to be understood, and to consider
its consequences in the context of the heavy quark expansion where 
we can vary the heavy quark mass. 
Using the mass formulae we have 
\beq 
\Delta M^2_Q\!=\! \Delta M_Q (2M_Q \!+\!\Delta M_Q)=\!
\frac43 c_G(\mu,m_Q) \mu_G^2 +\frac23 \frac{ \rho_{\pi G}^3+\rho_A^3-\rho_{LS}^3+2\La \mu_G^2}{m_Q}+
{\cal O}\!\left(\!\frac1{m_Q^2}\!\right)\!, \!
\label{dm2}
\eeq
where $c_G$ is the Wilson coefficient of the 
chromomagnetic operator in the heavy quark Hamiltonian. 
If the perturbative $m_Q$-evolution of $c_G$ accounted completely 
for the observed difference between $\Delta M^2_B$ and $\Delta M^2_D$,
the $m_Q$-independence of
$\Delta M_Q^2$ to the first nontrivial order in $1/m_Q$ 
would imply 
\beq
-(-\rho_{LS}^3+ \rho_{\pi G}^3+\rho_A^3) \simeq 2\La \mu_G^2\,.
\label{74}
\eeq
This relation will guide our analysis below.  

Numerically using  $\La \!=\! 600\MeV$, $\mu_G^2\!=\!0.3\GeV^2$ 
and $\rho^3_{LS}\!=\!-0.1\GeV^3$ we would get
\beq
-(\rho_{\pi G}^3+\rho_A^3) \approx  0.45\GeV^3 
\label{164}
\eeq
which indicates that the nonlocal
correlators $\rho^3$ are in general sizable. The negative sign for
$\rho_{\pi G}^3 \!+\! \rho_A^3$ is expected from   BPS arguments
\cite{chrom}. 

To proceed more quantitatively we introduce a factor $\kappa$ in
Eq.~(\ref{74}), 
to account for the actual  mismatch between the observed mass dependence  and 
the dependence stemming from $c_G$:
\beq
-(-\rho_{LS}^3+\rho_{\pi G}^3+\rho_A^3) = 2(1\!+\!\kappa) \,\La \mu_G^2\;.
\label{76}
\eeq
$\kappa$ does not include higher power corrections, which we  address later, and 
can be  defined through 
\beq
\kappa \, \La \mu_G^2= \lim_{m_Q\to
\infty} m_Q^2 \frac{\rm d}{{\rm d} m_Q} \left[\frac34 \Delta M_Q^2-  c_G(m_Q)
\mu_G^2  \right], 
\label{160}
\eeq
where the logarithmic derivative of $c_G \mu_G^2$ is related to the exact 
anomalous dimension of the chromomagnetic operator, times $\Delta M^2_Q$.

In simple words, the  soft part  of $\Delta M^2_Q$, identified by subtracting the 
(Wilsonian) short-distance renormalization factor, approaches a finite 
limit as $m_Q \tto \infty$. 
The assumption underlying the approximation of small $|\kappa|$ is that such
 a soft part is nearly a constant in  a wide  range 
of heavy quark masses, as Eq.~(\ref{72}) would naively suggest.

Let us now  look at the perturbative renormalization $c_G$. 
The one- and two-loop \cite{czargroz} contributions are known. At first glance
in the evolution from beauty to charm, namely in the ratio $c_G(m_c)/c_G(m_b)$, 
the two-loop contribution is quite sizable. 
However, the bulk of the higher-loop perturbative enhancement comes from
growing strong coupling at small momenta of virtual gluons. This
large-coupling domain must be removed  from the perturbative corrections to
avoid double counting. The subtraction piece is power-suppressed yet 
important for charm, especially in the effect of running of the strong coupling.

\thispagestyle{plain}
\begin{figure}[t]
\vspace*{-7pt}
 \begin{center}
\includegraphics[width=8.5cm]{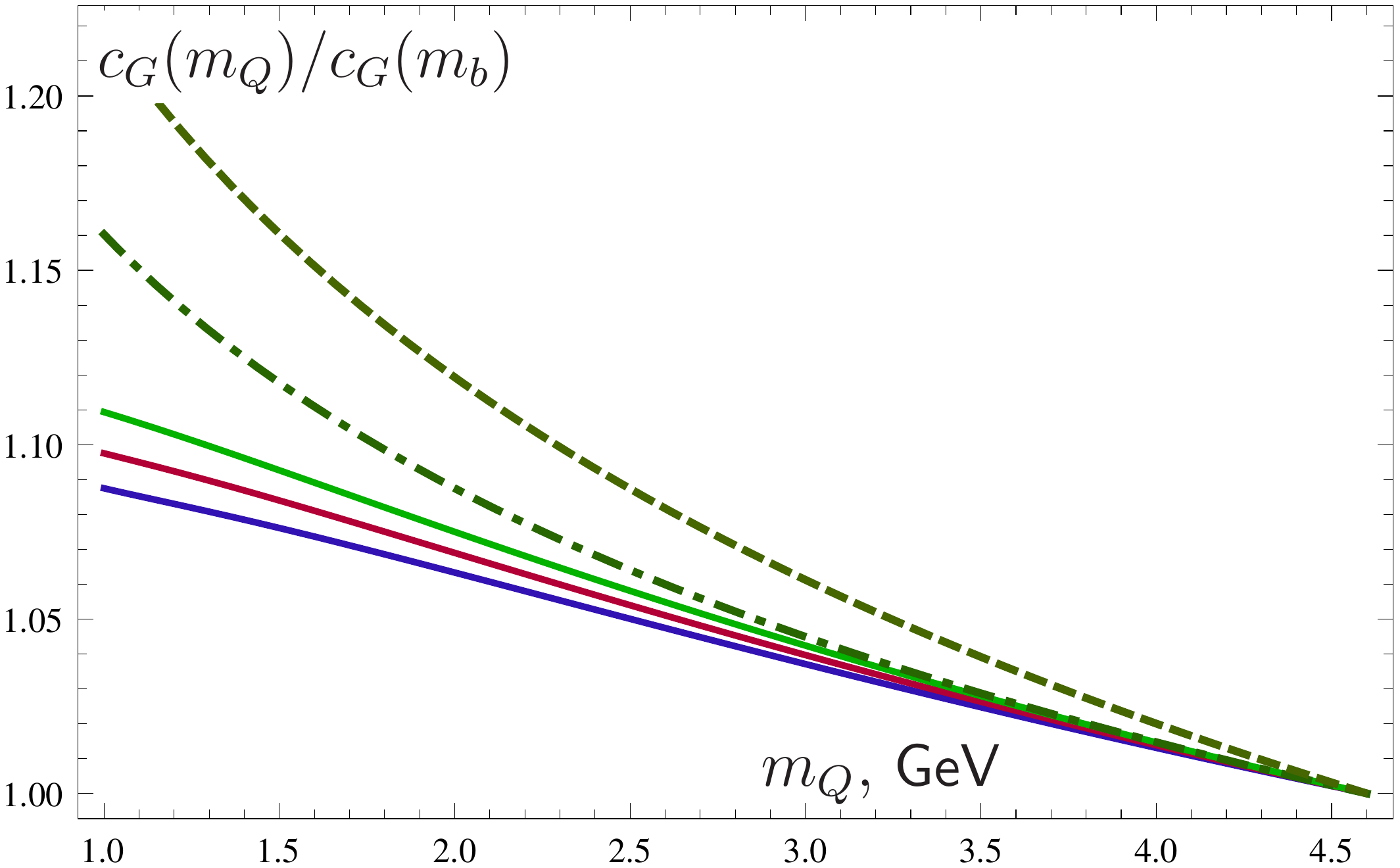}
 \end{center}\vspace*{-5pt}
\caption{ \small
The effect of removing the infrared piece belonging to the nonperturbative
expectation values, from the one-loop perturbative
evolution from beauty to charm $c_G(m_Q)/c_G(m_b)$, as a functon of
$m_Q$. The solid curves correspond to
$\mu\!=0.8\GeV$ (blue), $0.7\GeV$ (red) and $0.6\GeV$ (green),
respectively; the dashed curve shows 
the naive perturbative result with no cutoff, all at $\alpha_s\!=\!0.3$. For
comparison the dashed-dotted line shows the no-cutoff result at $\alpha_s\!=\!0.22$.
}
\label{pertrenwils}
\end{figure}

The numeric aspects are illustrated in Fig.~\ref{pertrenwils} in the case
of one-loop corrections, using $\alpha_s\!=\!0.3$. 
The cutoff effects are more important than 
higher-loop corrections. The two-loop calculation of 
Ref.~\cite{czargroz}, which strongly enhanced the
first-order renormalization, should not be used in the 
present context \cite{chrom}.
To get an accurate estimate in the following we will 
employ the one-loop 
calculation of $c_G(m_c)/c_G(m_b)$ with Wilsonian cutoff evaluated 
at $\alpha_s\!=\!0.3$.

\thispagestyle{plain}
\begin{figure}[t]
\vspace*{-7pt}
 \begin{center}
   \includegraphics[width=8cm]{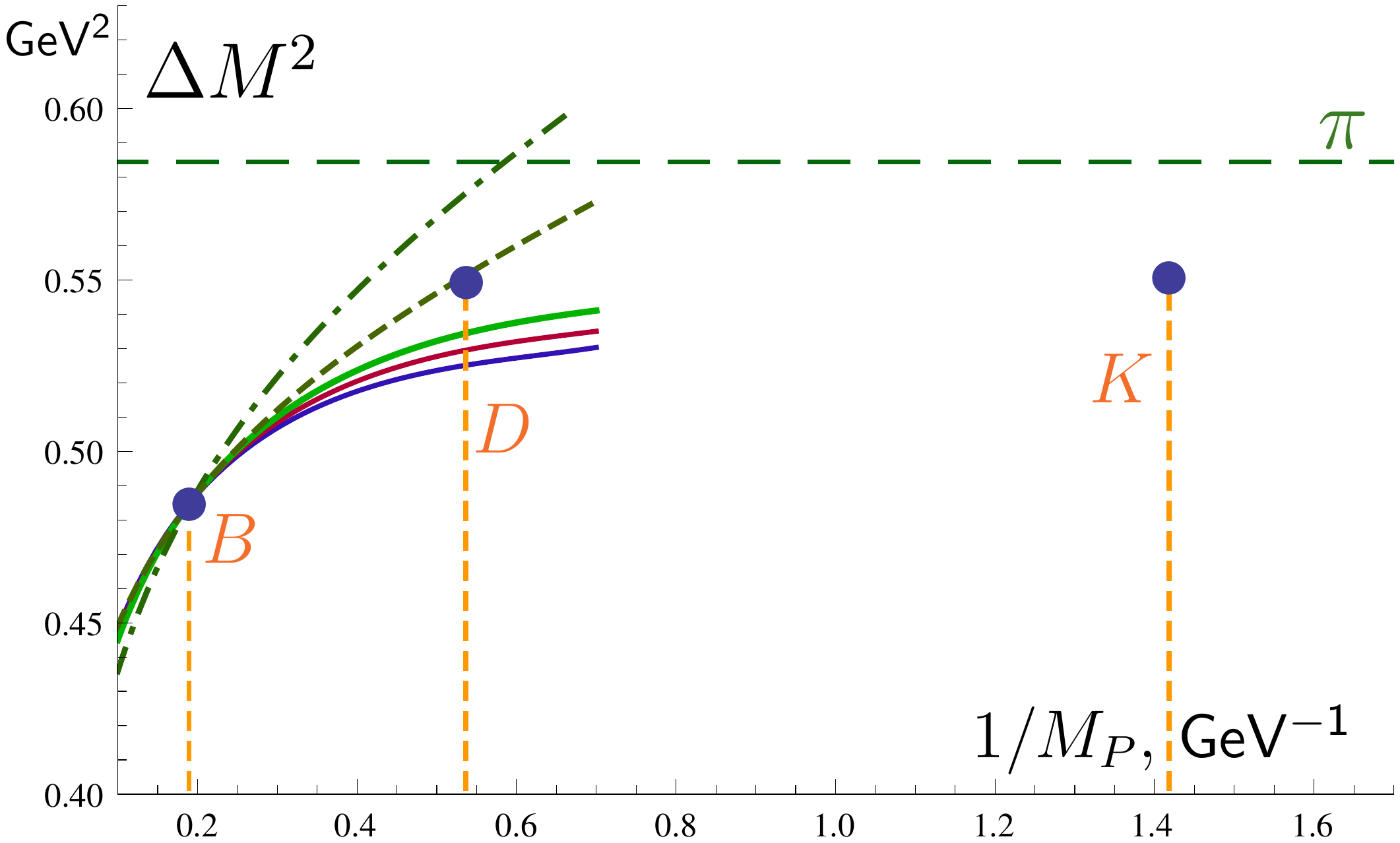}
 \end{center}
\caption{ \small
The difference $\Delta M^2\!=\!M_V^2\!-\!M_P^2$ plotted as a function of the inverse meson mass
$1/M_P$. The three data points stand for beauty, charm and strangeness,
the horizontal line shows the asymptotic value given by the $\rho$-$\pi$
splitting. (The $K$ point has been moved leftward.) Solid curves
represent the expected perturbative evolution
in the case all power corrections are neglected, using the Wilsonian perturbative
renormalization factor; they correspond to
$\mu\!=0.8\GeV,\,0.7\GeV,\,0.6\GeV$, respectively. The dashed curves shows 
the naive perturbative result without cutoff. 
}
\label{DM2pertcut}
\end{figure}

A precise determination of $\kappa$ is not easy  since only two
data points exist on the  $\Delta M^2$ curve in the heavy quark regime. 
Moreover, the perturbative treatment of charm  may have
insufficient precision and  higher power corrections for $D^{(*)}$ may
be  significant. The situation is illustrated   in Fig.~\ref{DM2pertcut}, 
where only the change in the
perturbative coefficient $c_G$ is considered. In other words, the continuous lines
show the quantity $
(M_{B^*}^2\!-\!M_B^2) \, \frac{c_G(m_Q)}{c_G(m_b)}$,
without power corrections associated with using the meson masses.\footnote {The physical masses depend on the
pseudoscalar meson mass $M_P$, viz.\ $M_B,\,M_D,$ ...\ whereas the 
perturbative renormalization is expressed through the quark masses. To put
them on the same plot and to compare we evaluate the perturbative calculations at 
$m_Q=M_{P_Q}-\La$.
The dependence on the precise value of $m_Q$ is minor.}
Clearly, this approximation is valid for sufficiently large masses.

Let us now discuss the implications for the nonlocal correlators that follow from 
Fig.~\ref{DM2pertcut}.
The {\sl solid} curves correspond to $\kappa\!=\!0$, and 
do not include any power correction.
For $\kappa\neq 0$  a power correction appears according to Eqs.~(\ref{dm2})
on top of the perturbative renormalization. 
Therefore, if the actual $\Delta M^2$  below beauty goes
lower than the perturbatively continued dependence, $\kappa$ is
positive; conversely, a negative $\kappa$ corresponds to the case 
where the actual $\Delta M^2$
increases steeper than the one computed perturbatively from the beauty point. 
We also note that what matters here is actually the relative
position of the curves in the vicinity of the beauty point, or, more generally
the difference in the slope of the two curves at any sufficiently large mass
where the $1/m_Q^3$ and higher terms in the masses can be neglected.

It looks improbable that the actual $\Delta M^2$ dependence on the inverse
meson mass may be significantly steeper around the $B$ meson than the 
 one-loop $\alpha_s\!=\!0.22$ perturbative curve, because that 
would require an unnatural shape, perhaps with
a maximum higher than the charm point. Moreover,
in that case sizable power corrections would be necessary to hit the $\Delta
M^2_D$ point, and there would be no reason to expect a small value of 
the nonlocal correlators.
Since the $\Delta M^2_D$ point in Fig.~\ref{DM2pertcut} is above the 
Wilsonian perturbative curves, 
a somewhat  steeper dependence on $1/M_P$ is observed.
Therefore, $\kappa$ must be relatively small and negative in the 
Wilsonian approach. 

In order to obtain a numerical prediction for $\Delta M^2$ and refine 
Eq.~(\ref{164}) 
we use the expansion for $\Delta M$, Eq.~(\ref{70}), 
and rewrite it, neglecting, as usual, the perturbative 
corrections to power corrections, as
\beq
\Delta M(m_Q)=  \frac{m_b \,c_G(m_Q;\mu)}{m_Q \,c_G(m_b;\mu)}
\frac{1-\frac{\delta}{m_Q}}{1-\frac{\delta}{m_b}}\,\Delta M(m_b) 
\left[1+{\cal O}\left(\mbox{$\frac{1}{m_Q^2}$} \right) \right], 
\label{70.02}
\eeq
with the shortcut 
$$
\delta=  \frac{-(\rho_{\pi G}^3+\rho_A^3-\rho_{LS}^3)}{2\mu_G^2}, \mbox{ ~or~
} 1\!+\!\kappa = \frac{\delta}{\La}\,.
$$

Higher order power terms modify the $m_Q$-dependence of $\Delta M^2$.
Therefore, in order to obtain definite  numerical values for the 
nonlocal correlators from $\Delta M^2$ in beauty and charm 
we need to make assumptions on the
higher-order terms ${\cal O}(1/m_Q^2)$ in Eq.~(\ref{70.02}). The simplest
option is to discard $1/m_Q^2$ and higher
terms in the mass difference. An alternative way is to write the power
correction factor as 
\beq
\Delta M(m_Q)= \frac23 c_G(m_Q) \frac{\mu_G^2}{m_Q} \cdot \frac{1}{1+\frac{\delta}{m_Q}}
\label{70.04}
\eeq
as an ansatz for higher-order terms; it
expresses them in terms of $\mu_G^2$ and of the powers of $\delta$. The two forms
have identical $\mhad^3/m_Q^2$ corrections but differ at higher orders. The latter
ansatz has the advantage of yielding a reasonable finite value even at small $m_Q$.

\thispagestyle{plain}
\begin{figure}[t]
\vspace*{-7pt}
 \begin{center}
   \includegraphics[width=16cm]{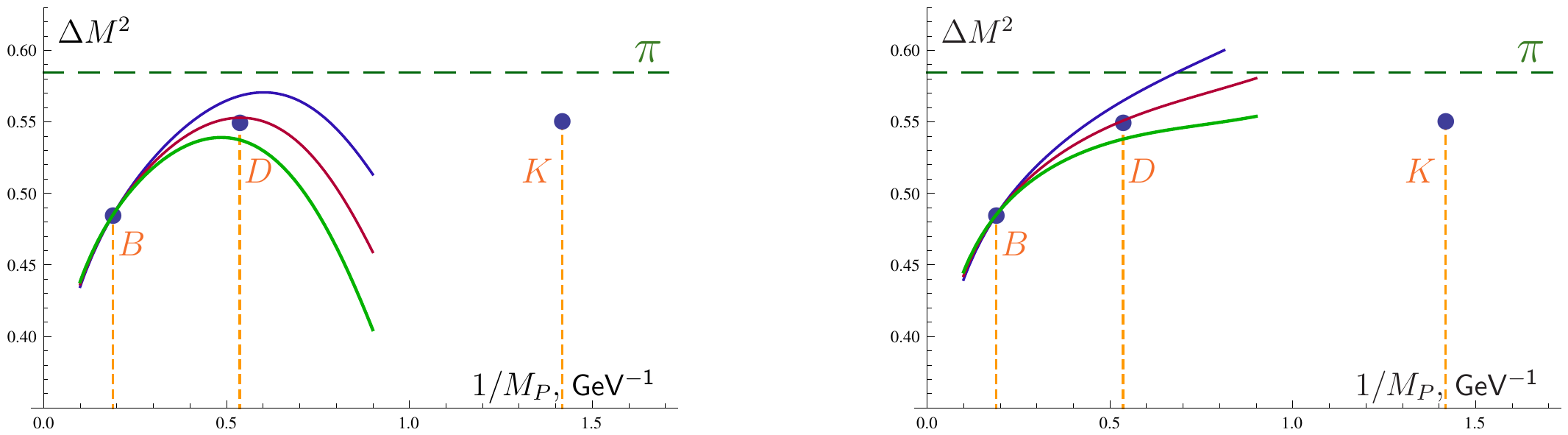}
 \end{center}\vspace*{-1pt}
\caption{ \small
 $\Delta M^2$ for different heavy quark masses computed under different
assumptions for higher-order power terms: no higher power term (left) and 
the ansatz in Eq.~(\ref{70.04}) (right). The two families of curves assume
different values for $1\!+\!\kappa$  
from top to bottom: $0.5$, $0.55$, $0.6$ 
(left) and $0.7$, $0.8$, $0.9$ (right), respectively. $\La\!=\!0.65\GeV$, 
$\mu\!=\!0.8\GeV$ and $\alpha_s\!=\!0.3$.
}
\label{DM2AB}
\end{figure}

To depict results graphically as a function of $M_P$, for the pseudoscalar 
mass $M_P(m_Q)$ we use its conventional heavy quark 
expansion well described by just $m_Q \!+\!\La$ due to the proximity to the BPS
regime (for even lighter quark, to continue the plots below charm we use 
an ad-hoc extrapolation giving a reasonable behavior at small $m_Q$). 

Fig.~\ref{DM2AB} shows the expected $\Delta M^2$ for three values of $\kappa$ 
under the two assumptions about
higher-order terms. 
The difference illustrates
the sensitivity to the unknown power corrections in the meson masses. It is
modest around the $B$ mass scale, yet becomes significant for charm. 
 We have also tried alternative ans\"atze
for power corrections and found that $\kappa$ always varies between   $-0.5$ and 0.
The extrapolation of the curves below charm seems to suggest a preference for
the smaller $|\kappa|$,  as in the right of Figs.~\ref{DM2AB}.
Moreover, higher order perturbative
effects are qualitatively expected to increase $1+\kappa$.
We will therefore use $\kappa=-0.2\pm 0.2$, arriving at
\beq
2(1+\kappa)\La\, \mu_G^2 \approx 2(0.8\pm 0.2) 
\,0.65\cdot 0.35\GeV^3 \simeq (0.35\pm 0.1) \GeV^3.
\label{70.06}
\eeq
Combining this estimate with the expected  value $\rho_{LS}^3\approx -0.1\GeV^3$
we end up with 
\beq
-(\rho_{\pi G}^3+\rho_A^3) \approx  0.45\GeV^3  ,
\label{164a}
\eeq
as in the original estimate, Eq.~(\ref{164}). One of the reasons is
that a larger value for $\La$ must be used in the Wilsonian approach, which
offsets a negative $\kappa$.

The main assumption employed in the above estimate is the possibility to
quantitatively use the mass expansion for charm particles, assuming a
reasonable magnitude of the higher-order power corrections in a physical
scheme.  This assumption could be avoided in  the case an additional input
from lattice calculations were available, with the squared mass splitting
reliably computed at one or more intermediate points for quarks heavier than
charm.  The short-distance expansion there is in a better shape, perturbative
corrections under better control, and higher-order power corrections are less
important.  This would allow to use the differential version of the constraint
on the hyperfine splitting in the large-$m_Q$ limit, namely Eq.(\ref{160}),
fixing directly the slope.  Even if a first-principle direct measurement of
$\Delta M^2$ with sufficient precision turns out difficult for large mass, any
intermediate point -- and even a point somewhat below charm -- 
would constrain the shape of the curve thereby narrowing
down the interval of possible values for the derivative.

 Eq.~(\ref{164a}) is a strong constraint, especially useful in the analysis of the inelastic
contribution to ${\cal F}(1)$. 
This will be discussed in the next subsection. Here we only note that 
a determination of $-(\rho_{\pi G}^3\!+\!\rho_A^3)$ gives a
 model-independent constraint on the spectral densities 
$\rho_{g}^{(\frac{1}{2}^+)}$, $\rho_{pg}^{(\frac{1}{2}^+)}$ and 
$\rho_{g}^{(\frac{3}{2}^+)}$, which are informative due to the 
factorization property and to the relation to the spectrum of physical
states. Since the value in Eq.~(\ref{164a})  is quite large,
$\;\rho_{pg}^{(\frac{1}{2}^+)}(\omega)$ together with 
$-\rho_{\pi G}^3$ is  most probably positive, unless the correlators 
are strongly
dominated by the $\frac{3}{2}^+$ states.
The predictive power sharpens significantly if additional dynamic
information is used, for instance the proximity to the BPS limit or the
saturation by a limited number of the heavy quark multiplets. 

We conclude this section noting that another interesting constraint on the
nonlocal correlators can be obtained from the difference of the spin-averaged
$B$ and $D$ meson masses, namely from
\beq
m_b\!-\!m_c = {\overline{M}_B} \!-\!  {\overline{M}_D} + 
\frac{\mu^2_\pi}2 \left(\frac1{m_c}\!-\!
\frac1{m_b}\right) + \frac{\rho_D^3\!-\!\bar\rho^3}{4} \left(\frac1{m_c^2}\!-\!
\frac1{m_b^2}\right)+{\cal O}\!\left(\frac1{m_Q^3}\right)\!,
\label{massdiff}
\eeq
where $\bar \rho^3 \!=\! \rho_{\pi\pi}^3 \!+\!\rho_S^3$. The use of this relation to 
determine $\bar \rho^3$ requires a good control of the 
heavy quark mass difference and a precise knowledge of $\mu_\pi^2$ and $\rho_D^3$, 
which can in principle be provided by the  global fits to semileptonic moments. 
We have considered the fits employed in Sec.~\ref{powercor};  they typically give
$\bar \rho^3 \!=\! (0.33\pm 0.17)\GeV^3$.  Notice that the semileptonic fits 
determine $\mu_\pi^2$
in the actual $B$ meson, while its static approximation appears in 
Eq.~(\ref{massdiff}).
As the difference is BPS and $1/m_b$ suppressed, it is certainly smaller than
the fit uncertainty on  $\mu_\pi^2$,   and we have neglected it.
The value of $\bar\rho^3$ can be linked to  $(\rho_{\pi G}^3+\rho_A^3)$
since the sum $\rho_{\pi\pi}^3\!+\!\rho_S^3\!+\!\rho_{\pi G}^3\!+\!\rho_A^3$  
is positive definite,
vanishes in the BPS limit, and receives only second order corrections 
to the limit. Therefore,  $-(\rho_{\pi G}^3\!+\!\rho_A^3)\!\lsim\! 
(0.33 \!\pm\! 0.17)\GeV^3$, which is compatible with our
primary estimate, Eq.~(\ref{164a}).

\subsection{The hyperfine constraint and the excited states}
\label{hyper}

The hyperfine splitting constraint Eq.~(\ref{164a}) fixes the
overall scale of the correlators which are  the focus of our study:
\beq
\int \frac{{\rm d} \omega}{\omega}\;  \left(
\frac{\rho_{g}^{(\frac{3}{2}^+)}(\omega)}{2}+
2\rho_{pg}^{(\frac{1}{2}^+)}(\omega)-\frac{2\rho_{g}^{(\frac{1}{2}^+)}(\omega)}{3}
\right)\simeq 0.45\GeV^3+(\kappa\!+\!0.2) \cdot 0.36\GeV^3.
\label{2320} 
\eeq
The normalization point enters here as the upper cutoff in the integral over
$\omega$; we assume it is around $0.8\GeV$.  At  first glance we have a single
relation among four spectral functions; nevertheless this relation turns out
to be very useful thanks to the positivity and
factorization properties discussed in Sec.~6.1.1.

The principal observation is that according to the estimate Eq.~(\ref{164a})
the nonlocal correlators are numerically large.  The particular combination
constrained significantly exceeds the estimated size of the expectation values
of local heavy quark operators of $D\!=\!6$, most notably $\rho_D^3$ which by
itself is large. Alternatively, the value in Eq.~(\ref{164a}) can be compared
to $\La^3$:  it exceeds it, even though $\La$ 
is already numerically large. Likewise, the square of the kinetic
expectation value $(\mu_\pi^2)^2$ can be used to gauge the scale of hadronic
parameters of mass dimension 3 if it is divided by a typical energy of the
radial excitations, yielding again a value close to $0.35\GeV^3$.

For more quantitative statements it is advantageous to separate the resonant
and the soft continuum contributions to the spectral densities: the inclusion
of the continuum makes the effective value of $\omega$ identified with
$\varepsilon_{\rm rad}$ more uncertain numerically.  The separation can be
done since the $\,\Omega_0\! +\!\pi$ states are readily analyzed in the soft-pion
technique; for instance, Sect.~\ref{sectdpi} estimated the  continuum part
of $I_{\rm inel}$ at actual values of $m_c$ and $m_b$. This analysis is
performed in Sect.~\ref{pionloop} and suggests that a relatively small
fraction should be subtracted from the value in Eq.~(\ref{2320}) if we want to
keep only the resonant contribution.

Once only the resonant contribution are retained, a meaningful approximation
well illustrating the physics of the constraints is a model where a
single multiplet in each channel involved is considered. This model has a
very few parameters essentially reduced to the residues $P$, $G$ and $g$ ($f$ and
$h$ do not enter), and all the constraints can be easily analyzed.

To gauge the size of the related hadronic parameters we take the sum of
the two spin-singlet nonlocal correlators $\rho_{\pi\pi}^3$ and $\rho_S^3$ as
a measure of their overall significance; each of these correlators is positive.
On the other hand, we have a relation
\beq
\rho_{\pi\pi}^3+ \rho_{S}^3 = -(\rho_{\pi G}^3+ \rho_{A}^3) + 
\int \frac{\rm d \omega}{\omega}
\left[\rho_{p}^{\frac{1}{2}+}
- 2\rho_{pg}^{\frac{1}{2}+} + \rho_{g}^{\frac{1}{2}+}\right].
\label{2222}
\eeq
The last bracket on the right is positive; it can be represented as a correlator 
of a certain operator. This shows that the l.h.s.\ of (\ref{2222}) 
always exceeds the combination 
$-\!\rho_{\pi G}^3\!-\!\rho_{A}^3$  fixed by the `hyperfine' constraint; the
minimum is attained at the BPS point where the two combinations coincide. The 
inequality actually holds at arbitrary spectral parameter $\omega$; it is the
numeric hyperfine constraint that applies only upon integration over the energy.

Eq.~(\ref{2222}) shows that the $1/m_Q^2$
spin-averaged meson mass shift must always be larger than 3/4 times the $1/m_Q^2$ 
correction to the hyperfine splitting. This is not difficult to trace directly:
\beq
\frac{3\delta M_B^*+ \delta M_B}{4}= \frac{3}{4}(\delta M_B^*- \delta M_B) +
\delta M_B; 
\label{2224}
\eeq
$\delta_{1/m_Q^2} M_B$ receives a positive contribution
$(\rho_D^3+\rho_{LS}^3)/4m_b^2$ from the local piece
and a negative contribution from the $T$-product of two $(\sigma \pi)^2$. 
Both vanish at the BPS point. 

The last positive term in Eq.~(\ref{2222}) is of the
second order in the deviation from BPS, therefore it may be a good
approximation to neglect it  in numerical  estimates unless 
the BPS is strongly violated in the actual $B$ mesons. For what follows we
do not need this additional assumption, the positivity is sufficient. 

Now we can see that the sum of the two correlators 
$\rho_{\pi\pi}^3$ and $\rho_{S}^3$ is also quite large numerically; 
this implies  large $1/m_Q$ corrections to the meson states, in particular
averaged over the spin multiplet.  
However, experiment tells us  that the heavy quark symmetry 
works reasonably well even in charm.
This means the correlators driving power
corrections should not be excessively large.
In order to have  both $\rho_{\pi\pi}^3$ and $\rho_{S}^3$
as small as possible, besides the BPS condition, we can assume 
that $\rho_{\pi\pi}^3$ and $\rho_{S}^3$ are equal, in which case
$\rho_{g}^{\frac{1}{2}^+}= \frac{3}{4}\rho_{g}^{\frac{3}{2}^+}$ must hold upon
corresponding integration over energy. 

On the other hand, since the r.h.s.\ of Eq.~(\ref{2320})
is large, it is reasonable to expect  that $\rho_{pg}^{(\frac{1}{2}^+)}$ is
positive, and  that $\rho_{p}^{(\frac{1}{2}^+)}$ is not much smaller
than $\rho_{g}^{(\frac{1}{2}^+)}$. In this case the left-hand side is a sum of
two positive contributions from $\frac{1}{2}^+$ and $\frac{3}{2}^+$ channels.

It makes sense to parameterize the
relative contribution of the $\frac{1}{2}^+$ and $\frac{3}{2}^+$ states, introducing a parameter $\nu$:
\beq
\int \frac{{\rm d} \omega}{\omega}\;  
\frac{\rho_{g}^{(\frac{3}{2}^+)}(\omega)}{2}\equiv  
\nu \int \frac{{\rm d} \omega}{\omega}\;  \left(
\frac{\rho_{g}^{(\frac{3}{2}^+)}(\omega)}{2}+
2\rho_{pg}^{(\frac{1}{2}^+)}(\omega)-\frac{2\rho_{g}^{(\frac{1}{2}^+)}(\omega)}{3}
\right), \qquad \nu >0.
\label{2226nu}
\eeq
The large numeric value of the the r.h.s.\ of Eq.~(\ref{2320})  points at
$\nu$  between $0$ and $1$ as the most natural solution; 
$\nu$ exceeding unity is highly improbable.

\subsubsection{Estimate of $I_1$}
\label{I1est}

The spectral representation together with the hyperfine constraint 
allows one to analyze the possible values of
$I_{\rm inel}$. Ignoring a possible spread in the values of average excitation
energy we equate $I_{\rm inel}$ with $I_1/\varepsilon_{\rm rad}$ or,
equivalently, express the  $\tilde \rho$ correlators in Eq.~(\ref{150}) through
the conventional $\rho^3$. 
The evaluation of  $I_{\rm inel}$ then depends directly on the size 
of the corresponding nonlocal correlators.
For instance, the last positive term in Eq.~(\ref{124}) is
immediately recognized as the square bracket in  Eq.~(\ref{2222}); it is given
solely by the BPS-violating transitions into the $\frac{1}{2}^+$ multiplet. 

All the terms in $I_1$ but the leading BPS piece fixed by the
hyperfine constraint are  independent of  the  $\frac{3}{2}^+$ contributions, 
and the l.h.s.\ of Eq.~(\ref{2320})
gets a positive piece from them. Therefore the minimum
of $I_1$ is attained at vanishing $\rho_g^{(\frac{3}{2}^+)}$ which  also 
corresponds to $\rho^3_A= 2 \rho^3_S$. 

We now refer to  Eq.~(\ref{152.02}) for the structure of $I_{\rm inel}$ and express it 
in terms of the transition amplitudes $P,G, g$.
At fixed  $G$ and $g$,  $I_{\rm inel}$  is a quadratic
trinomial in the ratio $P/G$, with the positive
leading power coefficient proportional to $(1/m_c\!-\!1/m_b)^2$. 
We recall that in the BPS limit $P/G\!=\!1$.  The hyperfine splitting implies
an additional  constraint on $P$, $G$, $g$, which can be taken into account.
The result is that 
\beq
\frac{3m_c^2 I_1}{-(\rho_{\pi G}^3\!+\!\rho_A^3)}
\ge 1-(1-\nu)\frac{m_c^2}{m_b^2}\simeq 1-0.07(1-\nu), 
\label{173.02}
\eeq
and $I_1$ has a minimum at a value of $P/G\!>\!1$, whose
exact position depends only on the heavy quark mass ratio but not on $g$.
The minimum is attained, with the inequality saturated, at
\beq
\frac{P}{G}=\frac{1}{3}+ \frac{2}{3}\,\frac{m_b\!+\!m_c}{m_b\!-\!m_c}\simeq 1.47,
\label{173.04}
\eeq
where the numeric values correspond to $m_c/m_b\!=\!1.2/4.6$. This result was
referred to in Sect.~\ref{winelnumer}.  For $\nu=0$ it implies  $G\approx 0.37\GeV^2$.

The minimum, however, is rather shallow, see Fig.~\ref{I1}, 
because the coefficient of the term quadratic in $P/G$ is large while 
the linear term is small. While 
$I_1$ may, in principle, significantly exceed  its BPS value 
$I_1^{\rm BPS}$, Eq.~(\ref{172}), for
strong BPS violation if $P/G$ is negative, at more natural positive  $P/G$ 
(a regime where both $\rho_{\pi\pi}^3$ and $\rho_{s}^3$ are not too large) 
the value of $I_1/I_1^{\rm BPS}$ is typically slightly smaller than unity,
as illustrated in  Fig.~\ref{I1}.

\subsubsection{The \boldmath inclusive $\frac{1}{2}^+$ and 
$\frac{3}{2}^+$ yield}
\label{yield}

The hyperfine constraint implies relatively large values of the transition
matrix elements to $\frac{1}{2}^+$ and/or $\frac{3}{2}^+$ states and an
enhancement of the overall yield of the corresponding excited charm states 
in the semileptonic decays of actual $B$ mesons. This is quantified by the
estimate of $I_{\rm inel}$ constituting, roughly speaking, 
$15\%$ of the $D^*$ probability, in the zero-recoil
kinematics. This comparison does not include the phase-space suppression
of excited states resonances which becomes significant when one 
integrates over all available phase space. Other kinematic effects
may work in the opposite direction, however, and a
more substantiated estimate is desirable.

Our preceding analysis constrains directly the $1/m_Q$ squared transition amplitudes
into the corresponding hadronic states at zero velocity transfer.  At
first glance, this seems too crude an estimate at non-zero recoil,  
where contributions not suppressed by  $1/m_Q$ appear. 
However, since we consider states with the quantum numbers of radials or of
$D$-waves,  the leading-order  amplitude is proportional to the second power
of the velocity  of the charmed meson --- roughly speaking, it is generated by replacing the 
momentum operators $\pi_k$ with $m_Q v_k$.  
Since for excited mesons $v_k$ is always relatively small, the zero-recoil
amplitude receives only a minor correction when integrated over the whole phase space. 
Therefore, we can retain only the $1/m_Q$ part of the transition amplitude.  For
comparison, the leading heavy-quark  transition amplitude into the
$P$-waves provides only  a suppressed correction to the $1/m_Q$ term
\cite{leib}, even though the leading amplitude in that case is of the 
first order in the velocity, and  the $P$-wave states are lighter.

We will therefore relate the transition probabilities into radially excited
states directly to the 
zero-recoil observables we have studied in the previous sections. 
The total yield is also fed by vector-current transitions, even though
we do not expect them to provide a large contribution, as
their amplitudes vanish in the BPS
limit. These transitions can  be considered in the same way as  
the axial timelike component, see Appendix~\ref{aprad}.

The total yields are roughly proportional to $(M_B\!-\!M)^5\!\equiv \!\Delta^5$
where $M$ is the mass of the corresponding charm multiplet.
Numerically we equate $\Delta$ with $M_B\!-\!M_D\!-\!\epsilon_{\rm rad}$,
leading to $\Delta\!\simeq \!2.6\GeV$. However, for $\Delta/M_B \!\gsim\! 0.5$ 
relativistic effects  modify significantly the nonrelativistic $\Delta^5$ dependence.
A more accurate approach is the following (see Appendix~\ref{aprad} for  details). 
Instead of considering the transition
amplitudes between the  $B$ meson and all the excited mesons belonging to a 
multiplet, we evaluate the decay rate of the $\Omega_0$
spin-$\frac{1}{2}$ heavy state into the corresponding excited
half-integer spin multiplets. The weak current coupling of these fictitious hadrons
is then fixed by the corresponding transition probabilities near zero recoil.

In practice, we parameterize the transition amplitudes to $\frac{1}{2}^+$ and $\frac{3}{2}^+$ states, for axial and vector currents, 
and  compute the decay rate integrating  over all the available phase
space. The overall normalization of the amplitudes is fixed at zero recoil by 
the corresponding nonlocal correlator.
Let us illustrate this in the case of the transitions into
$\frac{1}{2}^+$ states.

The most general vector or axial vertices in terms of full Lorentz spinor 
wavefunctions have the form
\bea
J^A_\mu&=& g_A \bar\chi \gamma_\mu\gamma_5 \Psi_0 +
b_A\bar\chi i\sigma_{\mu\nu}\frac{q_\nu}{M_B} \gamma_5 \Psi_0 +
c_A \frac{q_\mu}{M_B} \bar\chi  i\gamma_5 \Psi_0, \nonumber\\
J^V_\mu &=& g_V\bar\chi \gamma_\mu \Psi_0 +b_V\bar\chi
\sigma_{\mu\nu}\frac{q_\nu}{M_B} \Psi_0 +c_V \frac{q_\mu}{M_B}\bar\chi \Psi_0\,.
\label{2710.02}
\eea
At zero recoil the effect of $b_{A}$ reduces to a change in $g_{A}$ of
$(M_B\!-\!M)/M_B \,b_A$, while the term proportional to $c_A$ vanishes. 
However,  for generic recoil  these are all independent
structures. Similar considerations apply to the vector current. 
We neglect the additional contributions at non-zero recoil and effectively 
set  the formfactors $b_{A,V}$, $c_{A,V}$ to zero. Likewise we neglect 
the velocity-dependence of the
formfactors. The corresponding decay rates are then given by
\beq
\Gamma^{(\frac{1}{2})}_{V,A} = |g_{V,A}|^2\, \frac{G_F^2 |V_{cb}|^2 M_B^5}{192\pi^3 } \,z_{A,V}(r),
\label{2240.02}
\eeq
where 
\bea
\label{992}
z_{A,V}(r)=\frac{z_0(r)\pm \tilde z_0(r)}{2}, && 
\qquad \quad z_0(r)=1\!-\!8r \!+\!8r^3\!-\!r^4 \!-\!12r^2\ln{r},\\
\quad r=\frac{M^2}{M_B^2},\quad\qquad&&
\qquad \quad
\tilde z_0(r) = 2\sqrt{r}\left[1+9r-9r^2-r^3+6r(1\!+\!r)\ln{r}\right], \nonumber
\label{2243}
\eea
are the weighted phase space factors for the axial and vector transitions,
 respectively;  $z_0(r)$ is the standard kinematic factor for 
$V\!-\!A$ decays. 

The value of $|g_A|^2$ is directly given by the $\frac{1}{2}^+$ contribution in
$w_{\rm inel} (\omega)$ in Eq.~(\ref{152.02}) for one multiplet, and the
transition amplitudes $G$ and $P$ are constrained by our
analysis of the hyperfine splitting.
 $|g_V|^2$ follows from the analogous relation for vector-induced
probabilities, Eq.~(\ref{152v}). The vector-induced probability is
significantly suppressed compared to the axial one. 

As already mentioned,  we neglect additional recoil corrections. 
A justification for this is
the `extended' SV regime relevant in the context of a large-$5$ expansion
\cite{five} where the enhanced corrections reside in the lepton phase
space. The deviations from the small-recoil kinematics in the amplitude are
not enhanced, but rather suppressed by the large power of energy release decreasing
the average recoil.

The case of the decays into the $\frac{3}{2}^+$ is treated similarly --- the details 
are given in Appendix~\ref{aprad}. 
We describe these states by 
complete relativistic Rarita-Schwinger wavefunctions at arbitrary velocity,
and calculate the corresponding contribution to the 
(unpolarized) structure functions of $\Omega_0$. 
Their integration yields
the total decay rate, and we fix the normalization of the
formfactors at zero recoil. 

In this way the decay rates are calculate separately for the $\frac{1}{2}^+$
and  $\frac{3}{2}^+$ states and separately for the axial and the vector
transitions. Numerically we get for the combined yield
\beq
\frac{\Gamma_{\rm rad}}{\Gamma_{\rm sl}} \approx 0.07 ,
\label{2244.02}
\eeq
with the axial part strongly dominating (we have included  perturbative corrections in  the denominator). While  the relative weight of   $\frac{1}{2}^+$ and
$\frac{3}{2}^+$ states varies depending on their couplings,  their sum is 
approximately fixed by  the hyperfine condition (\ref{2320}).

In our approach the total decay rate is computed in the heavy 
quark limit. We included the $1/m_Q$ symmetry-breaking terms which mediate the transitions in
question, and exploited the fact   
that the total decay rate does not depend on the exact mixing of the final
heavy-quark eigenstates in the actual QCD hadrons. 

To apply the estimates to the yield of actual individual charmed mesons in QCD
one would need to properly construct the corresponding states in terms of
$\frac{1}{2}^+$ and  $\frac{3}{2}^+$ states.
While straightforward in
practice, this may yield unreliable predictions in practice since the heavy
quark symmetry is probably quite strongly violated for the excited charm
states.  We expect much more robust predictions for the total yield summing up
all the associated channels. Such inclusive probabilities are insensitive to
the details of the strong Hamiltonian in the final states, are not affected by
possible degeneracies and altogether enjoy smaller preasymptotic corrections.
Therefore, we view  Eq.~(\ref{2244.02}) as a good 
starting estimate of the overall yield of the descendants of the `radial'
states. 
It refines the earlier estimate given in Ref.~\cite{f0short}, and probably
represents a natural lower limit. A compatible number has recently been
suggested in Ref.~\cite{bernturcz}.

It is worth noting that the semileptonic phase space factor strongly
suppresses the yield of the states with higher mass; in particular, for a wide
resonance with mass $M\!=\!\bar M \!+\!\delta \,m$ one has
\beq
(M_B\!-\!M)^5= (\Delta\!-\!\delta \,m)^5 
\approx  \Delta^5\left (1- \frac{\delta \,m}{600\,{\rm MeV}}+\dots\right)
\label{2246}
\eeq
This shows that for broad resonances the phase space
factor averaged over the whole decay kinematics significantly distorts the
Breit-Wigner shape and shifts the apparent peak towards lower
mass. The factor in Eq.~(\ref{2246}) applied to a wide resonance may easily
mimic a typical non-resonant continuum yield with a threshold suppression.

The above estimate suggests that the total yield of the discussed `radial'
excitations with mass below $M_D\!+\!1\GeV$ is expected to be,  in terms of 
$\Gamma_{\rm sl}$,  at the $7\%$ level. This is close  to what is 
observed in experiment, yet
traditionally is attributed to the `wide' $\frac{1}{2}$ $P$-wave states. The
conventional allocation creates a problem: theory  predicts that 
the  $\frac{3}{2}$-states must strongly dominate among the $P$-waves; they
have been measured at the right rate of about $10\%$ 
of $\Gamma_{\rm sl}$, see {\it e.g.\   }\cite{memorino} for a discussion.
We are therefore  led to argue that the decays into the $\frac{1}{2}^-$ states are
indeed suppressed, while the bulk of the experimentally observed ``wide''
structure is actually the result of the significant fraction of the `radial'
states. 

\subsubsection{\boldmath The radial and $D$-wave excited states}

The phenomenological analysis earlier in this section suggests, as the
most natural solution that both the $\frac{1}{2}^+$ true radial
excitations and the $\frac{3}{2}^+$ $D$-wave states contribute
significantly the spectral functions involved 
and are produced at appreciable rate in the
semileptonic $B$ decays. At present we do not have accurate enough
data to state which of the two channels dominate; we have only observed that the
solutions where one of them, say, the $D$-wave, is small is
disfavored. 

In fact, the identification of the two families of states with the $S$-
or $D$-waves is not strict in the actual mesons in QCD even in the
limit of a large mass $m_Q$. For instance,  the $\frac{3}{2}^+$
states can instead be a result of excitation of gauge degrees of
freedom. Whether this is so or not is an open question, and the fact
that we expect a numerically large transition matrix element of the
chromomagnetic field,
\beq
\matel{\mbox{$\frac{3}{2}^+$}}{\bar{Q}B_k Q}{\Omega_0}_{\vec q=0}= 
\frac{g}{4} \, \chi_{k}^\dagger \Psi_0 \,,
\label{2160}
\eeq
is not directly related to it. It appears that so far the radial
states attracted more attention in the literature, while the $D$-waves
were marginally considered \cite{fazio}. At the same time we expect the dominant
mechanism for the production of the corresponding charmed states to be
the $1/m_c$-component of the amplitude, and it is qualitatively
similar for both of them. Their masses are also expected to be in the
same range about $700\MeV$ above the ground state. 

A challenging question is how one can disentangle these states
in experiment. In the simplest constituent quark model one expects the
hyperfine splitting inside the $D$-wave multiplet to be particularly
suppressed. However, it is not evident to which extent this would be the
property of the actual QCD states. Moreover, the hyperfine splitting may
well happen to be suppressed within the radially excited multiplet as
well. 

Some differences are expected in the decay pattern. (We reason in
terms of the asymptotic states deprived of the heavy
quark spin; the translation into the actual mesons is standard.)
The radially excited states can decay into the ground state $\Omega_0$
and a single pion in the $P$-wave, or into $\Omega_0$ plus two pions
in $S$-wave. We expect the dominant channel to be the latter where two
pions have a $\sigma$-meson enhancement; that is, they must 
predominantly be in the isospin-singlet $S$-wave state. 

This particular two-pion channel is not allowed for the
$\frac{3}{2}^+$ state which should then decay mostly into $\Omega_0$
and a pion. A weaker two-pion channel cannot show the
resonance enhancement associated with $\sigma$-meson. We 
naturally expect the $\frac{3}{2}^+$ states to have a smaller
width. It is possible that the excited mesons recently reported by
BaBar \cite{radials} with mass around $2.75 \GeV$ are related to 
these states. The states with the lower mass around $2.6\GeV$ may be
the radial states.

Another decay chain where the first decay proceeds into a single pion in
the $S$-wave and a $P$-wave charm state, either $\frac{1}{2}^-$ or
$\frac{3}{2}^-$, can be competitive and may provide an additional
handle through the identification of the $P$-wave state via its subsequent decay.
These questions deserve further consideration. Since the
two multiplets are expected to have close masses, the actual excited
vector $D^*$ states may show significant mixture, which has to be
considered.

\subsection{\boldmath Nonresonant $D^{(*)}\pi$ in the spectral representation}
\label{pionloop}

The special role of the nonresonant  $D^{(*)}\pi$ states manifested itself 
already in the analysis of ${\cal F}(1)$. In the heavy quark limit we consider their
counterpart, the non-resonant states $\Omega_0 +\pi$ (kaon or $\eta$ may also
be included). They are of special interest for a relatively soft pion
where its energy is essentially below the resonance excitation gap. This gap
depends on the orbital momentum of the pion: for the $P$-wave states it is
about $400\MeV$. Our focus is on the radial (or $D$-wave) excitations 
where it is about $700\MeV$.

The traditional classification over spin-parity of the light degrees of freedom is
equally applicable to multi-particle states, including the 
$\Omega_0 \pi$ continuum. They can be classified in a way similar to the ground-state
excitations, the analogies of $P$-waves etc., and only have to be additionally
labeled by the continuum excitation energy. In the static limit the latter 
is equal to the pion energy.  The quantum numbers of $\Omega_0 \pi$
are not fixed a priori: a continuum state is generally a
mixture with different quantum numbers  
depending on the production amplitude. For
instance, this would apply to the relative weight of the 
$\frac{3}{2}^-$ and $\frac{1}{2}^-$ states in $P$-wave.

The advantage of the expansion we employ is that only the heavy quark states 
with vanishing total spatial momentum are involved, and they are considered in
the static limit. This fixes the structure of the transition amplitudes appearing to a
particular order in the $1/m_Q$ expansion; the relative weight of different
spin-orbit multiplets is then determined as well. In this section we 
obtain the decomposition into  $\frac{1}{2}$,  $\frac{3}{2}$ etc.\ states for
$\Omega_0 \pi$ and calculate the corresponding spectral densities 
as a function of the 
pion energy. They  then determine the contributions to $|\tau_{1/2}|^2$,
$|\tau_{3/2}|^2$, $\rho_{\pi\pi}^3$ etc.
For instance, we will see that the pion loop has a $\tau_{1/2}\!=\!\tau_{3/2}$
property \cite{habil}. The associated continuum states do not contribute 
to $\mu_G^2$ or $\rho^3_{LS}$, yet they change the IW slope or $\La$ in a
predictable way, and mediate a positive contribution to $\mu_\pi^2\!-\mu_G^2$.

The $\Omega_0 \pi$ states at rest are uniquely characterized by the energy and by 
the pion orbital momentum $L$. Indeed, the  total
angular momentum $j$ consists of $\frac{1}{2}$ of $\Omega_0$ and of $L$ of
pion:   $L\!=\!j\pm\frac{1}{2}$. Its parity relative to
parity of $\Omega_0$ is $(-1)^{L+1}$. Therefore, the combination of $j$ and
parity unambiguously specifies $L$. For instance, $L\!=\!0$ are
$\frac{1}{2}^-$ $P$-wave states, $L\!=\!1$ give $\frac{1}{2}^+$ and
$\frac{3}{2}^+$ `radial' excitations and the $\frac{3}{2}^-$ $P$-wave states
require $L\!=\!2$.

The combination of the `$P$-waves' 
appearing to the  leading order in either $1/m_Q$ or in velocity is mediated
explicitly by the operator $\bar{Q} i\vec D Q$; the relative mixture is determined
by a concrete form of the amplitude which involves the spin of $\Omega_0$. 
The most straightforward approach is to 
consider the $\Omega_0\pi$ contribution to the zero-recoil correlation 
function of operators  $\bar{Q} iD_j Q$ and $\bar{Q} iD_k Q$, as we did in 
Sect.~\ref{modind}, Eq.~(\ref{2270.02}).

\begin{figure}[t]
\vspace*{-1pt}
 \begin{center}
\includegraphics[width=15cm]{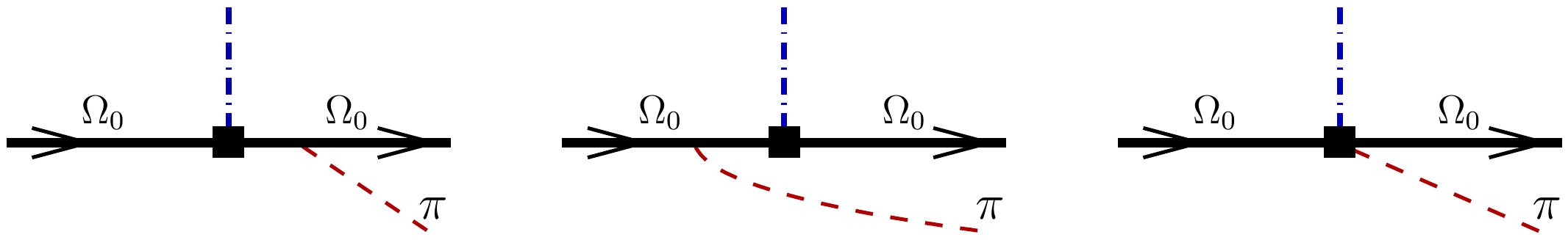}
\end{center}\vspace*{-5pt}
\caption{ \small The pion emission diagrams in the static
limit. The solid blocks denote the $\bar{Q}iD_j Q$  or 
$\bar{Q}iD_jiD_l Q$ operators with $\vec q\!=\!0$. Pion carries spatial momentum
$\vec{k}$; incoming $\Omega_0$ have vanishing spatial
momentum and outgoing $-\vec k$.
}
\label{omega0pi}
\end{figure}

The effective low-energy Lagrangian of the $\pi \Omega_0 \Omega_0$-interaction
corresponding to Eqs.~(\ref{1130}) and (\ref{1162}) in relativistic notations is
\beq
{\cal L}_{\rm chi} = -g_{B^*B\pi} \bar \Omega_0 \gamma_\mu\gamma_5 \Omega_0
\partial^\mu \pi = -2g_{B^*B\pi} \,M_\Omega\, \bar \Omega_0 i\gamma_5 \Omega_0 \pi
\label{2274} 
\eeq 
(total derivatives are omitted). The diagrams to be calculated 
are shown in Fig.~\ref{omega0pi} where the heavy hadron lines 
now all refer to $\Omega_0$ and the solid vertex stands for the operator 
$\bar{Q} iD_j Q$. The vertex is simple:
\beq
\frac{1}{2M_{\Omega_0}}\matel{\Omega_0(p_2)}{\bar{Q} 
(i\!\stackrel{\leftarrow}{D}_j \pm
i\!\stackrel{\rightarrow}{D}_j )Q}{\Omega_0(p_1)}= (p_2 \pm p_1)_j 
\Psi^\dagger_0 \Psi_0,
\label{2278} 
\eeq 
where we have generally distinguished the left and right derivatives for the case of
different momenta of $\Omega_0$. Since in our case the spatial momentum
flowing into the vertex vanishes, this specification is superfluous. 

\thispagestyle{plain}

The diagram {\bf a)} vanishes, while {\bf b)} yields an
amplitude with a simple spin structure: 
\beq
\frac{1}{2M_{\Omega_0}}\matel{\Omega_0 \pi}{\bar{Q} i{D}_j Q}{\Omega_0(0)}= 
-g_{B^*B\pi}\,\frac{k_j}{\omega} \,\Psi^\dagger_0 i\vec \sigma \vec k\Psi_0,
\label{2280} 
\eeq 
where $\vec k$ is the pion momentum. The resulting correlator
$P_{jk}(\omega)$ obtained by squaring the amplitude and summing over
polarizations of intermediate $\Omega_0$ has only a part symmetric in $j,k$:
\beq
\frac{1}{\pi} {\rm Im}\, P_{jk}(\omega)= 
\frac{g^2_{B^*B\pi}}{12\pi^2} \frac{|\vec k|^5}{\omega^2} \,\delta_{jk} 
\Psi^\dagger_0 \Psi_0.
\label{2282} 
\eeq 
From this we read off 
\beq
{\cal T}^{(\frac{1}{2}^-)}(\omega) = {\cal T}^{(\frac{3}{2}^-)}(\omega)=
\frac{g^2_{B^*B\pi}}{36\pi^2 }
\theta(\omega\!-\!m_\pi)(\omega^2\!-\!m_\pi^2)^{3/2} 
\left(1\!-\!\frac{m_\pi^2}{\omega^2}
\right), 
\label{2284} 
\eeq 
where the overall factor refers to a single charged pion loop contribution. The
neutral pion additionally contributes a half of that. Relations
(\ref{2272.02}) give the values of the corresponding $|\tau_{1/2}|$
and  $|\tau_{3/2}|$, which are equal. 

It is worth noting that the equality $\tau_{1/2}\!=\!\tau_{3/2}$ is not an
automatic property of quantum numbers in the $D\pi$ system. It rather
follows from the form of the soft-pion amplitude for heavy mesons. For
instance, a structure $\bar \Psi_0 \sigma_j \Psi_0$ would have produced only
$\frac{1}{2}^-$ but not $\frac{3}{2}^-$.
The relation generally changes already due to the final state interaction (FSI) in the
$\Omega_0\pi$ system, although the latter is suppressed by extra powers of 
pion momentum. Before the three-pion threshold the amplitude is
completely characterized by the pion-$\Omega_0$ scattering phases which are
different in the $\frac{1}{2}^-$ and  $\frac{3}{2}^-$ states,
$\delta_{1/2}(\omega)$ and $\delta_{3/2}(\omega)$:
\beq
\frac{1}{2M_{\Omega_0}}\matel{(\Omega_0 \pi)_j}{A}{\Omega_0(0)} \propto 
|A| e^{i\delta_j(\omega)} 
\label{2286} 
\eeq 
for arbitrary operator $A$. These phases, strictly speaking, should be included
into $\tau_{1/2}(\omega)$ and $\tau_{3/2}(\omega)$. The FSI phases also 
depend on the isospin state,
$I\!=\!\frac{1}{2}$ or $I\!=\!\frac{3}{2}$, but we now abstract  from the light
flavor symmetries. 

The scattering phases emerge via iterations of
the $\Omega_0$-pion interaction along with renormalization of the coupling
constant, and they must vanish in the limit of small pion 
four-momentum due to the pseudoscalar nature of the interaction. This fixes
the physical value of the coupling as its value renormalized at the threshold.
We identify this coupling with the one in the `bare' chiral Lagrangian
Eq.~(\ref{2274}). Although the higher-order terms in pion momentum, including 
both scattering phases, can
formally be obtained in perturbation theory in $g_{B^*B\pi}$ from the Yukawa-type
Lagrangian in Eq.~(\ref{2274}), this would make no sense  for a
number of evident physical reasons. What is relevant here is  that
even a naive perturbative expansion yields different scattering phases 
$\delta_{1/2}(\omega)$ and  $\delta_{3/2}(\omega)$. 

The net effect of FSI is to introduce an effective phase $e^{i\delta_j(\omega)}$
in $\tau_j$ and
to replace the threshold $g_{B^*B\pi}$ in Eq.~(\ref{2284}) by an
energy-dependent coupling. In principle, the phases together with the couplings are
constrained by analyticity of the amplitude and its unitarity property which
are restrictive before the higher thresholds open. However, these may not fix
the amplitude completely since the resulting relations are not local in energy and
depend on the multiparticle domain. Likewise, the solution generically admits
resonances in a particular channel whose number, position in energy and
residues may vary.

We shall neglect all such effects in what follows, assuming that the bulk of
them is included in the resonance contributions, and subtracting the latter
largely results in an effective cutoff of the soft-pion amplitudes at a
certain scale near or below the lowest resonance. In particular, we attribute
the difference between the $\frac{1}{2}^-$ and  $\frac{3}{2}^-$ channels to
the resonant states. 

Next on our list is $L\!=\!1$ producing the true
$\frac{1}{2}^+$ radial and  $\frac{3}{2}^+$ states; in our treatment 
they will again appear along with the $\frac{5}{2}^+$ `$D$-wave' states.
For the heavy quark expansion we need to calculate the matrix
elements for the operator with two derivatives, 
\beq
\frac{1}{2M_{\Omega_0}} \matel{\Omega_0 \pi}{\bar{Q} i{D}_j
  i{D}_l Q}{\Omega_0(0)},\\
\label{2290.02} 
\eeq 
contracted with $\delta_{jl}$, $\epsilon_{mjl}$ and 
$(\delta_{mj}\delta_{nl}+\delta_{ml}\delta_{nj}\!-
\mbox{$\frac{2}{3}$}\delta_{mn}\delta_{jl})$ to yield the spin-$0$,  spin-$1$
and  spin-$2$ operators, respectively.

To evaluate the pole diagrams for them we need the couplings 
analogous to Eq.~(\ref{2278}), at zero momentum transfer. 
These expectation values, however, are required over the
heavy hadron moving with a small momentum of order $\omega \!\sim\! \mhad$. 
They are related by Lorentz invariance to the generic rest frame matrix elements
with up to two spatial derivatives. Namely, the matrix elements of a product of
any number of full covariant derivatives $m_Q v_\mu\!+\!\pi_\mu$ form a 
Lorentz tensor of
the corresponding rank, and its value in an arbitrary frame is obtained,
regardless of $m_Q$, from the rest frame
components by the corresponding Lorentz transformation. In
the case at hand we get 
\bea
\nonumber
\frac{1}{2M_{\Omega_0}} \matel{\Omega_0(\vec p)}
{\bar Q \vec \pi^2Q}{\Omega_0(\vec p)} 
\msp{-4}&=&\msp{-4} (\vec p^{\,2}\!+\!\mu_\pi^2)\Psi^\dagger_0 \Psi_0, \\\nonumber
\frac{1}{2M_{\Omega_0}} \matel{\Omega_0(\vec p)}
{\bar Q (\pi_j\pi_l\!-\!\pi_l\pi_j)Q}{\Omega_0(\vec p)} 
\msp{-4}&=&\msp{-4} -\frac{\mu_G^2}{3}\Psi_0^\dagger \sigma_{jl}\Psi_0, \\
\frac{1}{2M_{\Omega_0}} (\delta_{mj}\delta_{nl}\!+\!\delta_{ml}\delta_{nj}\!-
\!\mbox{$\frac{2}{3}$}\delta_{mn}\delta_{jl})
\matel{\Omega_0(\vec p) }{\bar{Q} \pi_j  \pi_l Q}{\Omega_0(\vec p)}  
\msp{-4}&=&\msp{-4}
\!(2 p_m p_n \!-\!\mbox{$\frac{2}{3}$}\delta_{mn} \vec p^{\,2}) \Psi^\dagger_0 \Psi_0. \quad
\label{2292} 
\eea 
Additional $p$-independent terms are absent since no spin-$2$ or higher 
current can
be constructed with a spin-$\frac{1}{2}$ particle at rest. Similar
relations hold for any higher order product as well; basically, the
result for a non-zero momentum is obtained by incrementing 
each $\vec \pi$ by $\vec p$:
\beq
\matel{\Omega_0(\vec p) }{\bar{Q} \pi_{j_1} ... \pi_{j_k} Q}{\Omega_0(\vec p)}=
\matel{\Omega_0(\vec p\!=\!0) }{\bar{Q} (\pi_{j_1}\!\!+\!p_{j_1}) ... 
(\pi_{j_k}\!\!+\!p_{j_k}) Q}{\Omega_0(\vec p\!=\!0)}.
\label{2293}
\eeq
Otherwise the calculation of the transition amplitude proceeds
exactly like for $P$-waves and yields
\beq
\frac{1}{2M_{\Omega_0}}\matel{\Omega_0 \pi}{\bar{Q} \pi_j\pi_l Q}{\Omega_0(0)}= 
g_{B^*B\pi}\,\frac{1}{\omega}\, \,\Psi_0^\dagger \,i \!\left( 
\mbox{$\frac{\mu_G^2}{3}$}(k_j\sigma_l\!-\!k_l\sigma_j)+ (\vec \sigma \vec k)
k_j k_l \right)\Psi_0.
\label{2294} 
\eeq 

The amplitude above calculated for spin-$0$ and spin-$1$ operators
has a nontrivial `diagonal' $\Omega_0$ matrix element at rest, with the
diagram {\bf a)} not vanishing. 
The `diagonal'
piece of the matrix elements in Eqs.~(\ref{2292}) (in this case it is the 
value at $\vec p\!=\!0$) is independent of the momentum. 
The propagators in diagrams {\bf a)} and {\bf b)} have opposite
sign; this contribution 
then enters universally as a commutator with the pion interaction Hamiltonian,
in accord with stationary perturbation theory:
\beq
A_{mn} = \frac{[\delta {\cal H}, {\cal H}]_{mn}}{(E_m\!-\!E_n)^2}\,.
\label{2295}
\eeq
For instance, the kinetic
expectation value $\mu_\pi^2$ always drops out, 
but a contribution remains
proportional to the chromomagnetic interaction $\mu_G^2$ which does not
commute with the spin-dependent pion vertex. This yields the most IR-singular
contribution for a soft pion due to the double pole in Eq.~(\ref{2295}).

Strictly speaking, the pole diagrams with virtual $\Omega_0$ -- they have a pole
at $\omega\!=\!0$ ($\vec k$ is assumed to be fixed, and the overall $\vec k$
from the pion interaction is factored out ) -- does not 
describe the pion emission amplitude completely. It contains a piece finite
at $\omega\!=\!0$ coming from the contact interactions, see
Fig.~\ref{omega0pi}{\bf c}. These
contact vertices assume some values in QCD; we only know they are proportional
to $\vec k$ since the heavy quark operators we consider are chirally sterile
so that their (light flavor) axial charge vanishes:
\beq
\lim_{k_\mu \to 0}\,\matel{\Omega_0 \pi^a(k)}{\bar Q iD_j iD_l Q}{\Omega_0} =
-\frac{1}{F_\pi} \int\! {\rm d}^3 x\; 
\matel{\Omega_0 \pi^a}{[\bar Q iD_j iD_l Q(0),J_0^{5(a)}(0,\vec{x})]}{\Omega_0}=0.
\label{2295b}
\eeq
Otherwise
they are arbitrary, and we can only parameterize them by four constants of
dimension mass:
\bea
\nonumber
\frac{1}{g_{B^*B\pi}}\matel{\Omega_0 \pi}{\bar Q iD_j iD_l Q}{\Omega_0} 
\msp{-4} & = & \msp{-4} \alpha\,
\Psi_0^\dagger(k_j \sigma_l\!-\!k_l \sigma_j)\Psi_0+
\beta\, i\epsilon_{jlm}k_m
\Psi_0^\dagger\Psi_0 
+ \gamma\, \delta_{jl} \Psi_0^\dagger i \vec \sigma \vec k \Psi_0 
+ \\
& & \msp{20}
\tau \,i\Psi_0^\dagger (\sigma_j k_l+ \sigma_l k_j-
\mbox{$\frac{2}{3}$} \delta_{jl}\vec \sigma \vec k) \Psi_0,
\label{2295a}
\eea
plus higher terms in $k$.
Such terms are induced, in particular, by intermediate excited heavy quark
meson resonance states propagating in diagrams  Fig.~\ref{omega0pi}, a),b), with
their relative size depending on spin.
Unequal meson-to-pion couplings in Sect.~\ref{sectdpi} correspond to particular values
of the constants  $\alpha$ to $\gamma$; based on the QCD sum rule analysis 
Ref.~\cite{khod} they appear large. 
While these terms are parametrically smaller than the pole
amplitudes for small pion momenta, in general they contribute, 
especially the quantities regular in the chiral limit. 

There are also terms next order in $p_\pi/\mhad$. They would have 
the same scaling as the last term in Eq.~(\ref{2294}).
Nevertheless, 
we shall neglect them in what follows when address the pion loop effects
proper: we relegate the corresponding contributions
to the effects of resonances. This parallels the 
pole-dominance assumption employed in the calculation of the $D^{(*)}\pi$ amplitudes
in Sect.~\ref{sectdpi}, however the $Z$-diagrams included in the relativistic
propagators of intermediate $B^*$ and $D^*$ mesons also induce such terms. If
an argument can be put forward that one should retain only the $D^*$ or $B^*$
in the normal diagrams, there is no physical reason to exclude the higher
states in the $Z$-diagrams: they all have the same large
virtuality. Therefore, it may be natural to assume that in aggregate 
the $Z$-induced contact terms are suppressed or vanish in the nonrelativistic
expansion.

The result for the pion loop in $R(\omega)$ takes the
following form:
\bea
\nonumber
R^\pi_{ijkl}(\omega) &\msp{-4}=\msp{-4}&  
\frac{g^2_{B^*B\pi}}{12\pi^2} \,\theta(\omega\!-\!m_\pi)\,
\frac{|\vec k|^3}{\omega^2} \Psi^\dagger_0 
\left[
\frac{(\mu_G^2)^2}{9}(2\delta_{ik}\delta_{jl}\!-\!2\delta_{il}\delta_{jk}+
\delta_{ik}\sigma_{jl}\!+\!\delta_{jl}\sigma_{ik}\!-\!
\delta_{il}\sigma_{jk}\!-\!\delta_{jk}\sigma_{il})\right. \\
& & \msp{-4}
\left.
+\frac{\mu_G^2}{3}\frac{2\vec k^{\,2}}{5}
(\delta_{ik}\sigma_{jl} -\delta_{kl}\sigma_{ij} + \delta_{ij}\sigma_{kl}-
\delta_{jl}\sigma_{ik})
+\frac{\vec k^{\,4}}{5}(\delta_{ij}\delta_{kl}+
\delta_{ik}\delta_{jl}+ \delta_{il}\delta_{jk}) 
\right]\Psi_0 .\msp{9}
\label{2296} 
\eea 
This can be decomposed into the invariant structures; introducing the common factor
$$
W \equiv \frac{g^2_{B^*B\pi}}{12\pi^2} 
\,\theta(\omega\!-\!m_\pi)
$$
we get 
\bea
\nonumber
\rho_p^{(\frac{1}{2}^+)}(\omega)  &\msp{-4}=\msp{-4}& 
W\, \frac{3|\vec k|^7}{\omega^2} , \quad 
\rho_{pg}^{(\frac{1}{2}^+)}(\omega) = 
-W \frac{4\mu_G^2|\vec k|^5}{\omega^2}
,\quad
\rho_{g}^{(\frac{1}{2}^+)}(\omega) = W \frac{16(\mu_G^2)^2|\vec
  k|^3}{3\omega^2} , \qquad
\\
\label{2298}
\rho_f^{(\frac{3}{2}^+)}(\omega)  &\msp{-4}=\msp{-4}& 
W\, \frac{16|\vec k|^7}{\omega^2}
, \quad 
\rho_{fg}^{(\frac{3}{2}^+)}(\omega) = W \frac{8\mu_G^2|\vec k|^5}{3\omega^2}
, \quad 
\rho_{g}^{(\frac{3}{2}^+)}(\omega) = W \frac{16(\mu_G^2)^2|\vec
  k|^3}{9\omega^2},\\
\nonumber
\rho^{(\frac{5}{2}^+)}(\omega) &\msp{-4}=\msp{-4}& W\, \frac{2|\vec
  k|^7}{5\omega^2} .
\eea 
The factorization properties stated in Sect.~\ref{interm} are manifest here. 

For practical applications it is important to include in the amplitude the
next-to-leading `contact' terms described by $\alpha,\beta$ and $\gamma$ 
in Eq.~(\ref{2295a}); they can actually 
be derived in a model-independent way and do not depend on the form of the 
resonance ansatz. Calculating the full $R^\pi_{ijkl}(\omega)$ with them
 is straightforward, but we pospone it to future work.

The above calculation of the pion loop can be readily generalized to 
higher-dimensional heavy quark momentum operators, including the
case of different number of derivatives in the two vertices. The expressions are
particularly simple where no timelike momentum or antisymmetric spatial 
indices are involved; then the corresponding vertices simply
amount to products of pion momentum and the resulting integrals 
can be simply calculated.

We now return to our practical need.  We are concerned with only the
zero-momentum correlators of the $1/m_Q$-terms in the heavy quark Lagrangian, and we
use the above derived spectral densities to estimate 
\bea
\delta \rho^3_{\pi\pi} \approx 0.02 \GeV^3 , \qquad 
\delta \rho^3_{\pi G}\approx 0.065\GeV^3, 
\qquad \delta \rho_S^3=\delta \rho_A^3\approx 0.03\GeV^3, 
\label{2310} 
\eea 
and 
\bea
\delta \tilde\rho^3_{\pi\pi} \approx  0.03 \GeV^2 , 
\qquad  \delta \tilde\rho^3_{\pi G}\approx 0.1\GeV^2,
\qquad \delta \tilde \rho_S^3=\delta \tilde \rho_A^3\approx 0.07\GeV^2, 
\label{2311} 
\eea 
where we have assumed the upper cutoff at $\omega\!=\!700\MeV$ and have
adopted $g_{B^*B\pi}=4\GeV^{-1}$. The loops with charged and neutral
pions are included, but not the kaon and $\eta$ contributions. Of
special interest is the `hyperfine' combination of Eq.~(\ref{164a}) for which we
get 
\beq
\delta (-\rho^3_{\pi G }-\rho_A^3) \approx -0.09\GeV^3;
\label{2310a}
\eeq
as expected, the pion contribution is suppressed, but it is negative and, taken at
face value, would strengthen the lower bound on the resonant
contributions. This is due to the positive sign of $\delta \rho^3_{\pi G}$,
opposite to the BPS regime. 

With the same choice for $g_{B^*B\pi}$ we would have
\bea
\nonumber
\delta \varrho^2  \msp{-4} & \approx & \msp{-4} 0.015
\mbox{$\left(\frac{\omega_{\rm max}}{0.5\,{\rm GeV}}\right)^2$}, 
\qquad \qquad \delta \La\approx 
\mbox{$\left(\frac{\omega_{\rm max}}{0.5\,{\rm GeV}}\right)^3$}
\,10\MeV, \\
\delta \mu_\pi^2 \msp{-4} & \approx & \msp{-4} 
\mbox{$\left(\frac{\omega_{\rm max}}{0.5\,{\rm  GeV}}\right)^4$}\, 0.006\GeV^2, 
\qquad \delta \rho_D^3\approx 
\mbox{$\left(\frac{\omega_{\rm max}}{0.5\,{\rm GeV}}\right)^5$}\,
0.0025\GeV^3
\label{2312} 
\eea 
where we have anticipated a lower effective cutoff in the $P$-wave channel
(the spin-triplet counterparts are not affected).
We see that generally the pion loop contributes only a small fraction of the
nonlocal correlators and a tiny amount of the local expectation values. 
 A possible exception are quantities vanishing in
the BPS limit where 
the pion loop may constitute a significant part of the deviation. 

The first calculation  \cite{randwise}
of the chiral correction to the formfactor to order $1/m_Q^2$ accounted
for the terms proportional to $(\mu_G^2)^2$ from the HQS-breaking 
masses in the meson propagators and picked up only the $\tilde\rho_S^3$ and  
$\tilde\rho_A^3$ pieces in Eq.~(\ref{2311}), yet leaves out
$\tilde\rho_{\pi\pi}^3$ and $\tilde\rho_{\pi G}^3$. We see, however that the latter
contribution has the same size even though it is not singular
in the chiral limit. 

One can combine the above description of the nonresonant $D^{(*)}\pi$ states with
the method of Sect.~\ref{yield} for the inclusive yield, to get an
estimate of the total continuum contribution to the inclusive yield. To that
end one only needs to integrate the expressions in Eq.~(\ref{2298}) 
over the mass of the $\Omega_0 \pi$  state;
due to the strong phase-space suppression the integral is effectively cut off at
relatively soft pions. In this way we arrive at the yield in the ball park of
$1\%$ of $\Gamma_{\rm sl}$. This, however, refers to only the specified
$j^P$ of the $D^{(*)}\pi$ states. The $P$-wave continuum is an independent
channel. 

\subsection{Nonfactorizable contributions to higher-dimensional local expectation values}

The size of the expectation values of local heavy quark operators of $D\!=\!7$
and $D\!=\!8$ is important to estimate the impact of
higher-order power corrections 
both in beauty and charm, to assess the accuracy of the OPE predictions and
to study the convergence of the OPE series. The relevant operators here 
are those which emerge in the calculation of power corrections at  tree level;
 they are sometimes called `color-through' operators. Ref.~\cite{hiord} 
illustrated their effect in inclusive $B$ decays for
the semileptonic $b\tto c$ transitions and for $B\tto X_s\!+\!\gamma$ used
in measuring $|V_{cb}|$. They have also 
appeared in Sect.~\ref{powercor} in our analysis of 
$B\tto D^* \ell\nu$ at zero recoil. 

In order to estimate the significance of the expectation values
a ground-state factorization method has been devised in Ref.~\cite{hiord}
which contains a derivation of the
formalism and  the explicit expressions for the factorization
contributions to all nine dimension-$7$, $m_{1-9}$, and eighteen dimension-$8$,
$r_{1-18}$, $B$-meson expectation values. 
Using the intermediate state saturation representation of
Ref.~\cite{hiord} together with the relations elaborated in 
the previous sections we can now
supplement the ground-state factorization values with the contributions from 
the excited states: 
 the analysis of hyperfine splitting allows  to quantify their effect.
This enables us to assess the accuracy of the factorization
ansatz, the potential scale of the corrections to factorization and, ultimately, 
 may elucidate the pattern of the higher-order effects in a more 
quantitative manner.

Once again we start with infinitely heavy spinless quarks. The
intermediate state representation for the operators with four spatial
derivatives reads
\beq
\frac{1}{2M_Q} \matel{\Omega_0} {Q^\dagger \pi_i\pi_j \pi_k\pi_l Q(0)}{\Omega_0} =
\int\! {\rm d}\omega \; R_{ijkl}(\omega),
\label{2400}
\eeq
with $R_{ijkl}(\omega)$ introduced in Eq.~(\ref{128}). The factorization
contribution is located at $\omega\!=\!0$; the upper limit of integration
over  $\omega$ is determined by the normalization point assumed for the operator.
The non-factorized pieces take a different form for each class of  intermediate
states.

Contracting indices in $R_{ijkl}(\omega)$ and multiplying the tensor, for the
spin-triplet operators,  by the spin matrix with an appropriate index, we obtain
for the actual $B$ mesons
\bea
\nonumber
\delta^{\rm nf} m_1 \msp{-4}&=&\msp{-4} \frac{5}{9}\rho_p^{(\frac{1}{2}^+)} +
\frac{1}{30}\rho_f^{(\frac{3}{2}^+)} + 2\rho^{(\frac{5}{2}^+)}
\quad \qquad \delta^{\rm nf} m_3 = -\frac{2}{3}\rho_g^{(\frac{1}{2}^+)} -
\rho_g^{(\frac{3}{2}^+)}\\
\nonumber
\delta^{\rm nf} m_4 \msp{-4}&=&\msp{-4} \frac{4}{3}\rho_p^{(\frac{1}{2}^+)} +
\rho_g^{(\frac{1}{2}^+)} -\frac{1}{10}\rho_f^{(\frac{3}{2}^+)} +
\frac{3}{2} \rho_g^{(\frac{3}{2}^+)} -6\rho^{(\frac{5}{2}^+)}\\
\nonumber
\delta^{\rm nf} m_6 \msp{-4}&=&\msp{-4} \frac{2}{3}\rho_g^{(\frac{1}{2}^+)} 
-\frac{1}{4}\rho_g^{(\frac{3}{2}^+)} 
\qquad\qquad\qquad
\delta^{\rm nf} m_7 = -\frac{8}{3}\rho_{pg}^{(\frac{1}{2}^+)} 
-2\rho_{fg}^{(\frac{3}{2}^+)} \\
\nonumber
\delta^{\rm nf} m_8 \msp{-4}&=&\msp{-4} -8\rho_{pg}^{(\frac{1}{2}^+)} 
\\
\delta^{\rm nf} m_9 \msp{-4}&=&\msp{-4} -\frac{10}{3}\rho_{pg}^{(\frac{1}{2}^+)} +
\rho_g^{(\frac{1}{2}^+)} -\frac{3}{20}\rho_f^{(\frac{3}{2}^+)} -
\rho_{fg}^{(\frac{3}{2}^+)}-\frac{3}{4}\rho_{g}^{(\frac{3}{2}^+)} +
6\rho^{(\frac{5}{2}^+)}\,.
\label{2402}
\eea
In the above equations  the integration over $\omega$ is assumed, it has not
been shown explicitly for compactness. We remind that $m_2$ and $m_5$ are
given by the fourth moment of the standard SV ($P$-wave) structure functions
and have no conventional ground-state factorizable  contributions. 

For numeric estimates one simply considers the contributions of the
individual multiplets of the excited states using their spectral densities in
Eq.~(\ref{2140}). 
The contribution of $\frac{1}{2}^+$ is obvious beforehand: it follows the
factorizable one, see Ref.~\cite{hiord}, and only requires 
to replace $(\mu_\pi^2)^2$ by $P^2$, $(\mu_G^2)^2$ by $G^2$ and 
$\mu_\pi^2\mu_G^2$ by $PG$. 
The effect of the higher-spin states has a different structure.  
Since the  basis $\left\{m_1-m_9\right\}$ has been selected arbitrarily,
the impact of 
nonfactorizable contributions should be gauged in specific cases; this is easily
done based on Eqs.~(\ref{2402}). 

To quantify the overall scale of 
the effect we consider here three representative combinations $M_1, M_2, M_3$
corresponding to the expectation values $\bar{b}\vec \pi^2 \vec\pi^2 b$, 
$\bar{b}(\vec \sigma \vec B) (\vec \sigma \vec B) b$ and $-\bar{b}(\vec \sigma
\vec B)  \vec\pi^2 b$, respectively:
\beq
M_1 = m_1+\frac{1}{2}m_3+\frac{1}{3}m_4, \qquad
M_2 = -m_3, \qquad
M_3 = -\frac{1}{8}m_8,
\label{2460}
\eeq
for which  we have
\bea
\nonumber
M_1 \msp{-4}&=&\msp{-4} (\mu_\pi^2)^2 + \int {\rm d}\omega 
\:\rho_p^{(\frac{1}{2}^+)}(\omega)\\
\nonumber
M_2 \msp{-4}&=&\msp{-4} \frac{2}{3}(\mu_G^2)^2+ \int {\rm d}\omega 
\:\Big(\mbox{$\frac{2}{3}$}\rho_g^{(\frac{1}{2}^+)}(\omega)
+\rho_g^{(\frac{3}{2}^+)}(\omega)\Big)  \\
M_3 \msp{-4}&=&\msp{-4} \mu_\pi^2\mu_g^2 + \int {\rm d}\omega 
\:\rho_{pg}^{(\frac{1}{2}^+)}(\omega).
\label{2462}
\eea
Numerically the corrections depend to some extent on the ratio of the
$\frac{3}{2}^+$ and $\frac{1}{2}^+$ contributions and on $P/G$ in the latter.
Taking, for instance,  
$\varepsilon_{\rm rad}\!\approx\!700\MeV$, $\;\int {\rm d}\omega 
\:\rho_g^{(\frac{1}{2}^+)}(\omega)\approx \int {\rm d}\omega 
\:\rho_{pg}^{(\frac{1}{2}^+)}(\omega) \approx \int {\rm d}\omega 
\:\rho_g^{(\frac{3}{2}^+)}(\omega)\;$ ($P\!\approx \!G$, see Sect.~\ref{hyper}) 
and using the hyperfine
constraint Eq.~(\ref{2320}) with $\kappa \!\approx\! -0.2$  we obtain
\bea
\nonumber
M_1 \msp{-4}& \approx & \msp{-4} 0.2\GeV^4_{\rm fact} + 0.17\GeV^4 _{\rm n-fact}
\\
\nonumber
M_2 \msp{-4}& \approx &\msp{-4} 0.08 \GeV^4_{\rm fact} + 0.28\GeV^4 _{\rm n-fact}
\\
M_3 \msp{-4}& \approx &\msp{-4} 0.15 \GeV^4 _{\rm fact} + 0.17\GeV^4 _{\rm n-fact}.
\label{2464}
\eea

A more definite value is obtained for the special combination
$$
-\mbox{$\frac{1}{5}$}m_3-\mbox{$\frac{6}{5}$}m_6 -\mbox{$\frac{1}{4}$}m_8
$$
which has the same structure as the hyperfine constraint; here we get
\beq
-\mbox{$\frac{1}{5}$}m_3-\mbox{$\frac{6}{5}$}m_6 -\mbox{$\frac{1}{4}$}m_8
\simeq
-\mbox{$\frac{2}{3}$}(\mu_G^2)^2 + 2\mu_\pi^2 \mu_G^2 -
\varepsilon_{\rm rad}(\rho^3_{\pi G}\!+\!\rho_A^3) \approx
(0.23 + 0.32_{\rm nf})\GeV^4 .
\label{2466}
\eeq
From this brief comparison we conclude that the factorization ansatz
generally provides  no more than a reasonable starting approximation 
for the expectation values not affected by  cancellations.

The hyperfine constraint and the approximations
we complemented it with have nothing to say about the contribution of spin-$\frac{5}{2}$ states which require a different theoretical input.  Positivity and various 
relations between different nonfactorizable contributions are implicit in 
Eqs.~(\ref{2402}); 
these constraints should, in principle, be applied only after the contributions 
from the non-resonant continuum are subtracted. 

The $D\!=\!8$ operators with five derivatives were found to contribute at a
 lower level to the  inclusive moments in $B$ decays
\cite{hiord}; their precise expectation values are therefore less important in
practice. Nevertheless, in this case it would  also be useful 
to have at least a crude estimate
of the potential error in the factorization ansatz, to be more
confident in the assessment of the impact of $1/m_Q^5$ terms. 

The nonfactorizable effects for the $D\!=\!8$ 
operators can be analyzed along the same lines as the $D\!=\!7$ operators, identifying $iD_0$ adjacent to
the intermediate state in question with $-\varepsilon$. Most notably, the excited
states contribute to a few combinations of $r_i$ which
vanish in the ground-state factorization: 
\bea
\nonumber
r_5 \msp{-4}&=&\msp{-4} - \omega \rho_p^{(\frac{1}{2}^+)} (\omega) 
\approx -0.13\GeV^5
\qquad\quad
r_{15} = - \omega \frac{1}{2}\rho_{pg}^{(\frac{1}{2}^+)}(\omega) 
\approx -0.06\GeV^5
\\
\nonumber
r_6\!-\!r_1 \msp{-4}&=&\msp{-4} -\omega \left(
\frac{1}{3}\rho_p^{(\frac{1}{2}^+)} + \frac{1}{6}\rho_g^{(\frac{1}{2}^+)}
+\frac{1}{20}\rho_f^{(\frac{3}{2}^+)} + \frac{1}{4}\rho_{g}^{(\frac{3}{2}^+)} +
3\rho^{(\frac{5}{2}^+)}
\right)\\
\nonumber
r_7 \msp{-4}&=&\msp{-4} -\omega \left(
\frac{1}{3}\rho_p^{(\frac{1}{2}^+)} - \frac{1}{6}\rho_g^{(\frac{1}{2}^+)}
+\frac{1}{20}\rho_f^{(\frac{3}{2}^+)} - \frac{1}{4}\rho_{g}^{(\frac{3}{2}^+)} +
3\rho^{(\frac{5}{2}^+)}
\right)
\\
\nonumber
r_{16} \msp{-4}&=&\msp{-4}   -\omega \left(
-\frac{2}{3}\rho_{pg}^{(\frac{1}{2}^+)} - \frac{1}{6}\rho_g^{(\frac{1}{2}^+)}
-\frac{3}{40}\rho_f^{(\frac{3}{2}^+)} - \frac{1}{2}\rho_{fg}^{(\frac{3}{2}^+)}
+ \frac{1}{8}\rho_{g}^{(\frac{3}{2}^+)} + 3\rho^{(\frac{5}{2}^+)}
\right)
\\
\nonumber
r_{17}\!-\!r_8 \msp{-4}&=&\msp{-4}  -\omega \left(
\frac{2}{3}\rho_{pg}^{(\frac{1}{2}^+)} - \frac{1}{6}\rho_g^{(\frac{1}{2}^+)}
-\frac{3}{40}\rho_f^{(\frac{3}{2}^+)} + \frac{1}{2}\rho_{fg}^{(\frac{3}{2}^+)}
+ \frac{1}{8}\rho_{g}^{(\frac{3}{2}^+)} + 3\rho^{(\frac{5}{2}^+)}
\right)
\\
r_{18} \msp{-4}&=&\msp{-4}   -\omega \left(
\frac{1}{6}\rho_g^{(\frac{1}{2}^+)}
-\frac{3}{40}\rho_f^{(\frac{3}{2}^+)}
- \frac{1}{8}\rho_{g}^{(\frac{3}{2}^+)} + 3\rho^{(\frac{5}{2}^+)}
\right).
\label{2472}
\eea
In the above equations the integration over $\omega$ is again
assumed; 
the numeric estimates for $r_5$ and $r_{15}$ are obtained 
under the same assumptions as Eqs.~(\ref{2464}). 
 
The most general analysis of the nonfactorizable corrections for the other
operators (except for $r_1$ and $r_8$ given  by the fifth moment of the
generalized SV structure functions) requires considering
\beq
\tilde R_{ijkl}(\omega) =\frac{1}{2\pi}\int\! {\rm d}^3x \int\! {\rm d}x_0\; 
e^{-i\omega x_0} \frac{1}{2M_Q} \matel{\Omega_0} 
{Q^\dagger \pi_i\pi_j Q(x)\: Q^\dagger \pi_k \pi_0\pi_l Q(0)}{B}
\label{2482}
\eeq
which is  an analogue of the tensor spectral 
density $R_{ijkl}(\omega)$ in Eq.~(\ref{128}).
One of the operators now includes an extra time derivative.  The corresponding
decomposition is lengthier than  Eq.~(\ref{130}) since there is no symmetry between
pairs of indices. The factorization 
of Eqs.~(\ref{2140}) is modified for the new invariant structures:
  along with the transition matrix elements in
Eqs.~(\ref{2095g}), (\ref{2095}) and (\ref{2095c}) we need to introduce the
similar ones $\tilde P, \tilde G$, $\tilde f, \tilde g$ and $\tilde h$ for 
the operators $\bar{Q}\pi_l \pi_0\pi_l Q$ (it is assumed that
$\pi_0$ acts on the right which matters for non-diagonal matrix
elements). The analogue of Eq.~(\ref{2400}) for the operators with
five derivatives in terms of $\tilde R_{ijkl}(\omega)$ holds and 
the general relations for the remaining $r_i$ similar to 
Eqs.~(\ref{2402}) or (\ref{2462}) can readily be derived. 

The hyperfine splitting constraint cannot, however
be directly applied to $\tilde R_{ijkl}(\omega)$ and the corresponding tilded residues
remain largely unconstrained even in the single excited
multiplet approximation. The BPS approximation  in this case yields
$\tilde P \!=\!\tilde G$ and $\tilde f \!=\! \tilde g$, but it is not  too helpful.
 Therefore we do not quote here the corresponding expressions.

It is nevertheless possible to get a rough estimate by making the
assumption that the first $P$-wave excitation(s) approximately saturate, as an
intermediate state, the transition amplitudes into the radial or the $D$-wave states:
\beq
\matel{\rho}{\pi_j\pi_0\pi_k}{\Omega_0}\approx 
\matel{\rho}{\pi_j}{P_{\frac{3}{2}}^{(1)}}
\matel{P_{\frac{3}{2}}^{(1)}}{\pi_0\pi_k}{\Omega_0}+ 
\matel{\rho}{\pi_j}{P_{\frac{1}{2}}^{(1)}}
\matel{P_{\frac{1}{2}}^{(1)}} {\pi_0\pi_k}{\Omega_0},
\label{2410}
\eeq
where $\rho$ generically refers to the $\frac{1}{2}^+$,  $\frac{3}{2}^+$ or
$\frac{5}{2}^+$ states under consideration. 
This can also be regarded as an approximate relation obtained by truncating the
complete representation 
\beq
\pi_j\state{\Omega_0}= \sum_m  \sqrt{3} \epsilon_m \tau_{3/2}^{(m)} 
\,\state{\chi^{(m)}}_j
+\sum_n  \epsilon_n \tau_{1/2}^{(n)} \,\sigma_j
\state{\phi^{(k)}} \approx \sqrt{3} \epsilon_{3/2}^{(1)} \tau_{3/2}^{(1)} 
\, \state{\chi^{(1)}}_j
+ \epsilon_{1/2}^{(1)} \tau_{1/2}^{(1)} \,\sigma_j \state{\phi^{(1)}}
\label{2412}
\eeq
after the lowest $P$-wave families.
Such an assumption seems to work satisfactorily for the transition between the 
ground states, yet may be
expected to degrade with higher initial and/or final states. Identifying
$\epsilon_{1/2}$ and $\epsilon_{3/2}$ with $\tilde\epsilon$ this would yield 
\beq
\tilde P \approx -\tilde \epsilon P, \qquad 
\tilde G \approx -\tilde \epsilon G, \qquad 
\tilde f \approx -\tilde \epsilon f, \qquad 
\tilde g \approx -\tilde \epsilon g, \qquad 
\tilde h \approx -\tilde \epsilon h 
\label{2414}
\eeq
for individual residues, and
\beq
\tilde \rho^{(\frac{l}{2}^+)} \approx -\tilde \epsilon \,\rho^{(\frac{l}{2}^+)}
\label{2416}
\eeq
for all invariant tensor structures with $l\!=\!1$ and  $l\!=\!3$, 
and, most generally,
\beq
\tilde  R_{ijkl}(\omega) \approx -\tilde \epsilon\cdot R_{ijkl}(\omega).
\label{2418}
\eeq
 
Adopting for orientation  such an approximation we obtain
\bea
\nonumber
\delta^{\rm nf} r_2 \msp{-4}&  \approx &\msp{-4} -\tilde \epsilon
\:\rho_p^{(\frac{1}{2}^+)} \approx -0.07\GeV^5\\
\nonumber
\delta^{\rm nf} r_3 \msp{-4}& \approx &\msp{-4} -\tilde \epsilon 
\left(\frac{1}{3}\rho_p^{(\frac{1}{2}^+)} -\frac{1}{6}\rho_g^{(\frac{1}{2}^+)} 
+\frac{1}{20}\rho_f^{(\frac{3}{2}^+)} -\frac{1}{4}\rho_g^{(\frac{3}{2}^+)} 
+3 \rho^{(\frac{5}{2}^+)}
\right)\\
\nonumber
\delta^{\rm nf} r_4 \msp{-4}& \approx &\msp{-4} -\tilde \epsilon 
\left(\frac{1}{3}\rho_p^{(\frac{1}{2}^+)} +\frac{1}{6}\rho_g^{(\frac{1}{2}^+)} 
+\frac{1}{20}\rho_f^{(\frac{3}{2}^+)} +\frac{1}{4}\rho_g^{(\frac{3}{2}^+)} 
+3 \rho^{(\frac{5}{2}^+)}
\right)\\
\nonumber
\delta^{\rm nf} r_9 \!\approx\! 
\delta^{\rm nf}r_{10}  \msp{-4}& \approx &\msp{-4} -\tilde \epsilon 
\:\rho_{pg}^{(\frac{1}{2}^+)}  \approx -0.07\GeV^5 \\
\nonumber
\delta^{\rm nf} r_{11} \!\approx\!\delta^{\rm nf} r_{12} 
\msp{-4}& \approx &\msp{-4} -\tilde \epsilon  
\left(\frac{1}{6}\rho_g^{(\frac{1}{2}^+)} 
-\frac{3}{40}\rho_f^{(\frac{3}{2}^+)} -\frac{1}{8}\rho_g^{(\frac{3}{2}^+)} 
+3 \rho^{(\frac{5}{2}^+)}
\right)\\
\nonumber
\delta^{\rm nf} r_{13} 
\msp{-4}& \approx &\msp{-4} -\tilde \epsilon 
\left(-\frac{2}{3}\rho_{pg}^{(\frac{1}{2}^+)} -\frac{1}{6}\rho_g^{(\frac{1}{2}^+)}
-\frac{3}{40}\rho_f^{(\frac{3}{2}^+)} -\frac{1}{2}\rho_{fg}^{(\frac{3}{2}^+)} 
+\frac{1}{8}\rho_g^{(\frac{3}{2}^+)}  +3 \rho^{(\frac{5}{2}^+)}
\right)\\
\delta^{\rm nf} r_{14} 
\msp{-4}& \approx &\msp{-4} -\tilde \epsilon 
\left(\frac{2}{3}\rho_{pg}^{(\frac{1}{2}^+)} -\frac{1}{6}\rho_g^{(\frac{1}{2}^+)}
-\frac{3}{40}\rho_f^{(\frac{3}{2}^+)} +\frac{1}{2}\rho_{fg}^{(\frac{3}{2}^+)} 
+\frac{1}{8}\rho_g^{(\frac{3}{2}^+)}  +3 \rho^{(\frac{5}{2}^+)}
\right)
\label{2420}
\eea
where the integration over $\omega$ is assumed similar to Eqs.~(\ref{2402}).
Taking $\tilde \epsilon \!\approx\!400\MeV$ the numerical estimates for the
relevant combinations of the expectation values are then as straightforward as
those for the $D\!=7\!$ operators.  
The nonfactorizable corrections to the typical 
non-suppressed expectation values are of the order of
$50$ to $100\%$.

Bearing in mind the dependence of all the contributions on a 
 few poorly known hadronic parameters we cannot regard the 
calculation of general nonfactorizable
effects accurate. They should be used primarily to assess the potential scale of
the corrections to factorization and to clarify the expected sign
pattern. 
The corrections associated with the $\frac{5}{2}^+$ states, for instance
in $\aver{\vec \pi^2 \vec\pi^2}$ or in $\aver{\vec \pi^2 \pi_0\vec\pi^2}$ are
largely unconstrained since only physics related to the  $\frac{1}{2}^+$
and  $\frac{3}{2}^+$ have been considered. 

The formalism employed in this subsection could be used together with the 
results of Sec.~\ref{pionloop} to find 
the nonfactorizable contributions due to non-resonant states composed of the 
ground-state multiplet and a pion. However, we expect these 
contributions to be relatively small, as they
come with higher powers of the excitation energy which is lower for the
soft pion continuum than for the principal resonances, and therefore we do not consider them here.

\subsubsection{Ground-state factorization and $N_c$}

The expectation values like 
$\aver{\bar Q \vec \sigma \!\cdot\! \vec {\boldmath B} \!\times\!  
\vec{\boldmath B}  Q}$ or 
$\aver{\bar Q \vec \sigma \!\cdot\! \vec E \!\times\! \vec E Q}$ 
are possible due to the
non-Abelian nature of QCD; in the 
factorization approximation they are proportional to $\frac{2}{3}(\mu_G^2)^2$ or to
$-\tilde\epsilon^2 \mu_G^2$. Such expectation values must vanish, on the
other hand, in the bound states of  Abelian theories like QED. In QCD proper one
can consider  similar time correlators for the usual magnetic and/or
electric fields; they would enter, for instance, the electromagnetic
corrections. The correlators can be decomposed into the same set
of invariant spectral densities with  different residues. 
For such Abelian fields certain expectation values must vanish
reflecting the commutativity of the different components of the Abelian field
strength. For instance, this is the case for the counterpart of the $\rho_A^3$
structure:
\beq
\int_{\omega>0} {\rm d} \omega\, \rho_A^{\rm Abel}=
\int_{\omega>0} {\rm d} \omega\,  
\left(\mbox{$\frac{2}{3}$} \rho^{(\frac{1}{2}^+)}_{g,\rm Abel} 
- \mbox{$\frac{1}{2}$}  \rho^{(\frac{3}{2}^+)}_{g,\rm Abel} \right) = 
-\frac{2(\mu_{G,\rm Abel}^2)^2}{3}
\label{2430}
\eeq
must hold for the Abelian analogies of the spectral densities and of the
$B$-meson local expectation values. 
This may be reminiscent of Eqs.~(\ref{3130}), but is more general.
Such a sum rule shows that  
the ground-state saturation itself may not be a universally applicable
approximation. Eq.~(\ref{2430}) may even be regarded as an indication of
the importance of the $\frac{3}{2}^+$ state. 

In actual QCD such local antisymmetric products of the guon field strength do
not need to vanish; all the considered correlators and the factorized pieces
count as constants in the large-$N_c$ limit. One may think that the
ground-state factorization approximation is generally representative at not
too small $N_c$. 

\subsubsection{Perturbative normalization point dependence}

Having at our disposal the perturbative heavy quark spectral
functions of Eqs.~(\ref{4114}) and (\ref{mu2}) we
can easily incorporate the leading powerlike mixing for the operators to
order $\alpha_s$ and to any BLM order using the preceding analysis. 
The corrections to the expectation values are obtained 
from Eqs.~(\ref{2402}),  (\ref{2472}),  etc.; the $B$-meson spectral
densities themselves are given by Eqs.~(\ref{3130}).
This gives the one-loop renormalization-scale evolution of the expectation
values, except for the scale-dependence of the
factorizable contributions themselves, like $(\mu_\pi^2(\mu))^2$, 
which in practice may be significant.

As a typical example, the one-loop and two-loop BLM piece in the combination
$M_2$ in Eqs.~(\ref{2462}) is 
\beq
M_2^{\rm pert}(\mu) = 
C_F\frac{\alpha_s(\bar M)}{\pi}\mu^4 \left[1 + 
\frac{\beta_0\alpha_s}{2\pi}\left(\ln{\frac{\bar M}{2\mu}} + \frac{29}{12}\right)
\right], 
\label{2458}
\eeq
with $\bar M$ denoting the
normalization scale for $\alpha_s$ (in the $\overline{\rm MS}$ scheme). It is
obtained using Eqs.~(\ref{4128}).
The first-order correction to $M_1$ vanishes and the second-order term is
negative (although quite suppressed).\footnote{This shows a typical problem of
applying  naive  non-Abelization to cases where the order-$\alpha_s$ effect is absent; 
the full ${\cal O}(\alpha_s^2)$ contribution to $\rho^2_{\pi\pi}$ is, of
course positive, paralleling the similar term in the Abelian theory.}
Of course, the perturbative calculation is meaningful 
only for not too low  values of $\mu$.
At a numeric value of the strong coupling such estimates can be used to 
gauge the relative importance of the scale-dependence effects. 
The one-loop term with fixed $\alpha_s\!=\!0.3$ would yield 
$$
M_2^{\rm pert}(0.7\GeV) \approx 0.03\GeV^4,
$$
and is relatively small for $\mu$ between $0.7$ and $1\GeV$. 

To extend this to the complete set of $D\!=\!8$
operators including those in  Eqs.~(\ref{2420}) an additional class 
of the spectral functions would be needed where one of the pair products $\pi_k
\pi_l$ is replaced by $\pi_k \pi_0\pi_l$, cf.\ Eq.~(\ref{2482}).
The one-loop answer for it is simple, because the transition 
amplitude for such operators into a state with a single extra gluon amounts to 
$-\omega$ times the amplitude for the corresponding operator
without extra $\pi_0$:
$$
\matel{Qg}{\bar{Q}\pi_k \pi_0\pi_l Q}{Q}= -k_0\matel{Qg}{\bar{Q}\pi_k \pi_l
  Q}{Q} + {\cal O}(g_s^3).
$$
In other words, $\tilde
\epsilon\!=\!\omega$ in terms of Eqs.~(\ref{2420}) to this accuracy. This 
exactly parallels the result for the $P$-wave SV operators. In this way both
${\cal O}(\alpha_s)$ and the higher BLM corrections to the mixing are readily
obtained for the $D\!=\!8$ operators alongside the power mixing for $D\!=\!7$.

\section{Discussion}
\label{disc}

\subsection{On a resummation of the $1/m_c^k$ corrections} 

The accuracy in the estimate of ${\cal F}(1)$ is limited, in particular, by
significant higher-order power corrections in $1/m_c$. We
mention here a possibility to consider all these potentially dangerous 
corrections together and, therefore, in a certain sense, to resum them. 
The price to pay
is the appearance of a limited number of new hadronic expectation
values. 

The idea is to apply the OPE  to the reversed zero-recoil transition  $D^* \tto
B$ instead of $B \tto D^*$. At first glance this amounts only to exchanging  $c$
and $b$ in Eq.~(\ref{80}) and taking the expectation values over the vector
rather than pseudoscalar state.\footnote{The fact that $D^*$ is not a stable particle
is not an issue since the
width of $D^*$ is extremely small. To completely bypass the complication 
one may simply
assume that the pion mass is a few MeV larger than it is in reality; the
properly defined $B \tto D^*$ formfactor may not depend on this.}
However, the axial $\bar{c}b$ current produces not
only $B$ out of $D^*$, but also $B^*$. Therefore a scattering amplitude of 
two arbitrary spatial components
$j,l$ should instead be considered, and the corresponding indices contracted with
the polarizations of $D^*$,
\beq
\tilde T^{\rm zr}(q_0)=\int\! {\rm d}^3x
\int\! {\rm d}x_0\; e^{-iq_0x_0}
\frac{1}{2M_{D^*}} \matel{D^*_j}{\mbox{$\frac{1}{3}$}\:{ iT}\,
  \bar{c}\gamma_j\!\gamma_5 b(x)\:\bar{b}\gamma_k\!\gamma_5 c(0)}{D^*_k}\,.
\label{780}
\eeq 

The analogue of 
Eq.~(\ref{88}) will still contain terms  with powers of $1/m_c$. However,
they only come from the nonrelativistic  expansion of the full-QCD charm quark 
operators $\bar{c}O_k c$, starting with the leading $\bar{c}c$. Consequently, 
the full set
of the $1/m_c^k$ corrections originate from the heavy charm expansion 
of the finite-$m_c$ expectation
value over actual $D^*$ states, $\matel{D^*}{\bar c c}{D^*}$. The latter is
a physical quantity and can in principle be measured on the lattice. 
Similarly, for a
given power $l$ of $1/m_b$ all the terms $1/m_b^l 1/m_c^k$ come from the the
heavy charm expansion  of the finite-$m_c$ expectation
value of the corresponding $\bar{c}c$ operator with $l$ derivatives.
Likewise, the $1/m_c$ power effects in the inelastic
transition amplitudes combine to yield directly the non-diagonal transition
probabilities for the finite-$m_c$ charm states. The nontrivial explicit OPE 
corrections to the corresponding sum rule may only depend
on powers of $1/m_b$ since they come from the dynamic expansion of the
intermediate-quark propagator. 
For what concerns  the perturbative corrections, they are the same
as in the sum rule Eq.~(\ref{80}), modulo the 
$m_c \!\leftrightarrow \! m_b$ replacement. 

An attractive element of such an approach is that the $m_c$-dependence 
of these matrix elements must be regular in the whole  $m_c$ range
and it should be possible to construct accurate interpolating functions,
which  would replace a large number of
different coefficients appearing in the expansion in $1/m_c$.  Theory-wise, the
regularity puts constraints on the higher-order behavior of the $1/m_c$ series
and  suggests that the series is sign-alternating, implying numerical
cancellations between successive orders for actual charm mass.

The series in $1/m_b$ is not  resummed in this approach. As was discussed in
Sect.~\ref{powercor}, numerically these corrections can be discarded to a good
approximation, and retaining the known 
leading $1/m_b^2$ terms for them would be sufficient for all practical purposes.  

Guided by the quantum-mechanical interpretation of the sum rules it is not 
difficult to verify that the leading, $1/m_Q^2$
corrections are identical to those in the direct approach, separately for the local
OPE piece and for the inelastic contribution. Therefore the actual difference between the 
expansions  would appear when the higher-order power corrections are addressed.

\subsection{Vector formfactor in $B\tto D$ transitions}

The present study focused on the $B\tto D^*$ decay mediated by the axial
current. A similar analysis may be applied to the $B\tto D$ transitions where
only the vector current contributes. Two different aspects can be addressed
here.

The direct zero-recoil vector-current analogue $F_D$ of ${\cal F}_{D^*}$ 
is related to the matrix element
$\matel{D}{\bar{c}\gamma_0 b}{B}$ and does not determine the semileptonic decay
rate near zero recoil for massless leptons, unlike  $B\tto D^*$. 
It can be measured in the decay $B\tto D \,\tau \nu_\tau$ whose amplitude is 
proportional to $m_\tau$ at $\vec q\!=\!0$, and this may represent
an interesting opportunity  for a new generation Super-B facility. 

The more conventional decays $B\tto D \,\ell \nu$ with nearly massless leptons
are $P$-wave at small recoil and are more difficult to measure in this 
corner of the phase space. 
 Of the two general vector formfactors $f_+(q^2)$ and $f_-(q^2)$
the latter does not contribute for massless leptons; 
the former, on the other
hand, depends on both the time and the spatial components of the 
current (see,  e.g.\ Ref.~\cite{BPS}). The
spatial component assumes a change of the heavy meson velocity, 
is not related to a conserved Noether current in the heavy
quark limit,  and generally suffers from  linear power corrections
${\cal O}(1/m_Q)$. This fact fed, for a long time, a theoretical prejudice against
the precision evaluation of the corresponding formfactor
$$
{\cal F}_+ \equiv \frac{2\sqrt{M_B M_D}}{M_B +M_D} f_+((M_B\!-\!M_D)^2).
$$
It was nevertheless argued later \cite{BPS}, based on the BPS expansion, that in
fact the power corrections in ${\cal F}_+$ are smaller, and may even enjoy a better
numeric control than in ${\cal F}_{D^*}$:
\beq
{\cal F}_+=1.04 \pm 0.01_{\rm pert} \pm 0.01_{\rm power}.
\label{790}
\eeq

The analysis developed in the present paper can be applied to both
formfactors, $F_D$ and ${\cal F}_+$. For $F_D$ the required 
modifications are minimal. The case of ${\cal F}_+$ is somewhat different both
technically and conceptually; in particular, the physical interpretation is
different, and there is no simple probabilistic interpretation that would mean
positivity already for the $1/m_Q^2$ corrections. The positivity holds for the
leading $1/m_Q$ power correction, however it can simply be regarded as known
within the required numeric precision.

The analysis of the  perturbative corrections follows closely that of
Sect.~\ref{pert} in the case of $F_D$, with corrections generally 
smaller.  This kind of analysis cannot be directly applied to ${\cal F}_+$ due
to subtleties  at nonvanishing recoil; the treatment is more complicated here
and has to be analyzed ad hoc \cite{BPS}.

For what concerns the power corrections, we do not expect improvements
in either cases. 
The main reason is that the power corrections (in the kinetic
scheme we consistently use) are numerically small to start with, since they all
vanish in the exact BPS limit. Moreover, to any order in $1/m_Q$ the terms are
of the second order in the deviation from the BPS limit \cite{BPS}, an
analogue of the Ademollo-Gatto theorem \cite{ademgatto} which applies to the
BPS expansion for both $F_D$ and  ${\cal F}_+$. 
As a consequence, any numeric result
strongly depends on the degree of proximity of the actual-QCD dynamics in
$B$ mesons to the BPS limit, for instance,  on the excess of
$\mu_\pi^2$ over $\mu_G^2$. 
 This appears as a significant cancellation between terms 
belonging to different spin structures. It is therefore
difficult to expect an increase in defendable accuracy in $F_D$
and  ${\cal F}_+$
unless the numerical aspects of the BPS breaking are experimentally
scrutinized. Once this aspect of the strong dynamics is studied, we would
have more theory constraints to improve the accuracy of the nonperturbative
predictions, in particular if the BPS regime turns out a good 
starting approximation.

\subsection{Lattice determination of ${\cal F}_{D^*}$}

The recent PDG policy has been to rely solely on the lattice evaluation of
${\cal F}_{D^*}$ for the exclusive extraction of $|V_{cb}|$ from the $B\tto
D^*\ell \nu$ differential rate extrapolated to the no-recoil kinematics. The
lattice values for ${\cal F}_{D^*}(1)$ have always been on the higher side,
well above $0.9$ and carried small error bars, in particular since unquenched
simulations were first employed \cite{laiho}: 
\beq 
{\cal F}_{D^*}(1)=0.924\pm 0.012 \pm 0.019.
\label{796}
\eeq
An update of this result was  presented by the FNAL-MILC collaboration
\cite{FNALnew} after our first publication \cite{f0short}. It has  
a lower central value,\footnote{The last FNAL paper included the electroweak
decay enhancement factor $1.007$ into $F$; we have removed it  in 
Eq.~(\ref{796a}). We thank A.~Kronfeld for the communication.} 
\beq 
{\cal F}_{D^*}(1)=0.902 \pm 0.005\pm0.016
\label{796a}
\eeq
and is closer to our number. 
The other recent lattice result is based on a quenched simulation by the 
Tor Vergata 
group \cite{Rome-TOVstar}, ${\cal F}_{D^*}(1)=0.924\pm 0.008 \pm 0.005$
and has an even smaller nominal error.
 
Confronting our evaluation of ${\cal F}_{D^*}(1) $ with the 
lattice ones, we should first emphasize that the latter are not 
direct calculations of this formfactor. The lattice theory
with heavy quarks and continuum QCD are two different theories, and there is
no limit at $m_Q \,a \!\sim \!1$ ($a$ is the lattice spacing) where they 
would coincide nonperturbatively. Present simulations do not reach values 
of $m_c a$ below  about $0.3$-$0.4$.
The fact that the two theories share the same heavy quark symmetry was
emphasized as the key point behind the approach pursued by 
the FNAL group \cite{FNAL1}.
However, it is the power-suppressed deviations from the symmetry that 
matter in this case, and in principle they are  different.

This becomes transparent if one takes a closer  look at the $1/m_Q$
corrections. In QCD, as a consequence of Lorentz symmetry,  the mass entering
the nonrelativistic kinetic energy $\vec p^{\,2}/2m_Q$ is the same rest-energy
mass $m_Q$ controlling also the power-suppressed terms in the currents (from
  equations of motion and from the Foldy-Wouthuysen transformation), the 
role of the correlators of the subleading operators, etc. No symmetry,
however, enforces their equality on the lattice, and all these 
terms are driven by different effective masses. 

The FNAL lattice approach  
appreciates this complication \cite{FNAL2}.
To handle it the heavy quark sector is modified by adding {\it ad hoc}
power-suppressed terms  allowing to change the corresponding effective
masses; in turn this changes the power corrections.  The outcome for the $1/m_Q^k$
effects in ${\cal F}(1)$ is then determined by the {\it ad hoc} constants
which have to be specified through a matching. 
However, such a matching has only been performed at tree level. 
 Furthermore, the matching was performed only to the leading power effects,
corresponding to the $1/m_Q^2$ corrections in ${\cal F}(1)$.  All $1/m_Q^3$
and higher effects are therefore not under control,\footnote{Earlier FNAL
evaluations \cite{FNALold} claimed to extract the principal $1/m_Q^3$ terms, 
however they were determined in an effective theory essentially different 
from the actual QCD. The later analyses considered only $1/m_Q^2$
corrections.}  although it can be argued that these discretization effects are somewhat
 suppressed for charm, at low $a$.

In view of these practical limitations the FNAL approach cannot be regarded a
first-principle evaluation of the zero-recoil $B\tto D^*$ formfactor in QCD.
Some of the related potential biases are included in the error budget detailed
in the publications. However, the error assignment may not be realistic. 
In fact,  the recent value in Eq.~(\ref{796a}) has a reduced discrepancy with the estimate
Eq.~(\ref{c24});  a more conservative treatment of the systematic errors 
would make it compatible with the central value of the present analysis. 

We also note that, comparing the earlier and the 
more recent FNAL lattice simulations,
 the group did not find a noticeable effect of the light quark unquenching. 
This differs from the estimated size of the chiral loop contributions discussed 
in Sect.~\ref{sectdpi} related to the nonresonant states with light dynamic
pions, although there may be no formal contradiction. 

The above reservations apply to the lattice determination of the
vector $B\tto D$ formfactor ${\cal F}_+$ as well. In this case larger corrections
to the symmetry limit were  found \cite{fpluslattice} 
$$
{\cal F}_+=1.074 \pm 0.018 \pm 0.015
$$
compared to the theoretical prediction in  Eq.~(\ref{790}). Here, however, 
the disagreement is less
significant than for ${\cal F}_{D^*}$.\footnote{More recently, the Tor Vergata
group reported the value ${\cal F}_+=1.026 \pm 0.017$, 
based on  quenched simulations \cite{Rome-TOV} and 
a new preliminary value ${\cal F}_+=1.058 \pm 0.009_{stat}$ 
\cite{Bailey:2012rr}
has been presented, where the systematic uncertainty still needs to be evaluated.
}

\section{Conclusions}
\label{conc}

The present study has been motivated by the need for an updated
evaluation of the phenomenologically important $B\tto D^*$ semileptonic
transition formfactor near the zero-recoil point, ${\cal F}(1)$, that could  account for the
latest progress in  heavy quark theory. The main numeric outcome has been
reported in  \cite{f0short} and the details have been given here. 

Numerically we conclude that the unitarity upper bound for the formfactor is 
\beq
{\cal F}(1) < 0.92
\label{c20}
\eeq
assuming only positivity of the inelastic contributions. Including the soft
$D^{(*)}\pi$ continuum the bound becomes
\beq
{\cal F}(1) < 0.90\,.
\label{c21}
\eeq
These numbers  refer to values of $\mu_\pi^2$ close to its lower bound. The
bounds on ${\cal F}(1)$  become stronger if the actual  
$\mu_\pi^2$ value is larger, which is
more natural on theory grounds. Our analysis of the inelastic transitions
into the low-lying channels incorporating the constraints following from the
observed amount of the hyperfine splitting in $B$ and $D$ mesons allows us to
go beyond the unitarity upper bound and to obtain an estimate. The actual
value for ${\cal F}(1)$ comes out about
\beq
{\cal F}(1) \approx 0.86
\label{c24}
\eeq
at  low values of $\mu_\pi^2$; it somewhat decreases at larger $\mu_\pi^2$
and/or $\rho_D^3$.

The quoted numbers for the unitarity bounds Eqs.~(\ref{c20}), (\ref{c21})  carry
a theoretical uncertainty of about $1\%$. The estimated central value has a 
larger uncertainty, of around $2\%$. 
We quote here the value literally obtained in our estimate (for low $\mu_\pi^2$, 
$\rho_D^3$);  it is not implied that this value peaks the 
expectation probability. 

Thus, ${\cal F}(1)$ in excess of $0.9$ would be consistent with unitarity and the
short-distance expansion of the QCD amplitude only under 
contrived assumptions. Values larger than $0.92$ should be viewed in violation
of unitarity assuming that the conventional short-distance expansion in QCD
works in the case of the zero-recoil scattering amplitude off heavy quarks. 

In earlier analyses of 
the power corrections to ${\cal F}(1)$ the ``wavefunction
overlap'' effect used to be uncertain and, essentially was only parameterized; in
the language of the heavy quark sum rules it was 
\beq
I_{\rm inel}=\chi \cdot \Delta
\label{c30}
\eeq
with $\Delta$ the power corrections in the sum rule, cf.\ Eq.~(\ref{88}), setting the scale of the
nonperturbative effects in ${\cal F}(1)$. It was simply guessed following 
Refs.~\cite{vcb, optical} that $\chi$ is somewhere between $0$ and $1$ leading
to an assumption $\chi\!=\! 0.5 \! \pm \! 0.5$. On the
other hand we have linked $I_{\rm inel}$ to measured hadronic parameters in the
heavy mesons and found $\chi$ to be large, 
$$
\chi \approx 1.3\div 1.7.
$$
This is the main factor driving the prediction for ${\cal F}(1)$ down compared to
earlier estimates, along with a shift due to the higher-order power corrections.

Since our conclusion appears in some  conflict with the lattice results for ${\cal F}(1)$, 
one may examine how robust this conclusion is. We found that the phenomenology
of the heavy mesons suggests that the inelastic contributions, as well as the
$1/m_Q^3$ and higher-order power corrections are numerically significant and
that they lower the expected value for ${\cal F}(1)$. Our derivations, of
course, use the power expansion for charm, and this may be a vulnerable point
for a precision prediction.

We emphasize, however, that a scenario with smaller corrections from the
nonlocal correlators or high orders would not be consistent. If the 
neglected higher-order effects in charm are  relatively small, our analysis of
the hyperfine splitting is reliable. 
If, on the contrary, the higher-order corrections in charm are too significant
to affect the credibility of the  analysis, this would imply a higher overall
mass scale of nonperturbative QCD in heavy quarks.  Then there would be no
reason to expect small corrections to ${\cal F}(1)$ either.

Our numeric  predictions involved a number of theoretical improvements.
The first to mention
concerned the calculation of the Wilsonian perturbative renormalization factor
with a hard cutoff. We derived a general ansatz applicable to the one-loop
level as well as to arbitrary BLM order, 
which yields the
perturbative correction in the kinetic scheme with a full dependence on
$\mu/m_Q$. It appears to be different from the naive prescription for having a
cut on the gluon momentum. For the zero-recoil transitions the difference
emerges starting with the terms ${\cal O}(1/m_Q^3)$. Numerically it 
is important whenever the hard scale is determined by the charm mass.

The other direction is the treatment of the inelastic contributions, along
with a detailed analysis of the continuum soft-pion states in the
context of the heavy quark expansion.

We have presented a novel model-independent analysis of the transitions into
the radially excited (or $D$-wave) states near the rest kinematics. So far the
excited states considered were mainly the $P$-wave states. The important new
phenomenological constraint  comes from the hyperfine splitting in $B$ and
$D$: qualitatively, the latter tells us that the $D\!=\!3$ zero-momentum
nonlocal correlators are numerically large  in actual QCD, and this enhances
the predicted size of the inelastic probabilities (the `overlap deficit' in
the formfactor) over the naive expectations.
The  analysis of the spin-averaged $B$ and $D$ meson mass difference 
supports the same conclusion, albeit with larger uncertainty.
Using the hyperfine splitting, one can then set a lower bound 
on  the inelastic contribution.
The bound  is very close to the value $I_{\rm inel}$ assumes in
the BPS limit where a single
combination of the four general correlators determines both $I_{\rm inel}$ and
the $1/m_Q$ dependence of the hyperfine splitting.

A related implication of the hyperfine splitting analysis is the enhancement of 
the transition amplitudes into the excited `radial' charm states which
must be dominated by the $1/m_c$-suppressed terms rather than by the
velocity-dependent component. This leads to an increased yield of the wide
charm hadronic structures from the decays of the `radial' states, and would
eliminate the `$\frac{1}{2}\!>\!\frac{3}{2}$' puzzle if the observed wide
yields are dominated by them rather than by the $\frac{1}{2}$ $P$-waves
predicted to be suppressed.  We emphasize that the $D$-wave states must be
produced along with true radially excited mesons, and may even
dominate. Some possibilities to distinguish them
in experiment were discussed. 

Therefore, we have identified a link among  three apparently unrelated
physics points: the size of the hyperfine splitting in charm
and beauty, the reduction in ${\cal F}(1)$ through the enhanced nonlocal power
corrections, and the resolution of the `$\frac{1}{2}\!>\,\frac{3}{2}$' puzzle
in the semileptonic $B$ decays. 

On the theoretical side, yet another implication is manifest: we find
significant corrections to the factorization  for higher-dimension heavy-quark
expectation values. Our model-independent analysis provides the basis
for a dedicated account of the nonfactorizable effects from higher orders in
$1/m_Q$ \cite{hiord}, in particular for the semileptonic $B$-decay fits. 

With respect to the $D^{(*)} \pi$ states, we have expanded the treatment in a
few aspects. Our approach allowed us to account for their effect in the
analysis of ${\cal F}(1)$ beyond the leading order $1/m_Q^2$, relying instead
on the soft-pion approximation. The subleading 
terms turned out significant. In particular, the heavy quark symmetry-breaking
corrections in the heavy meson-to-pion couplings seem to yield the dominant
effect. 

Theoretically, we have presented a  consistent treatment of 
the $D^{(*)} \pi$ states in the heavy quark approximation within
the soft-pion approach. The decomposition of the $P$-wave into the
$\frac{1}{2}$- and the $\frac{3}{2}$-components has been addressed before yet
remained largely unknown. We briefly recapitulated it in Sect.~\ref{pionloop}
and extended it to radial/$D$-wave states, with the decomposition
into the corresponding $\frac{1}{2}^+$-, $\frac{3}{2}^+$- and 
$\frac{5}{2}^+$-channels. This made it explicit 
that earlier studies of  
 the soft-pion corrections were incomplete: they accounted only for 
the most singular effect at
the softest pion momentum, however not dominant in the typical configuration
with $|\vec k_\pi|\!\sim \! \mhad$. Typically, the pion loops 
yield negligible contributions, with the notable exception of 
the contribution to the axial sum rule for ${\cal F}(1)$. 

We conclude with the following remark. 
The pattern of the hyperfine splitting in $B$ and $D$ mesons, in particular its
precise mass dependence, is important for a few different
phenomena and draws novel qualitative conclusions for our understanding 
of the  heavy meson states. 
The precision interpretation of the splitting, on the other hand, 
may potentially be hindered by higher-order effects in charm.
A sufficiently accurate measurement at
a different heavy quark mass would radically improve the credibility of the
hyperfine analysis. We have pointed out that a  first principle
lattice determination of the hyperfine splitting at better than $5\%$
accuracy, if possible, can provide
this information, and have discussed how it can be used for mesons either
heavier or lighter than charm.

\section*{Acknowledgements} 

We are grateful to S.~Turczyk for computing the higher-order power
corrections for the axial-vector decays during work on Ref.~\cite{hiord}.
It is our pleasure to thank Ikaros Bigi and Alex Khodjamirian for
numerous discussions. 
The work of P.G.\ is supported in part by the Italian Ministry of
Research (MIUR) under contract  2008H8F9RA$\_$002.
The work was supported by the German research foundation DFG under
contract MA1187/10-1 and by the German Ministry of Research (BMBF), contracts
05H09PSF; it enjoyed a partial support from the NSF grant PHY-0807959 and
from the grant RSGSS 4801.2012.2.

\section{Appendices}

\setcounter{equation}{0}
\renewcommand{\theequation}{A.\arabic{equation}}
\renewcommand{\thetable}{\Alph{table}}
\renewcommand{\thesubsection}{\Alph{subsection}}
\setcounter{section}{0}
\setcounter{table}{0}

\subsection{BLM corrections and summation}
\label{blmstuff}

The technique that allows the computation and summation of BLM corrections 
of order $\beta_0^{n} \as^{n+1}$  
has been concisely reviewed in Ref.~\cite{blmvcb}; 
Ref.~\cite{imprec} focussed upon its application to  Wilsonian OPE calculations.
To BLM-dress a one-loop result one needs to evaluate the generic
one-loop correction $A_1$
$$
{\cal A} = 1+\frac{\alpha_s}{\pi}A_1 + ...
$$
with a fictitious gluon mass $\lambda$, $A_1(\lambda^2)$, so that the conventional
$A_1$ is $A_1(0)$. The BLM series in terms of the $\overline{\rm MS}$ coupling
$\alpha_s$ normalized at the arbitrary scale $M$ takes  the form
\bea
\nonumber
A^{\rm BLM}\msp{-3}& =& \msp{-3}1\;+\; A_1(0)\frac{\alpha_s(M)}{\pi}
\;+\;
\sum_{n=0}^{\infty}\: 
\mbox{$\frac{4}{\beta_0}
\left(\frac{\beta_0\alpha_s(M)}{4\pi}\right)^{n+2}$} \times
\;\\
& & \msp{-3}
\sum_{k=0}^{\frac{n}{2}}\, 
(-\pi^2)^k \;\mbox{\large$\raisebox{-.5mm}{$C$}_{_{n+1}}^{^{2k+1}}$}
\cdot\int
\!\frac{{\rm d}\lambda^2}{\lambda^2}\:\left[\ln{\frac{M^2}{\lambda^2}}
\!+\!\frac{5}{3}\right]^{n-2k}
\!\mbox{$\left(A_1(0)\mbox{\large$\frac{M^2}{M^2+ 
e^{\mbox{\tiny -5/3}}\lambda^2}$}\!-\! 
A_1(\lambda^2) \right)$}\,
,\qquad\;\;\;\;
\label{blmseries}
\eea
with $C$ denoting the binomial coefficients. Let us remind that the integral
over $\lambda^2$ here typically has two (or more) domains. The last term
depending on $A_1(\lambda^2)$ is integrated from $0$ to the threshold value of
the gluon mass if there is a threshold (in particular, it would be 
set by $\mu$ in the Wilsonian calculations),
whereas the first term in the same brackets should always be integrated over
all values of $\lambda^2$ regardless of a cutoff or of kinematic details.

When the infrared part is removed from the one-loop diagram to leave a
genuinely short-distance correction, the corresponding $A_1(\lambda^2)$ is a 
real analytic function in the vicinity of zero. This allows to write an
integral representation for the resummed series,
\bea
\nonumber
A^{\rm BLM}\msp{-3}& =& \msp{-3} 1+
A_1(0)\frac{\alpha_s(M)}{\pi}\\
\nonumber
& & \msp{-.7}+
\int_{-\infty}^{\infty} \; {\rm d}t\:  
\frac{\frac{\beta_0}{4}\,\left(\frac{\alpha_s}{\pi}\right)^2}{\left(1+ 
\frac{\beta_0\alpha_s}{4\pi}(t\!-\!\frac{5}{3})\right)^2+
\left(\frac{\beta_0}{4}\alpha_s\right)^2} 
\:\left( A_1(0)\frac{1}{1+e^{t-\frac{5}{3}}}\!-\! A_1(e^{t}M^2)\right)
\\
& & \msp{-.7}
- \frac{4}{\beta_0}\left[\frac{M^2}{M^2\!-\!\Lambda_{\rm QCD}^2}A_1(0)
-A_1(-\Lambda_V^2)
\right],
\label{blmresum}
\eea
without any ambiguity associated with the last term; here  
\beq
\Lambda_{\rm QCD}^2 = M^2 \:e^{-\frac{4\pi}{\beta_0\alpha_s(M)}}, \qquad
\Lambda_V^2= e^{\frac{5}{3}} \Lambda_{\rm QCD}^2 =
M^2 \:e^{-\frac{4\pi}{\beta_0\alpha_s(M)}+\frac{5}{3}}\;.
\label{lambdav}
\eeq

\subsection{One-loop perturbative calculation with a Wilsonian cutoff}
\label{oneloopwils}

In this Appendix we discuss  aspects of the one-loop calculation of the
leading Wilson coefficient  $\xi_A$. The reasoning is quite
general and is applicable to other observables as well. As in 
Sect.~\ref{pert}, we need to distinguish between
$\varepsilon_M$ and Wilsonian $\mu$, therefore we will deal explicitly 
with $\xi_A^{\rm  pert}(\varepsilon_M,\mu)$ as it appears in Eq.~(\ref{88}). 
The dependence of $\xi_A^{\rm  pert}(\varepsilon_M,\mu)$ on $\varepsilon_M$ is
given by Eq.~(\ref{51a.4}). 
Since the right-hand side of Eq.~(\ref{88}) must be $\mu$-independent, and 
$\xi_A^{\rm pert}$ must satisfy the `boundary condition'  
Eq.~(\ref{51a.8}),  one way to determine $\xi_A^{\rm  pert}(\varepsilon_M,\mu)$ 
for arbitrary values of $\mu$ and (in principle) any order of perturbation 
theory is to  use the $\mu$-dependence of the matrix elements and masses
appearing in Eq.~(\ref{88}).

Indeed, in perturbation theory 
the expectation values $\matel{B}{O_k}{B}_\mu$ in Eq.~(\ref{88}) typically
depend on $\mu$ in a powerlike way, 
\beq
\frac{{\rm d} \matel{B}{O_k}{B}_\mu}{{\rm d} \mu} \propto 
\alpha_s\, \mu^{d_k-4},
\label{51.012}
\eeq
where $d_k$ is the dimension  of $O_k$; the same applies to
the heavy quark masses on which the Wilson coefficients generally 
depend. In this way  the $\varepsilon_M$-dependence of 
$\xi_A^{\rm  pert}(\varepsilon_M,\mu)$ is calculated explicitly while 
its $\mu$-dependence 
emerges as an expansion in $\mu/m_Q$ which is necessarily truncated after a few terms. 
This is not a serious limitation if the hard OPE scale is $O(m_b)$, but 
in our case  the charm mass sets the lower hard scale in the
problem and the expansion in $\mu/m_Q$ shows poor convergence.

To overcome this drawback we have devised a method allowing  to directly compute the
leading Wilson coefficient $\xi_A^{\rm pert}(\varepsilon_M,\mu)$ as a
function of $\mu/m_Q$, without an expansion, 
in the one-loop approximation. 
The method  is readily generalized to  higher-order BLM corrections. 
We have described its main points in Sect.~\ref{pert}; below we
provide  additional explanations. 

The idea behind the  approach is that in one-loop calculations there is a
simple connection between the normalization point of the heavy quark operators
in the kinetic scheme and the hard cutoff on the gluon momentum $k$ in
the diagram. In order to preserve 
the analiticity and the unitarity of the Feynman integrals no limit  on integrations
over $k_0$ of the gluon is imposed in the kinetic scheme: the separation of
scales is performed based on $|\vec{k}|$. 
In this way one computes the one-loop $\xi_A^{\rm pert}$ introducing an infrared cutoff on
$|\vec{k}|$, instead of taking the full integral ${\rm d}^4 k$.
In the $b$-quark static limit the step-function cutoff factor 
\beq
\theta(|\vec{k}|-\mu)
\label{24}
\eeq
in the Feynman integrand yields precisely the normalization 
at the scale $\mu$.
The remaining part, the integral with $\theta(\mu\!-\!|\vec{k}|)$ constitutes 
the power-suppressed terms described by higher-dimension matrix elements in the OPE.
In order to go beyond the static approximation
certain modifications of the cutoff  in Eq.~(\ref{24}) are required. 

Let us remind why $\xi_A$ is related to $\eta_A^2$ calculated with a
cutoff. The reasoning is based on 
considering the OPE relations in an ensemble of gluons with the
spatial momentum  limited by
$\mu$ in the $b$ rest frame; the non-Abelian nature does not play a
role at one-loop level. All these gluons can be considered soft, and
they satisfy the OPE sum rule where the coefficients assume the
tree level values.  Having in mind how the sum rules are derived (for pedagogical reviews see
Refs.~\cite{rev,ioffe}), the integration in the sum rule must run
over all excitation energies, from $0$ to $\infty$. However, 
in the soft gluon ensemble with  $|\vec k|<\mu$ 
the spectral density vanishes above the excitation energy
\beq
\varepsilon_M(\mu) = \mu+\sqrt{m_c^2\!+\!\mu^2}\!-\!m_c \,;
\label{24.10}
\eeq
below $\varepsilon_M(\mu)$ it is the usual one-loop spectral density
of QCD. At the same time, the operator expectation values in such an
ensemble are just the one-loop QCD expectation values in the kinetic
scheme normalized at $\mu$.

We now turn to the subtleties beyond the static limit, and  
focus on the soft contribution $\delta\eta_A^{\rm soft}$ 
to be subtracted from the one-loop $\eta_A$:  
\beq
\eta_A \longrightarrow \eta_A-\delta\eta_A^{\rm soft} = \eta_A - C_F g_s^2 
\int \frac{{\rm d}^4 k}{(2\pi)^4 i} \:\theta(\mu\!-\!|\vec{k}|) \dots
\label{24.04}
\eeq
Here the ellipses denote the same propagators and vertices 
encountered in the calculation of $\eta_A$ itself; $k$ is the gluon momentum
in the diagram. There are three one-loop 
diagrams -- the vertex correction and the wavefunction renormalization for
both $b$ and $c$ quark. 

Let us consider the vertex diagram of Fig.~\ref{diagr}a as an example.
The integrand has a general structure 
\beq
\int \frac{{\rm d}^4k}{(2\pi)^4i}\: 
\frac{{\rm numerator}}{(k_0^2\!-\! \vec{k}^{\,2}\!+\!i0) 
(m_b^2\!-\!(m_b\!-\!k_0)^2\!+\! \vec{k}^{\,2}\!-\!i0)
(m_c^2\!-\!(m_c\!-\!k_0)^2\!+\! \vec{k}^{\,2}\!-\!i0) } \, .
\label{36}
\eeq
At  given $\vec{k}$ the integral over $k_0$ is convergent and is
saturated at $|k_0| \!\sim\! |\vec{k}|$; the tails at large  $|k_0| \!\gsim\!
m_Q$ contribute a power-suppressed piece. 
Since no cut on $k_0$ is allowed, the integration over $k_0$ can be performed
by closing the integration contour in the lower half-plane, see Fig.~\ref{diagr}b. 
There are three pairs of poles in the $k_0$ plane,
\beq
k_0\!=\!\pm |\vec {k}|, \qquad k_0\!=\!m_b \!\pm\! \sqrt{m_b^2+\vec{k}^{\,2}}, 
\qquad k_0\!=\!m_c \!\pm \!\sqrt{m_c^2+\vec{k}^{\,2}}\,.
\label{38}
\eeq
With the standard Feynman prescription,  blue contour $\alpha$,  
the integration over $k_0$ results in the
sum of the three residues corresponding to the above poles. For the first
pole we have $k_0\!<\!\mu$; however for the two other,  distant,  poles  we have 
$k_0\!\gsim\! m_Q\!\gg\!\mu$,
regardless of the cutoff. Since the OPE generally corresponds to an expansion
in all components of the gluon four-momentum, it is clear that the
contributions to the integral associated with the distant (black) poles may not 
correctly describe the OPE power-suppressed terms. We recall that our
goal is just to subtract the piece of the one-gluon loop correction to
$\eta_A$ associated with the terms which have already been included 
in the  power-suppressed OPE. 

Indeed, it turns out that the OPE series for the soft piece correspond to the residue
 of only the `near'  pole at $k_0\!=\!|\vec {k}|$, 
while the two other resides should  be discarded. 
This means changing the bypass prescription for the
two distant poles, $-i0 \tto +i0$, which
moves the $k_0$ integration contour as shown by the green dashed line in
Fig.~\ref{diagr}b.

The $1/m_Q$ expansion in the Feynman diagrams leading to the OPE series
is essentially the Taylor expansion of the heavy quark propagators in the integrand 
for small  gluon four-momentum $k$. Eq.~(\ref{36}) then becomes
\beq 
\int \! \frac{{\rm d}^4 k}{(2\pi)^4 i} \;
\frac{\raisebox{-2pt}{\hspace*{-1pt}\scalebox{.84}{{\sf numerator}}}}{(k_0^2\!-\!\vec  k^2 \!+\!i0)} \,
\sum_{n=0} \frac{(k^2)^n}{(2m_b k_0\!-\!i0)^{n+1}}
\, \sum_{m=0} \frac{(k^2)^m}{(2m_c k_0\!-\!i0)^{m+1}}\, ;
\label{taylor}
\eeq
the poles at $k_0\!=\!0\!+\!i0$ are descendants of the nonrelativistic poles located
on the left of $k_0\!=\!0$.  To reproduce the nonrelativistic expansion
we calculate the integral for each term closing the contour in the lower
$k_0$ half-plane and picking up only the gluon pole  $k_0\!=\!|\vec
k|$. This cannot be done for the original integral in Eq.~(\ref{36}):
its value is not equal to the sum of the series, because large $k_0$
beyond the convergence radius of expansion (\ref{taylor})
contribute. However, with the modified bypass prescription for the
distant poles the integration contour $\beta$ in Fig.~\ref{diagr}b can be
shrunk to a contour $\gamma$ (maroon) on which $|k_0|\!<\!2m_c$ holds
everywhere. For the integral over contour $\gamma$ 
the series converges absolutely and uniformly, still embracing only the $k_0\!=\!|\vec k|$ pole. This
proves that the sum of the series obtained integrating term by term 
Eq.~(\ref{taylor}) gives the integral over $k_0$ of the original
expression in Eq.~(\ref{36}) yet with the modified bypass for the
distant poles along contour $\beta$.

Let us make the following general remark.
At first glance, the terms in the expansion of the Feynman diagrams
and in the OPE series 
assume a somewhat different form:
the expectation values of the operators in the kinetic scheme 
are given by  
three-dimensional integrals over spatial momentum of an on-shell gluon, 
$k_0=|\vec{k}\,|$, being defined through the heavy quark structure
functions:
\beq
\int \frac{{\rm d}^3 \vec k}{(2\pi)^3 2|\vec{k}\,|} 
\:{\cal P}(\vec{k})
\label{ope}
\eeq
with ${\cal P}(\vec k)$ a polynomial.
It is clear, however, that the two representations 
have the same form once the $k_0$ integration is performed closing the
contour in the lower half-plane. That is why integration over $k_0$
plays an important role in our reasoning. 

At this point it becomes transparent why the OPE series yield 
the expansion of the sole
contribution of the near pole with the on-shell gluon:
the heavy quark propagators appearing in the definition of the heavy quark 
expectation values in the effective theory are nonrelativistic (static) 
propagators which have a single pole. The second, distant, pole peculiar to
relativistic particles is absent from them. Therefore, at any finite order in
the OPE power expansion there is no contributions associated with the
distant poles, $k_0\!\approx\! 2m_Q$ in Eq.~(\ref{38}).
In this sense the distant singularities are
related to the divergence of the expansion in $k/m_Q$  rather than to
the discontinuity of the individual terms.

So far we have considered the vertex diagram. The other two Feynman diagrams with
the wavefunction renormalization for external quark legs have the  same
structure; they only depend on a single quark mass $m_Q\!=\!m_b$ or $m_c$, and 
the two pairs of the fermion propagator poles are degenerate. Consequently all
the above reasoning applies to them as well. 

Let us note that taking $\mu\!=\!\varepsilon_M$ the difference between
$\mu'$ and $\varepsilon_M$ becomes power-suppressed, the last term
of Eq.~(\ref{40.2}) becomes of order $1/m_Q^3$ and can be neglected to
the leading order $\mu^2/m_Q^2$. Therefore, it accounts for the recoil
correction in the relation between $\mu$ and $\varepsilon_M$ and 
becomes relevant where the terms $O(\alpha_s\mu^3/m_Q^3)$ are
included. Its form must already be clear from the preceding
derivation: in the soft gluon ensemble the emission of a gluon with
energy $\omega$ yields an excitation energy  
$\varepsilon\!=\!\omega\!+\!\sqrt{m_c^2\!+\!\omega^2}\!-\!m_c$ in the final
state, and the spectral density is 
\beq
w^{\rm soft}(\varepsilon)=w^{\rm pert}(\varepsilon)
\,\theta(\mu'\!-\!\varepsilon), 
\qquad \mu'\!=\!\mu\!+\!\sqrt{m_c^2\!+\!\mu^2}\!-\!m_c.
\label{32a}
\eeq
This explains the integration limit in the last term in Eq.(\ref{40.2}).

The above discussion of the one-loop corrections is directly extended to
incorporate any higher-order BLM corrections as well, 
or even to perform the complete BLM-summation. As detailed in 
Appendix \ref{blmstuff}, the same analysis must be repeated for the diagrams with an
arbitrary gluon mass $\lambda$. This has been stated already in Sect.~\ref{pert}
where the related technical modifications were listed. The necessary one-loop expressions
at non-zero $\lambda^2$ are given in Appendix \ref{pertdetails}.

\subsection{Details of the perturbative calculation}
\label{pertdetails}

In this Appendix  we provide  details 
omitted from the main text in Sect.~\ref{pert}. In the case of a massive gluon
the one-loop inelastic perturbative spectral density determining
the recoil correction is  given by 
\bea
\nonumber
  w^{{\rm pert}}(\varepsilon) 
& \msp{-5}= \msp{-5}&
C_F\frac{\alpha_s}{\pi}\,\frac{((M- m_c)^2  - \lambda^2)
  \sqrt{(M^2\!-\!(m_c\!+\!\lambda)^2)(M^2\!-\!(m_c\!-\!\lambda)^2)}}{12\,M^3 (M-m_c)^2 \left(m_b\,M^2 -M \lambda^2+m_b(\lambda^2-m_c^2)\right)^2} \, 
\theta(\varepsilon\!-\!\lambda)
\times\\
&& \nonumber
\Big[2 M^6 - 4 M^3 (4 m_b + m_c) \lambda^2 + 
 4 M  \lambda^2 (2 m_b + m_c)(m_c^2 - \lambda^2) 
 \\&&+ (3 m_b^2 + 2 m_b m_c + \nonumber
    m_c^2) (m_c^2 - \lambda^2)^2 + M^4 (3 m_b^2 + 2 m_b m_c - 3 m_c^2 + 4 \lambda^2)
    \\&& +
  M^2 (  4 m_c^2 \lambda^2 -4 m_b m_c^3+ 6 
\lambda^4 - 6 m_b^2 (m_c^2 - 2 \lambda^2)\Big].
\label{46}
\eea
Here $M\!=\!m_c\!+\!\varepsilon$ is the invariant mass of the final state. 
The one-loop zero-recoil
axial current renormalization without a cutoff $\eta_A\!=\!1\!+\!C_F
\frac{\as}{\pi} \eta_A^{(1)}(\lambda^2)+O(\as^2)$ is given by 
\bea
\eta_A^{(1)}(\lambda^2) \msp{-3.5}&\!=\!&
\!\!-\frac{9y + 2yz + 24 + 9yz^2}{24}-\frac{9y^2 - 7y^2z - 6y - 6yz
+ 18 + 18z}{24(1-z)}\,\ln z\nonumber\\
&&\!\! +\frac{y\left( 9y + 2yz - 6 + 2yz^2 - 12z +
     2yz^3 - 6z^2 + 9yz^4 \right)}{48}\,(\ln{y}\!+\!2\ln{z})\nonumber\\
&&\!\! -\frac{9y^3 - 7y^3z - 24y^2 + 8y^2z + 12y + 44yz -
96}{48y(1-z)\sqrt{1-4/y}}\,
\ln{\frac{1+\sqrt{1-4/y}}{1-\sqrt{1-4/y}}}
\label{etaalam}
\\
&& \!\!-\frac{-44y - 8y^2z^2 + 7y^3z^4 + 96/z - 12yz + 24y^2z^3 - 9y^3z^5}{48y
(1-z)\sqrt{1-4/yz^2}}\,\ln{\frac{1+\sqrt{1-4/yz^2}}{1-\sqrt{1-4/yz^2}}}\,.
\nonumber
\eea
Here we have used $y\!=\!\lambda^2/m_c^2$ and $z\!=\!m_c/m_b$. 
This expression coincides with the one in Eq.~(B3) of Ref.~\cite{bbbsl}, where it was 
derived  by a dispersion integral starting from the  Euclidean calculation of 
Ref.~\cite{neubflaw}.

The check of the $\mu$-independence of the OPE 
sum rule, Eq.~(\ref{88}), with the
perturbative factor calculated according to Eq.~(\ref{40.2}) can be accomplished
to an arbitrary BLM order at once: it suffices to establish it at a given value
of the gluon mass.  We demonstrate it here assuming $\varepsilon_M\!=\!\mu$.
The right-hand side of the sum rule depends on
$\mu$ through  $\xi_A^{\rm pert}(\mu)$ and  through
the heavy quark expectation values. Moreover, since all the $\mu$-dependent components 
of the perturbative calculation, including the expectation values and
the  inelastic integral, can
be written as three-dimension integrals 
\beq
\int \frac{{\rm d}^3 k}{(2\pi)^3 2k_0}\, \theta (\mu- k_0)\dots \,, \qquad k_0\equiv
\sqrt{\vec k^{\,2}\!+\!\lambda^2 } 
\label{int77}
\eeq
with the same cutoff  provided $\varepsilon_M\!=\!\mu$, we may simply
check the cancellation at the integrand
level at a given value of $\vec{k}^{\,2}$ and $\lambda^2$.

As the $\mu$-dependence of the individual contributions starts at
${\cal O}(1/{m_Q^2})$, the first check is provided by
the terms $\alpha_s \mu^2/m_Q^2$ \cite{blmope}. 
At this order the power-suppressed component of the OPE part of the sum rule (\ref{88})
includes only the expectation
values $\mu_\pi^2$ and $\mu_G^2$ which should be considered in the static
limit; the latter then vanishes. The required expressions have been given 
in Refs.~\cite{blmvcb,imprec}. In units of 
\beq
C_F g_s^2 \int \frac{{\rm d}^3 \vec k}{(2\pi)^3 2k_0}
\label{4030}
\eeq
we get 
\beq
2+\frac{\lambda^2}{k_0^2}\,, \qquad
\left(\frac{1}{m_c^2}\!+\!\frac{2}{3m_c m_b}\!+\!\frac{1}{m_b^2}\right)
\frac{\vec k^{\,2}}{2k_0^2} +\left(\frac{1}{m_c}\!-\!\frac{1}{m_b}\right)^2
\frac{\lambda^2\vec k^{\,2}}{4k_0^4} 
\label{mu2}
\eeq
for  $\mu_\pi^2$ and $w^{\rm pert}$, respectively.

Finally, the subtracted soft piece of $\eta_A$ depends on $\mu$. 
The expression for $\delta  \eta_A^{\rm soft}(\mu)$ given in Eq.~(\ref{44a.2}) 
must be expanded in $\vec k$ and $\lambda^2$:
\bea
\nonumber
\delta  \eta_A^{\rm soft} \msp{-4} & = & \msp{-4} C_F g_s^2 \!
\int \!\!\frac{{\rm d}^3 k}{(2\pi)^3 2k_0} \left\{\!
-\frac{1}{2}\left[\!\left(\frac{1}{m_c^2}\!+\!\frac{2}{3m_c
  m_b}\!+\!\frac{1}{m_b^2}\right)\! - \! 
\frac{\lambda^2}{k_0^2} \frac{2}{3m_c m_b} - 
\frac{\lambda^4}{4k_0^4} \left(\frac{1}{m_c}\!-\! \frac{1}{m_b}\right)^2
\right] \right. \\
& &\msp{-15}
\left. -
  \frac{\lambda^2}{2k_0}\left(\!\frac{1}{m_c}\!+\!\frac{1}{m_b}\!\right)
\left[\!\left(\!\frac{1}{m_c^2}\!-\!\frac{2}{3m_c  m_b}\!+\!\frac{1}{m_b^2}\!\right)  
\!-\!\frac{\lambda^2}{k_0^2}\frac{1}{3m_c m_b} \!-\!
\frac{\lambda^4}{4k_0^4}\left(\!\frac{1}{m_c}\!-\! \frac{1}{m_b}\!\right)^2
\right]\!+ ... \right\}\! , \,
\label{4084}
\eea
where the first square bracket upon integration gives the leading
$\mu^2/m_Q^2$ terms and the second line yields $\mu^3/m_Q^3$. 
Combining the above coefficients the $\mu$-independence of the 
sum rule is verified at order $\mu^2/m_Q^2$.

A far deeper check is encountered at the level of $\mu^3/m_Q^3$ terms. 
At this order the proper prescription to calculate the soft
virtual correction  $\delta \eta_A^{\rm pert}$ discussed 
in Appendix B becomes essential. At
this level one also needs to include the higher-dimension Darwin expectation value
as well as the $1/m_b$ effects in the kinetic and chromomagnetic expectation 
values. The latter are expressed in terms of the local $\rho_D^3$ and 
$\rho_{LS}^3$ and of the nonlocal expectation values. Perturbatively $\rho_{LS}^3$
vanishes as do $\rho_{\pi G}^3$ and $\rho_A^3$.
The subleading term in the continuum spectral density is also required and  
can be obtained directly expanding Eq.~(\ref{46}).

By virtue of the SV sum rules the perturbative component of the Darwin
expectation value amounts to an integral of the same form (\ref{int77}), 
with the integrand given by the product of the integrand for $\mu_\pi^2$ and
$k_0$, the excitation energy for a static quark:
\beq
k_0(2+\mbox{$\frac{\lambda^2}{k_0^2}$})\,.
\label{darwin}
\eeq
Since $\mu_G^2$  was defined in Eq.~(\ref{102}) to include only the magnetic
field piece $-\vec \sigma \cdot\vec B$ but not the lower-component term
in the complete chromomagnetic operator, to  order $\mu^3/m_Q$ it acquires a
contribution only from the nonlocal correlator $\rho_S^3$,
\beq
\delta_{\alpha_s} \mu_G^2 = \mbox{$\frac{1}{m_b}$}\delta_{\alpha_s} \rho_S^3.
\label{4110}
\eeq
The value of  $\mu_\pi^2$ is perturbatively corrected, instead, by the correlator 
$\rho_{\pi\pi}^3$:
\beq
\delta_{\alpha_s} \mu_\pi^2 = 2+\frac{\lambda^2}{k_0^2}-\mbox{$\frac{1}{m_b}$}\delta_{\alpha_s} 
\rho_{\pi\pi}^3,
\label{4112}
\eeq
where overal integration (\ref{4030}) is understood.

\thispagestyle{plain}
\begin{figure}[t]
 \begin{center}
 \includegraphics[width=10cm]{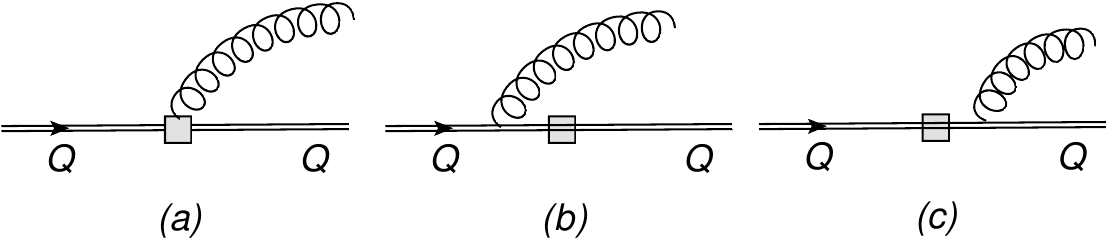}
 \end{center}\vspace*{-5pt}
\caption{ \small One-gluon amplitude diagrams involved in the calculation of the
 $\,{\cal O}(\alpha_s)\,$  perturbative spectral densities in the
 static limit. The square block denotes the operator in question.
}
\label{nlocdiagr}
\end{figure}

To find one-loop $\delta_{\alpha_s} \rho_{\pi\pi, S}^3$ or, more generally,
all the perturbative spectral densities in Eqs.~(\ref{130dec}), (\ref{3130})
one needs to square the sum of the
diagrams shown in Fig.~\ref{nlocdiagr}. The answer is obtained immediately, in
particular if a little trick \cite{blmope} is used, which eliminates  
two of the three diagrams, $b$ and $c$, for the transition amplitude:
\beq
\rho_0(\omega)= C_F\frac{\alpha_s}{\pi} 
\lambda^2 \frac{\left(\omega^2 \!-\!\lambda^2\right)^\frac{3}{2}}{\omega^2}, \quad
\rho_1(\omega)=C_F\frac{8\alpha_s}{3\pi} 
(\omega^2\!-\!\lambda^2)^\frac{3}{2}, \quad 
\rho_2(\omega)=C_F\frac{\alpha_s}{\pi} 
\frac{(\omega^2\!-\!\lambda^2)^\frac{5}{2}}{\omega^2}.
\label{4114}
\eeq
The trick uses gauge invariance to say that one can use 
the simple gluon propagator
\beq
\frac{\delta_{ij}-\frac{k_i k_j}{k_0^2}}{\lambda^2-k^2}
\label{4115}
\eeq
requiring to calculate only the spatial vertices; they come from the vertex
emission but not from the heavy quark lines.
Using Eqs.~(\ref{130dec}), (\ref{3130}) and (\ref{150}) we obtain at once,
for instance,
\beq
\delta_{\alpha_s} \rho_{\pi\pi}^3= \frac{\lambda^2 \vec k^{\,2}}{k_0^3}\,, \qquad
\delta_{\alpha_s} \rho_{S}^3=  2\frac{\vec k^{\,2}\!}{k_0}\,,
\label{4116}
\eeq
in the same units of Eq.~(\ref{4030}). 

Finally, one needs the ${\cal O}(\mu^3/m_Q^3)$ term in the expansion of 
$\delta\eta_A^{\rm soft}$;  it is given in the second line of Eq.~(\ref{4084}). 
Expanding Eq.~(\ref{46}) through  the next-to-leading order, on one hand, and 
collecting all
relevant terms from Eqs.~(\ref{4084}),  (\ref{darwin}) and
(\ref{4116}) with the explicit coefficients appearing in the sum rule, on the
other hand, we arrive at the same integrand in the both sides of 
the sum rule (\ref{88}) at  $O(1/m_Q^3)$, for arbitrary value of
$\lambda^2$.

For bookkeeping purposes we quote here the perturbatively calculated moments of the
spectral densities $\rho_0(\omega)$, $\rho_1(\omega)$ and
$\rho_2(\omega)$, as 
well as those of the $P$-wave spectral density $\rho^{\rm P}(\omega)$, defined 
in analogy to Eq.~(\ref{3130})
by the perturbative relation 
$$
{\cal T}^{\frac{1}{2}^-}(\omega) = {\cal T}^{\frac{3}{2}^-} (\omega)= \frac19\, \rho^P(\omega),
$$ 
see Eq.({\ref{2270.02}).
Their expressions to order $\alpha_s$ and $\beta_0\alpha_s^2$ are: 
\bea
\nonumber
\int_0^\mu \omega^k {\rm d}\omega\, \rho_0(\omega) \msp{-4} & = & \msp{-4} 
 -\frac{1}{5}C_F \frac{\beta_0}{2} \left(\frac{\alpha_s}{\pi}\right)^2 
\frac{\mu^{4+k}}{4+k}\\
\nonumber
\int_0^\mu \omega^k {\rm d}\omega\, \rho_1(\omega) \msp{-4} & = & \msp{-4} 
 \frac{8}{3}C_F \frac{\alpha_s(M)}{\pi}\frac{\mu^{4+k}}{4+k}
\left[1 +   \frac{\beta_0}{2} \frac{\alpha_s}{\pi}
\left(
\ln{\frac{M}{2\mu}}+ \frac{13}{6}+\frac{1}{4+k}
\right) 
\right]\\
\nonumber
\int_0^\mu \omega^k {\rm d}\omega\, \rho_2(\omega) \msp{-4} & = & \msp{-4} 
\;\;\,C_F \frac{\alpha_s(M)}{\pi}\frac{\mu^{4+k}}{4+k}
\left[1 +   \frac{\beta_0}{2} \frac{\alpha_s}{\pi}
\left(
\ln{\frac{M}{2\mu}}+ \frac{71}{30}+\frac{1}{4+k}
\right) 
\right]\\
\int_0^\mu \omega^k {\rm d}\omega\, \rho^{\rm P}(\omega) \msp{-4} & = & \msp{-4} 
\:2C_F \frac{\alpha_s(M)}{\pi}\frac{\mu^{2+k}}{2+k}
\left[1 + \frac{\beta_0}{2} \frac{\alpha_s}{\pi}
\left(
\ln{\frac{M}{2\mu}}+ \:\frac{5}{3}\: +\frac{1}{2+k}
\right) 
\right].
\label{4128}
\eea
The moments of $\rho^{\rm P}(\omega)$ determine, beyond the IW slope,
$\La$, the kinetic and the Darwin expectation values, the higher
expectation values $m_2$ and $r_1$.

For completeness, we give here also the perturbative dependence on $\mu$ 
of $\rho_{LS}^3$ associated with power mixing: 
\beq
\frac{\rm d}{{\rm d} \mu} (-\rho_{LS}^3) =  C_A \scalebox{1.25}{$\frac{\alpha_s}{\pi}$}
\mu_G^2 + 
C_A \scalebox{1.25}{$\frac{\alpha_s}{2\pi}$} \,
\frac{1}{\mu}\rho_{LS}^3  +{\cal O}\left(\alpha_s^2\right)
\label{165a}
\eeq
(the anomalous dimension of the spin-orbit operator coincides with that of the
full chromomagnetic one). This implies 
the relation between the 
extrapolated `pole-scheme' value $-\tilde \rho_{LS}^3$ and the 
Wilsonian $-\rho_{LS}^3(\mu)$:
\beq
-\tilde \rho_{LS}^3 =  - \rho_{LS}^3(\mu) -C_A \scalebox{1.25}{$\frac{\alpha_s}{\pi}$} \mu\,
\mu_G^2 +{\cal O}\left(\alpha_s^2\right).
\label{165}
\eeq

\subsection{Transitions to the `radial' states and the inclusive yield}
\label{aprad}

The inelastic zero-recoil spectral density for the {\sl vector} 
$\bar{c}\gamma_0 b$ current paralleling Eq.~(\ref{152.02}) is
\beq
\frac{1}{2\pi i}\,
{\rm disc}\,  T^{(V)}_{\rm zr}(\omega)
\equiv 
w^{(V)}_{\rm inel}(\omega)
\!=\! \left(\frac{1}{2m_c}\!-\!\frac{1}{2m_b}\right)^2 
\frac{\rho_{p}^{(\frac{1}{2}^+)}(\omega) \!-\! 2\rho_{pg}^{(\frac{1}{2}^+)}(\omega)
\!+\!\rho_{g}^{(\frac{1}{2}^+)}(\omega) }{\omega^2} ; 
\label{152v}
\eeq
no transitions into $\frac{3}{2}^+$ occur. It is manifestly BPS-suppressed to
the second order.\vspace*{3pt} 

As stated  in Section \ref{yield},
we evaluate the rate for a decay of the $\Omega_0$
spin-$\frac{1}{2}$ heavy state into the corresponding excited
half-integer--spin multiplets. The weak current coupling of these fictitious hadrons
is fixed by the corresponding transition probabilities near zero recoil.
Namely, the (unpolarized) zero-recoil structure functions are expressed 
through the effective transition
amplitudes; on the other hand, they are given by the $1/m_Q$ expansion, in our
case to the second order, of the actual $B$-meson zero-recoil semileptonic 
structure functions, cf.\ Eqs.~(\ref{152.02}), (\ref{152v}). 

There are more structure functions for a particle with spin than for actual
$B$ mesons. However, considering the unpolarized states (for instance,
averaging over spin) they are reduced to the standard ones, and we use the
same notations for them for decays of $\Omega_0$ as for $B$ mesons in QCD.
$V$-$A$ interference
(which is not relevant here) would require a nonvanishing recoil kinematics.

The unpolarized structure functions for the transitions into the 
$\frac{1}{2}^+$ states are the same as the tree-level ones for $B$ decays 
\cite{koyrakh}:
\beq\label{9010}
 w_1 \!=\! g_A^2\frac{(m_1\!+\! m_2)^2\!-\!q^2}{2}+
g_V^2\frac{(m_1\!-\! m_2)^2\!-\!q^2}{2}, 
\quad  w_2 \!=\!2(g_A^2\!+\!g_V^2)m_1^2,
\quad  w_3 \!=\!2g_A g_V m_1, 
\eeq
with the overall factor $\frac{\pi}{m_1^2}\delta(q_0\!-\!\frac{m_1^2-m_2^2+q^2}{2m_1})$,
where we have used the notation of Eq.~(\ref{2710.02}), namely 
\beq
\matel{\mbox{$\frac{1}{2}^+$}}{\bar c \gamma_\mu\gamma_5 b}{\Omega_0}= 
g_A\,\bar \chi \gamma_\mu\gamma_5 \Psi_0, \qquad
\matel{\mbox{$\frac{1}{2}^+$}}{\bar c \gamma_\mu b}{\Omega_0}= 
g_V\,\bar \chi \gamma_\mu \Psi_0\,.
\label{9012}
\eeq
Note that here we use the full bispinors and assume their relativistic
normalization. In this context the mass $m_1$ refers to the ground 
state in the beauty sector,
$m_1\!\simeq \! m_b\!+\!\La$ while $m_2$ to the excited state for charm, 
$m_2\!\simeq \! m_c\!+\!\La\!+\!\varepsilon_{\rm rad}$. Only 
$w_{1,2}$ contribute to the total width;  it is given in
Eqs.~(\ref{2240.02}), (\ref{992}).

For transitions into $\frac{3}{2}^+$ radial excitations we employ
\beq
\matel{\mbox{$\frac{3}{2}^+$}}{\bar c \gamma_\mu\gamma_5 b}{\Omega_0}= 
g_A\,\bar \chi_\mu  \Psi_0, \quad
\matel{\mbox{$\frac{3}{2}^+$}}{\bar c \gamma_\mu b}{\Omega_0}= 
g_V\,\bar \chi_\mu i\gamma_5 \Psi_0\,,
\label{9020}
\eeq
where $\chi_\mu$ are likewise fully relativistic Rarita-Schwinger
wavefunctions. (The vector matrix element  is considered
only for completeness; its transition amplitudes into $\frac{3}{2}^+$ states
involve further suppression: either more powers of velocity, or extra
$1/m_Q$ or an additional overall $\alpha_s$.) With this convention we
have 
\begin{alignat}{2}
\nonumber
& w_1 \!=\! \left(\!g_A^2\frac{(m_1\!+\! m_2)^2\!-\!q^2}{3m_1^2}+
g_V^2\frac{(m_1\!-\! m_2)^2\!-\!q^2}{3m_1^2}\!\right)\,
\pi\delta(q_0\!-\!\mbox{$\frac{m_1^2-m_2^2+q^2}{2m_1}$}), &
& \quad w_2 \!=\! \frac{m_1^2}{m_2^2}w_1, \\
& w_3 \!=\!0, \qquad\qquad\qquad \qquad  \qquad 
w_4 \!=\! m_1^2 \,w_1, & &
 \quad w_5\!=\!-\frac{m_1}{m_2^2}\,w_1. 
\label{9022}
\end{alignat}
Integrating over the full phase space we obtain the corresponding width
\beq
\Gamma^{(\frac32)}_{A,V}=\frac{G_F^2 M_0^5 |V_{cb}|^2}{192\pi^3} \, g_{A,V}^2 \:z^{(\frac32)}_{A,V}(r), 
\qquad r\!=\!\frac{M^2_{\frac{3}{2}}}{M_0^2}, 
\label{9024}
\eeq
with the phase space factors 
\bea
\nonumber
\msp{-10}  z^{(\frac{3}{2})}_{A,V}(r)\!=\! \frac{1}{2}\left(z^{(\frac{3}{2})}_0(r)\!\pm\! 
\tilde z^{(\frac{3}{2})}_0(r)\right)\!, \quad
z^{(\frac{3}{2})}_0(r) \msp{-4}&=&\msp{-4} 
\frac{1}{15r} -2r -\mbox{$\frac{2}{3}$}r^2 +3r^3-\mbox{$\frac{2}{5}$}r^4 -4r^2
\ln{r}, 
\\
\tilde z^{(\frac{3}{2})}_0(r) \msp{-4}&=&\msp{-4} \frac{1}{6\sqrt{r}} \!+\!
r^{3/2} \left(
12 \!-\!\mbox{$\frac{32}{3}$}r \!-\!\mbox{$\frac{3}{2}$}r^2 \!+\!(6\!+\!8r)\ln{r}
\right) \quad
\label{9026}
\eea
The  singularity at $r\tto 0$ reflects here the ultraviolet problems 
of point-like higher-spin particles. 
Note that all the components of the vector transition amplitude vanish 
at zero recoil; the zero-recoil analysis does not constrain $g_V$ for 
$\frac{3}{2}^+$. 
However, for this very reason it is generally suppressed 
by a higher power of the SV parameter $\Delta/M$:
$$
\frac{M_0^5}{192\pi^3} \,z^{(\frac{3}{2})}_{A}(r) \simeq \frac{\Delta^5}{30\pi^3}, \qquad
\frac{M_0^5}{192\pi^3} \,z^{(\frac{3}{2})}_{V}(r) \simeq \frac{11\,\Delta^2}{84\,M_0^2}\:
\frac{\Delta^5}{30\pi^3}
\quad \mbox{ at } \Delta \ll M_0.
$$
Numerically, the  yield in the vector channel  is negligible for
any relevant ratio $M_{\frac{3}{2}}/M_0$; the vector current does
not contribute to the production of $\frac{3}{2}^+$ in our approximation. 
 
The effective formfactors at zero recoil are obtained comparing the structure
functions in Eqs.~(\ref{9010}) or (\ref{9022}) in the zero-recoil kinematics
(which is $q^2\!=\!q_0^2\!=\!(m_1\!-\!m_2)^2$) with those in Eqs.~(\ref{152.02}),
(\ref{152v}). In this way the constants 
$g_A$, $g_V$ for  $\frac{1}{2}^+$ and  $g_A$ for
$\frac{3}{2}^+$ above are expressed through the amplitudes introduced in
Sec.~\ref{modind}: 
\beq
\varepsilon g_A^{(\frac{1}{2})}\!=\!  
\mbox{$\left(\!\frac{1}{m_c}\!-\!\frac{1}{m_b}\!\right)
\frac{P\!-\!G}{2}$}
+ \mbox{$\frac{2G}{3m_c}$}
, \qquad \varepsilon g_V^{(\frac{1}{2})} \!=\! 
\mbox{$\left(\!\frac{1}{m_c}\!-\!\frac{1}{m_b}\!\right)
\frac{P\!-\!G}{2}
$}, \qquad
\varepsilon g_A^{(\frac{3}{2})} \!=\! \mbox{$\frac{1}{\sqrt{6}m_c}$}g ,
\label{9030}
\eeq
where $\varepsilon$ is the mass gap for a particular state.


\begin{thebibliography}{99}

\bibit{f0short}
P.~Gambino, T.~Mannel and N.~Uraltsev, {\it Phys.\ Rev.\ }{\bf D81} 113002 (2010).

\bibit{optical}
I.I.~Bigi, M.~Shifman, N.~Uraltsev and A.~Vainshtein,
Phys.\ Rev.\  D {\bf 52} (1995) 196.

\bibit{hiord}
T.\,Mannel, S.\,Turczyk and N.\,Uraltsev,
{\it  JHEP} {\bf 1007} (2010) 109.

\bibit{five}
I.I.~Bigi, M.~Shifman, N.~Uraltsev and A.~Vainshtein,
Phys.\ Rev.\  D {\bf 56} (1997) 4017.

\bibit{dipole}
A.~Czarnecki, K.~Melnikov and N.~Uraltsev,
Phys.\ Rev.\ Lett.\  {\bf 80} (1998)  3189.

\bibit{chrom}
N.~Uraltsev,
Phys.\ Lett.\  B {\bf 545} (2002) 337.

\bibit{imprec}
D.\,Benson, I.\,Bigi, Th.\,Mannel and N.\,Uraltsev,
{\it Nucl.\ Phys.}\ {\bf B665} (2003) 367.

\bibit{blmvcb}
N.\,Uraltsev,
{\it Mod.\,Phys.\,Lett.}\ {\bf A17} (2002) 2317.

\bibit{xi2}
A.~Czarnecki, K.~Melnikov and N.~Uraltsev,
Phys.\ Rev.\  D {\bf 57} (1998) 1769.

\bibit{vcb}
M.A.~Shifman, N.G.~Uraltsev and A.I.~Vainshtein,
Phys.\ Rev.\  D {\bf 51} (1995) 2217.

\bibit{rev}
I.~Bigi, M~Shifman, N.G.~Uraltsev,
{\it Ann.\,Rev.\,Nucl.\,Part.\,Sci.}\ {\bf 47} (1997)
591.

\bibit{HFAG} 
Heavy Flavour Averaging Group, update for the Winter 2009 conferences, see \\
{\tt http://www.slac.stanford.edu/}.

\bibit{fit} 
P.~Gambino and C.~Schwanda,
 arXiv:1102.0210 [hep-ex].

\bibit{NNLOmoments}
A.~Pak and A.~Czarnecki,
Phys.\ Rev.\ D {\bf 78} (2008) 114015; \\
S.~Biswas and K.~Melnikov,
JHEP {\bf 1002} (2010) 089;\\
P.~Gambino,  
JHEP 9 (2011) 055.  

\bibit{newfit}P.~Gambino and C.~Schwanda, to appear.

\bibit{mcdet}
 K.G.~Chetyrkin 
 {\it et al.},
 Phys.\ Rev.\  {\bf D80 } (2009)  074010
 [arXiv:0907.2110 [hep-ph]] and  arXiv:1010.6157 [hep-ph].

\bibit{icsieg}
I.\,Bigi, T.\,Mannel, S.\,Turczyk and N.\,Uraltsev,  arXiv:0911.3322;   
{\it  JHEP} {\bf 1004} (2010) 073.

\bibit{tau32}
D.~Becirevic {\it et al.},
 Phys.\ Lett.\  B {\bf 609} (2005) 298.

\bibit{Belle}
A.~Kuzmin  et al.\ [Belle Collaboration],
Phys.\ Rev.\  D {\bf 76} (2007) 012006; 
Nucl.\ Phys.\ Proc.\ Suppl.\  {\bf 162} (2006) 228.

\bibit{BPS}
N.~Uraltsev,
Phys.\ Lett.\  B {\bf 585} (2004) 253.

\bibit{chiral}
 R.~Casalbuoni, A.~Deandrea, N.~Di Bartolomeo, R.~Gatto, F.~Feruglio and G.~Nardulli,
 Phys.\ Rept.\  {\bf 281} (1997) 145
 [hep-ph/9605342].

\bibit{khod}
V.M.~Belyaev, V.M.~Braun, A.~Khodjamirian and R.~Ruckl,
 Phys.\ Rev.\  D {\bf 51} (1995) 6177.
\bibit{latticeDpi}
 D.~Becirevic, B.~Blossier, E.~Chang and B.~Haas,
 Phys.\ Lett.\ B {\bf 679} (2009) 231
 [arXiv:0905.3355 [hep-ph]].

\bibit{ioffe}
N.~Uraltsev,
arXiv:hep-ph/0010328.  Published in the Boris Ioffe Festschrift `At the 
Frontier of Particle Physics\,/\,Handbook of QCD', 
eds.\ by M.~Shifman (World Scientific, Singapore, 2001), vol.\,3
p.\,1577.

\bibit{newsr}
N.~Uraltsev, Phys.\ Lett.\ B {\bf 501} (2001) 86.

\bibit{czargroz}
A.~Czarnecki and A.G.~Grozin,
Phys.\ Lett.\ B {\bf 405}, 142 (1997)  [Erratum-ibid.\ B {\bf 650}, 447 (2007)].

\bibit{leib}
A.~K.~Leibovich, Z.~Ligeti, I.~W.~Stewart and M.~B.~Wise,
{\it Phys.\ Rev.\ Lett.\ } {\bf 78} (1997) 3995; 
{\it Phys.\ Rev.\ } D {\bf 57} (1998) 308.
\bibitem{bernturcz}
 F.~U.~Bernlochner, Z.~Ligeti and S.~Turczyk,
 arXiv:1202.1834 [hep-ph].
 
\bibitem{memorino}
  I.~I.~Bigi, B.~Blossier, A.~Le Yaouanc, L.~Oliver, O.~Pene, J.~-C.~Raynal, A.~Oyanguren and P.~Roudeau,
  Eur.\ Phys.\ J.\ C {\bf 52} (2007) 975
  [arXiv:0708.1621 [hep-ph]].

\bibit{fazio} 
P.~Colangelo, F.~De Fazio, S.~Nicotri and M.~Rizzi,
Phys.\ Rev.\ D {\bf 77}, 014012 (2008).
\bibit{radials}
P.~del Amo Sanchez {\it et al.}  [BABAR Collaboration],
Phys.\ Rev.\ D {\bf 82}, 111101 (2010).
[arXiv:1009.2076 [hep-ex]].

\bibit{habil}
N.~Uraltsev,  ``Dynamic Heavy Quark Theory in Quantum
Chromodynamics'', Habilitation Thesis; St.\,Petersburg, 2007, 221p.

\bibit{randwise}
 L.~Randall and M.~B.~Wise,
 Phys.\ Lett.\ B {\bf 303} (1993) 135.
\bibit{ademgatto}
 M.~Ademollo and R.~Gatto,
 Phys.\ Rev.\ Lett.\  {\bf 13} (1964) 264.

\bibit{laiho}
J.~Laiho [Fermilab Lattice and MILC and MILC Collaborations],
 PoS LAT {\bf 2007} (2007) 358
 [arXiv:0710.1111 [hep-lat]];
  C.~Bernard {\it et al.}, 
 Phys.\ Rev.\ D {\bf 79} (2009) 014506
 [arXiv:0808.2519 [hep-lat]].

\bibit{FNALnew}
J.~A.~Bailey {\it et al.}  [Fermilab Lattice and MILC Collaboration],
 PoS LATTICE {\bf 2010} (2010) 311
 [arXiv:1011.2166 [hep-lat]].

\bibit{Rome-TOVstar}
G.~M.~de Divitiis, R.~Petronzio and N.~Tantalo,
 Nucl.\ Phys.\ B {\bf 807} (2009) 373
 [arXiv:0807.2944 [hep-lat]].

\bibit{FNAL1}
S.~Hashimoto  {\it et al.}, 
 Phys.\ Rev.\ D {\bf 66} (2002) 014503
 [hep-ph/0110253].

\bibit{FNAL2}
A.~S.~Kronfeld,
 Phys.\ Rev.\ D {\bf 62} (2000) 014505
 [hep-lat/0002008].

\bibit{FNALold}
S.~Hashimoto, A.~S.~Kronfeld, P.~B.~Mackenzie, S.~M.~Ryan and J.~N.~Simone,
Phys.\ Rev.\ D {\bf 66}, 014503 (2002).

\bibit{fpluslattice}
M.~Okamoto {\it et al.},
Nucl.\ Phys.\ Proc.\ Suppl.\  {\bf 140} (2005) 461
 [hep-lat/0409116].

\bibit{Rome-TOV}
G.~M.~de Divitiis, E.~Molinaro, R.~Petronzio and N.~Tantalo,
 Phys.\ Lett.\ B {\bf 655} (2007) 45
 [arXiv:0707.0582 [hep-lat]].
\bibit{Bailey:2012rr}
 J.~A.~Bailey  {\it et al.},
 arXiv:1202.6346 [hep-lat].
\bibit{bbbsl}
P.~Ball, M.~Beneke and V.~M.~Braun,
{\it Phys.\ Rev.\ }  D {\bf 52} (1995) p.\,3929-3948.


\bibit{neubflaw}
M.~Neubert, {\it Phys.\ Rev. } D{\bf 51} (1995) 5924.

\bibit{blmope}
N.G.~Uraltsev, {\it Nucl.~Phys.} {\bf B491} (1997) 303.
\bibit{koyrakh}
B.~Blok, L.~Koyrakh, M.~A.~Shifman and A.~I.~Vainshtein,
{\it Phys.\ Rev.\ } D {\bf 49} (1994) 3356
[Erratum-ibid.\  D {\bf 50} (1994) 3572].

\end{thebibliography}
\end{document}